\documentclass[12pt]{article} 
\usepackage{amsmath}
\usepackage{amssymb,amsfonts,euscript}
\usepackage{hyperref}
\renewcommand{\baselinestretch}{1.2}
\newcommand\nn{\nonumber}

\newcommand{\Ord}{{\cal{O}}}

\newcommand{\D}{{\cal D}}
\newcommand{\G}{{\cal G}}

\newcommand{\M}{{\cal M}}

\newcommand{\T}{{\cal T}}
\newcommand{\J}{{\cal J}}
\newcommand{\F}{{\cal F}}
\newcommand{\cL}{\cal{L}}
\newcommand{\lr}{{\cal{L}}_{\sgb(\s)} }

\newcommand{\bJ}{\bar{J}}
\newcommand{\bz}{\bar{z}}
\newcommand{\bw}{\bar{w}}
\newcommand{\bM}{\bar{M}}

\def\a{\alpha}
\def\be{\beta}
\def\l{\lambda}
\def\m{\mu}
\def\n{\nu}
\def\r{\rho}
\def\de{\delta}

\def\hQ{\hat{Q}}

\def\hg{\hat{g}}
\def\hgp{\hat{g}_+}
\def\hgm{\hat{g}_-}
\def\ga{\gamma}
\def\part{\partial}
\renewcommand{\sp}{,\hspace{.3in}}

\newcommand{\sa}{\mathop{\vtop{\ialign{##\crcr
  $\hfil\displaystyle{\longrightarrow}\hfil$\crcr\noalign{\kern-1pt\nointerlineskip}
  \hspace{.12in}$^\sigma$\hskip6pt\crcr\noalign{\kern3pt}}}}}
\newcommand{\slra}{\mathop{\vtop{\ialign{##\crcr
  $\hfil\displaystyle{\longleftrightarrow}\hfil$\crcr\noalign{\kern-1pt\nointerlineskip}
  \hspace{.12in}$^\sigma$\hskip6pt\crcr\noalign{\kern3pt}}}}}
\newcommand{\sat}{\mathop{\vtop{\ialign{##\crcr
  $\hfil\displaystyle{\longrightarrow}\hfil$\crcr\noalign{\kern-1pt\nointerlineskip}
  \hspace{.12in}$^\sigma$\hskip6pt\crcr\noalign{\kern3pt}}}}}
\newcommand{\pa}{\mathop{\vtop{\ialign{##\crcr
  $\hfil\displaystyle{\oplus}\hfil$\crcr\noalign{\kern+1pt\nointerlineskip}
  \hspace{.08in}$^{\alpha=0}$\hskip6pt\crcr\noalign{\kern3pt}}}}}
\newcommand{\pan}{\mathop{\vtop{ialgin{##\crcr
  $\hfil\displaystyle{\oplus}\hfil$\crcr\noaligan{\kern+2pt\nointerlinkeskip}
  \hspace{.03in} $^{\alpha}$\hskip6pt\crcr\noalign{\kern3pit}}}}}
\newcommand{\ka}{\mathop{\vtop{\ialign{##\crcr
  $\hfil\displaystyle{\longleftrightarrow}\hfil$\crcr\noalign{\kern-1pt\nointerlineskip}
  \hspace{.12in}$^K$\hskip6pt\crcr\noalign{\kern3pt}}}}}
\newcommand{\bp}{\mathop{\vtop{ialign{##\crcr
  $\hfil\displaystyle{}\hfil$\crcr\noalign{\kern-13pt\nointerlineskip}
  \big{(}\hskip0pt\crcr\noalign{\kern3pt}}}}}
\newcommand{\cbp}{\mathop{\vtop{ialign{##\crcr
  $\hfil\displaystyle{}\hfil$\crcr\noalign{\kern-13pt\nointerlineskip}
  \big{)}\hskip0pt\crcr\noalign{\kern3pt}}}}}

\newcommand{\+}{\hspace{-.03in}+\hspace{-.02in}}
\newcommand{\s}{\sigma}
\newcommand{\srange}{\sigma=0,...,N_c-1}

\renewcommand{\sp}{,\hspace{.3in}}

\newcommand{\p}{^\prime}

\newcommand{\w}{\omega}

\newcommand{\rmod}{\,{\textup{mod}}}
\newcommand{\sr}{\sqrt{\r}}

\newcommand{\hc}{$\hat{J}_{\gst}$}

\newcommand{\sgb}{{\mbox{\scriptsize{\gb}}}}
\newcommand{\sgbn}{{\mbox{\scriptsize{\gbn}}}}

\def\gb            {\mbox{$\hat{\mathfrak g}$}}

\def\gbn           {\mbox{$\mathfrak g$}}
\def\sm#1      {\mbox{\scriptsize $#1$}}

\def\sz        {\mbox{\scriptsize $\mathbb  Z$}}
\def\z         {\mbox{$\mathbb  Z$}}

\def\d             {\mbox{$\mathbb D$}}

\def\srac#1#2{\smal{\frac{#1}{#2}}}
\def\foot#1{\mbox{\footnotesize $#1$}}

\def\smal#1{\mbox{\small $#1$}}
\def\big#1{\mbox{\large $#1$}}
\def\Big#1{\mbox{\Large $#1$}}
\def\BIG#1{\mbox{\Huge $#1$}}
\def\g{{\cal{G}}}
\def\hsp#1{\hspace{#1in}}

\def\hjb{\hat{\bar{J}}}
\def\comment#1{\hsp{.3}\textup{#1}}
\newcommand{\mnb}{\overline{-n(r)}}
\newcommand{\nb}{\bar n(r)}

\def\hE{{\hat{E}}}
\def\hEb{{\hat{\bar{E}\hspace{.03in}}\hspace{-.03in}}}

\def\hh{\hat{H}}
\def\hhb{{\hat{\bar{H}\hspace{.01in}}\hspace{-.01in}}}
\def\hhbtwo{{\hat{\bar{H}\hspace{.02in}}\hspace{-.02in}}}

\def\bhl{\bar{\hat{l}\hspace{.02in}}}
\def\bhj{\bar{\hat{j\hspace{.03in}}}}
\def\hjs{{\hat{j\hspace{.03in}}\hspace{-.03in}}}
\def\hls{\hat{l\hspace{.03in}}}
\def\hjss{\hat{j\hspace{.01in}}}
\def\hms{\hat{m\hspace{.03in}}}
\def\hmss{\hat{m\hspace{.01in}}}
\def\hjbb{ \hat{\bar{J}}^\sharp }
\def\jbb{ \bar{J}^\sharp }
\def\hEbb{ \hat{\bar{E}}^\sharp }
\def\hHbb{ \hat{\bar{H}}^\sharp }
\def\sjbbh{{\hat{\bar{{\cal{J}^{\sharp}}}}} {} }
\def\gfrakh{\hat{\mathfrak g}}
\def\bfrak{\mathfrak b}
\def\dual{\underset{\s}{\longrightarrow}}

\def\ginv{g^{-1}}
\def\sg{\smal{\EuScript{G}}}
\def\sj{{\cal J}}
\def\sjb{{\bar{\cal J}}}
\def\sjbh{{\hat{\bar
{\cal{J}}}}}
\def\sjh{{\hat{\cal J}}}

\def\hc{^\dagger}
\def\hcj{\dagger}
\def\one{{\mathchoice {\rm 1\mskip-4mu l} {\rm 1\mskip-4mu} {\rm 1\mskip-4.5mu l} {\rm 1\mskip-5mu l}}}
\def\d{\delta}
\def\e{\eta}

\def\nnsrs{n+\srac{n(s)}{\r(\s)}}

\def\nrm{{n(r)\m}}
\def\mnrm{{-n(r),\m}}

\def\mnrn{{-n(r),\n}}
\def\nsn{{n(s)\n}}
\def\ntd{{n(t)\d}}

\def\nue{{n(u)\epsilon}}

\def\mnnrnsrs{{m+n+\srac{n(r)+n(s)}{\r(\s)}}}
\def\mnnrnsrsf{{m+n+\frac{n(r)+n(s)}{\r(\s)}}}
\def\mnrrs{{m+\srac{n(r)}{\r(\s)}}}

\def\mnrrsminus{{-m-\srac{n(r)}{\r(\s)}}}
\def\scf{{\cal F}}
\def\sG{{\cal G}}
\def\gfrak{\mbox{$\mathfrak g$}}
\def\goto{\longrightarrow}
\def\hj{\hat{J}}

\def\nsn{{n(s)\n}}
\def\schi{{\foot{\chi}}}
\def\schisig{{\foot{\chi(\s)}}}
\def\schizero{{\foot{\chi(0)}}}
\def\ntd{{n(t)\delta}}
\def\hc{^\dagger}
\def\st{{\cal T}}
\def\0b{\ }
\def\pl{\partial}
\def\Nr{{N(r)}}
\def\Nrm{{N(r)\m}}

\def\Nsn{{N(s)\n}}
\def\Ntd{{N(t)\d}}

\def\srange{\s=0,\ldots,N_c-1}
\def\sm{{\cal M}}
\def\sr{{\cal R}}

\makeatletter
\renewcommand{\@makefnmark}{\mbox{$^{\ddagger\@thefnmark}$}}
\renewcommand{\subsection}{\@startsection
  {subsection}{2}{0pt
}{-\baselineskip}{0.5\baselineskip}
   {\normalfont\normalsize\bf}}
\renewcommand{\section}{\@startsection
  {section}{2}{0pt
}{-\baselineskip}{0.5\baselineskip}
  {\bf\large}}

\makeatother
\numberwithin{equation}{section}
\numberwithin{table}{section}

\newcommand{\publititle}[8]
{ 
  \vspace*{-3cm}
  \begin{flushright} #1 \\ {\tt #2} \end{flushright}
  \vfill
  \begin{center}{\Large
    \bfseries #3}\end{center}
  \vskip 8mm
  \begin{center}{\large #4}\end{center}
  \begin{center}{\normalsize #5}\end{center}
  \vskip 8mm
  \nopagebreak
  \noindent #6
  \vfill
  \begin{flushleft} #7
  \end{flushleft}
  \hrule width 6.7cm \vskip.1mm
  {\small #8}
  \thispagestyle{empty}
  \clearpage
}

\begin{document}

\publititle{ ${}$ \\ ITFA-2001-18 \\ NSF-ITP-01-47 \\ UCB-PTH-01/17  \\ LBNL-47988 \\ SPIN-2001/13 \\
ITP-UU-01/21 }
{hep-th/0105305} {The Operator Algebra and Twisted KZ Equations \\ of
 WZW Orbifolds} {J. de Boer$^{\,a\S}$,
 M.B.Halpern$^{\,b\dagger}$ and N.A.Obers$^{\,c\ddagger{}}$}
 {$^a$Institute for Theoretical Physics, University of Amsterdam \\
 Valckenierstraat 65, 1018 XE Amsterdam, The Netherlands \\[2mm]
$^b$Department of Physics,
     University of California,
     Berkeley, California 94720, USA  \\ {\it and}
Theoretical Physics Group,  Lawrence Berkeley National
Laboratory \\
     University of California,
     Berkeley, California 94720, USA
\\[2mm]
$^c$Spinoza Institute {\it and} Institute for Theoretical Physics \\ Utrecht University,
Leuvenlaan 4, 3584 CE Utrecht, The Netherlands}
{We obtain the operator algebra of each twisted sector of all WZW orbifolds,
including the general twisted current algebra and the algebra of the
twisted currents with the twisted affine primary fields. Surprisingly,
the twisted
right and left mover current algebras are not a priori copies of each other.
Using the operator algebra we also derive world-sheet differential equations for the
twisted affine primary fields of all WZW orbifolds. Finally we include
ground state properties to obtain the twisted
Knizhnik-Zamolodchikov
equations of the WZW permutation orbifolds and the
inner-automorphic WZW orbifolds.  } {$^{\S}${\tt jdeboer@wins.uva.nl} \\
$^{\dagger}${\tt halpern@physics.berkeley.edu} \\ $^{\ddagger}${\tt  obers@phys.uu.nl}
}

 \clearpage

\renewcommand{\baselinestretch}{.4}\rm
{\footnotesize
\tableofcontents
}
\renewcommand{\baselinestretch}{1.0}\rm

\section{Introduction}

{\it Affine Lie algebra} [1-4],
or current algebra on the circle,
is the basis of a very large set of conformal field theories called the
affine-Virasoro constructions
\cite{Halpern:1989ss,Morozov:1990uu,Halpern:1990zy,Halpern:1996js} $A$.
The affine-Virasoro constructions include as important special cases the
affine-Sugawara constructions
[3,\ 8--11,\ 4]
$A_g$ on Lie $g$, the coset constructions
\cite{Bardakci:1971nb,Halpern:1971ay,Goddard:1985vk,Halpern:1996js}
$A_{g/h}$ and the $H$-invariant conformal field theories
\cite{Halpern:1992gb,Halpern:1996js} $A(H)$, where $H \subset {\rm Aut}(g)$ is
any finite symmetry group.

Orbifold theory [14--30]
has historically been approached at the level of examples, but recently
 a construction was given for the left mover stress tensors of all sectors
of the {\it general current-algebraic orbifold} \cite{deBoer:1999na,Halpern:2000vj}
\begin{equation}
\frac{A(H)}{H} \sp H \in {\rm Aut}(g)
\end{equation}
where $A(H)$ is any $H$-invariant CFT. This construction drew heavily on recent
advances \cite{Borisov:1997nc,Evslin:1999qb} in the theory of cyclic permutation
orbifolds, and the construction has been worked out in further detail for the
general cyclic permutation orbifold \cite{deBoer:1999na,Halpern:2000vj},
the general $S_N$ permutation orbifold \cite{Halpern:2000vj}, the general
inner-automorphic orbifold \cite{Halpern:2000vj}, the general WZW orbifold
\cite{deBoer:1999na,Halpern:2000vj}, the general coset orbifold
\cite{Halpern:2000vj} and the cyclic coset orbifolds
\cite{Evslin:1999ve}.

The principles of this construction are given in
Refs.~\cite{deBoer:1999na} and \cite{Halpern:2000vj}: \newline
$\bullet$ The {\it $H$-eigenvalue problem}, whose eigendata encodes the action
of $H$ on the untwisted currents (the action of $H$ in the adjoint of $g$).
\newline
$\bullet$ {\it Eigencurrents}, constructed from the eigendata of the
$H$-eigenvalue problem, with diagonal response to transformations in
the symmetry group.
\newline
$\bullet$ {\it Local isomorphisms}, which provide a map from the eigencurrents
and untwisted stress tensor to the {\it twisted currents} $\hj (\s)$ with
definite monodromy and the {\it orbifold stress tensor} $\hat T_\s$
\begin{subequations}
\begin{equation}
\hat T_\s (z) = {\cL}^{\nrm ; \mnrn} (\s) : \hj_{\nrm} (z,\s) \hj_{\mnrn} (z,\s):
\sp \s = 0, \ldots ,N_c -1
\end{equation}
\begin{equation}
\hj_\nrm (z e^{2 \pi i },\s) = e^{-2 \pi i \srac{n(r)}{\rho (\s)}}
\hj_\nrm (z,\s) \sp \hat T_\s (z e^{2 \pi i }) = \hat T_\s (z)
\end{equation}
\end{subequations}
in each sector $\s$ of the orbifold $A(H)/H$. Here
$N_c$ is the number of conjugacy classes of $H$, the $\s$-dependent integers
$\rho (\s)$ and $n(r)$ are included in the eigendata of the $H$-eigenvalue problem
and ${\cL} (\s)$ is the {\it twisted inverse inertia tensor} of sector $\s$.

The twisted inverse inertia tensor is an example of an {\it orbifold
duality transformation} \cite{deBoer:1999na,Evslin:1999ve,Halpern:2000vj}
\begin{equation}
L_H \dual {\cL} (L_H, \s)
\end{equation}
which expresses the twisted tensor in each sector of the orbifold as a discrete Fourier
transform of the $H$-invariant inverse inertia tensor $L_H$ of
 the symmetric theory $A(H)$. Other duality transformations include the
 {\it twisted structure constants} and the {\it twisted metric} of the general
 twisted left mover current algebra \cite{deBoer:1999na,Halpern:2000vj}.

In this paper, we restrict our attention to the special case of the
{\it general WZW orbifold}
\begin{equation}
\frac{A_g (H)}{H} \subset \frac{A(H)}{H} \sp H \subset {\rm Aut}(g)
\end{equation}
while extending the construction to obtain the full operator algebra of these
orbifolds. In particular (see Sec.~\ref{gwzw}), we obtain for each sector $\s$ of each $A_g (H)/H$:
\newline
$\bullet$ the twisted right and left mover current algebra, and the
corresponding right and left mover twisted affine-Sugawara constructions, \newline
$\bullet$ the algebra of the twisted currents and twisted affine-Sugawara
constructions with the {\it twisted affine primary fields}, \newline
$\bullet$ the WZW orbifold operator equations of motion, which are world-sheet
differential equations for the twisted affine primary fields. \newline
In this development we encounter some natural extensions of the principles
above, including: \newline
$\bullet$ the {\it extended $H$-eigenvalue problem} (see
Subsec.~\ref{EHeig}), which encodes the action of $H$ in any matrix
representation $T$ of $g$, \newline
$\bullet$ {\it eigenfields}  (see Sec.~\ref{Eigf}), including right and left
mover eigencurrents and now the eigenprimary fields, which are constructed
from the affine primary fields and the eigendata of the extended $H$-eigenvalue
problem. \newline
$\bullet$ extended local isomorphisms from the eigenfields to the twisted fields
with definite monodromy, including now the twisted right and left mover currents
and the twisted affine primary fields. \newline
The extended construction also includes
a number of new duality transformations associated to the twisted affine
primary fields. We mention in particular (see Subsec.~\ref{twrepmat}) the
{\it twisted representation matrices} $\T_\nrm \equiv \T_{\nrm}(T,\s)$,
\begin{equation}
T \dual \T (T,\s)
\end{equation}
which are the  duality
transformations in sector $\s$ of the untwisted matrix irreps $T$ of $g$.

One surprise in the full operator algebra is that {\it the twisted right and left
mover current algebras are not a priori copies of each other}
(see Subsecs.~\ref{MF}, \ref{quarg} and \ref{recpr}), a conclusion which we check both
at the level of characters (see Subsec.~\ref{quarg}) and by constructing the
corresponding {\it classical theory of WZW orbifolds} (see Subsec.~\ref{clarg}).
For permutation orbifolds (see Subsecs.~\ref{cyclorb}, \ref{snorb} and
\ref{unrec}) and inner-automorphic
orbifolds (see Subsec.~\ref{innerprop}) we are able to show by a mode
relabelling that the situation is equivalent to copies, but we do not have
a proof for the general case.

Finally, we include  ground state conditions to
obtain {\it twisted Knizhnik-Zamolodchikov (KZ) systems} for all sectors of
the general WZW
permutation orbifold (see Sec.~\ref{kzsec}) and the general inner-automorphic
WZW orbifold (see Sec.~\ref{innerorb}). These systems are a central result
of this paper.

As an example, we find the twisted KZ system
\begin{subequations}
\begin{equation}
\frac{A_g(H)}{H} \sp H (\mbox{permutation}) \subset {\rm Aut}(g)
\end{equation}
\begin{equation}
\part_\mu \hat A_+(\T,z,\s) = \hat A_+ (\T,z,\s) \hat W_\mu (\T,z,\s) \sp
\s = 0, \ldots ,N_c -1
\end{equation}
\begin{equation}
\hat W_{\mu}(\T,z,\s) = 2 \lr^{n(r)aj;-n(r),bl} (\s)
 \left[ \sum_{\nu \neq \mu}\left( \frac{z_{\nu}}{z_{\mu}}
\right)^{ \srac{\bar n(r)}{\r(\s)}} \frac{1}{z_{\mu \nu}}
\T_{n(r)aj}^{(\nu)} \T_{-n(r),bl}^{(\mu)}
- \srac{\bar n(r)}{\r(\s)} \frac{1}{z_{\mu}}
\T_{n(r)aj}^{(\mu)} \T_{-n(r),bl}^{(\mu)} \right]
\end{equation}
\end{subequations}
for the correlators of the twisted affine primary fields in
each twisted left mover sector of all WZW permutation orbifolds. Here
$\{ \T_{n(r) a  j }\}$ and $\lr (\s)$ are respectively the
twisted representation matrices of sector $\s$ and the twisted inverse inertia
tensor of the twisted affine-Sugawara constructions of that sector.
The integers $\bar n(r)$ are the pullback of the integers $n(r)$ into a
fundamental range.
Simple solutions of this system are discussed in Sec.~\ref{correl}, and the
corresponding twisted KZ system for the inner-automorphic WZW orbifolds is solved in
Subsec.~\ref{RUS}.

\section{The General Semisimple WZW Model}

\subsection{Notation}

In the following subsections, we will review the general semisimple WZW model
 $A_g$ in the composite notation
\begin{subequations}
\label{groupelement}
\begin{equation}
 (T_a)_\a{}^{\be}, \quad
 \left[T_a, T_b \right] = if_{ab}{}^c T_c,
  \quad M(k,T)_\a{}^\be = \left(\frac{k}{y(T)}\right)_\a \delta_\a{}^\be
\end{equation}
\begin{equation}
 {\rm Tr}(M(k,T)T_aT_b) = G_{ab},\quad G_{ac}G^{cb}=\d_a{}^{b} \quad,
 \left[M(k,T),T_a\right]  = 0 \label{MTA}
 \end{equation}
\begin{equation}
 a,b,c=1 \ldots {\rm dim }\0b g,\quad\;
 \a,\be = 1 \ldots {\rm dim }\0b T  \ .
\end{equation}
\end{subequations}
Here $T_a$ is any matrix representation of the compact semisimple Lie algebra
$g$, with metric $G_{ab}$ and structure constants $f_{ab}{}^{c}$.
The diagonal {\it data matrix} $M(k,T)$ stores information about the
 Dynkin indices $\{y \}$ of the matrix representations and the affine levels
 $\{k \}$ of each simple component of $g$.
When needed, the composite notation can be replaced by the explicit notation
\begin{subequations}
\label{semis}
\begin{equation}
    g=\oplus_I \gfrak^I ,\quad I=0,\ldots,K-1
\end{equation}
\begin{equation}
    a \goto a(I),I, \quad \a \goto \a(I),I
\end{equation}
\begin{equation}
 G_{ab} \goto G_{aI,bJ} = k_I  \e^I_{a(I),b(I)}\d_{IJ},
 \quad  G^{ab} \goto G^{aI,bJ} = k_I^{-1}  \e_I^{a(I),b(I)}\d^{IJ}
 \label{eq:invg}
 \end{equation}
 \begin{equation}
 f_{ab}{}^{c} \goto f_{aI,bJ}{}^{cK} = f^I_{a(I),b(I)}{}^{c(I)} \d_{IJ}\d_J{}^K
 \end{equation}
 \begin{equation}
 (T_a)_\a{}^\be \quad\goto\quad (T_{aI})_{\a J}{}^{\be K} =
 (T_a^I)_{\a(I)}{}^{\be(I)} \d_{IJ} \d_J{}^K \label{TaI}
 \end{equation}
\begin{equation}
 M(k,T)_\a{}^\be \goto M(k,T)_{\a I}{}^{\be J} = \frac{k_I}{y_I(T^I)}\d_{\a(I)}
 {}^{\be(I)} \d_I{}^J \label{2.2f}
 \end{equation}
 \begin{equation}
 M^{-1}(k,T)_\a{}^\be \goto M^{-1}(k,T)_{\a I}{}^{\be J} = \frac{y_I(T^I)}{k_I}
 \d_{\a(I)}{}^{\be(I)} \d_I{}^J
 \end{equation}
 \begin{equation}
 {\rm Tr}(T_a^IT_b^J) = y_I(T^I) \e^I_{a(I),b(I)}\d^{IJ} \ .
 \end{equation}
 \end{subequations}
Here $I$ is the semisimplicity index and $a(I)$, $\a(I)$ are indices associated
to the simple component $\gfrak^I$, with structure constants, Killing metric,
matrix representation and affine level $f^I$, $\e^I$, $T^I$ and $k_I$ respectively.
Note that the matrix representations $T$ of $g$ are direct sums of the matrix
representations $T^I$ of $\gfrak^I$.

We also need the left and right mover {\it affine-Sugawara constructions}
[3,\ 8--11,\ 4]
 on $g$
\begin{subequations}
\begin{equation}
 T (z) = L_g^{ab} : J_a (z) J_b (z) : \sp
\bar{T} (\bz)  = L_g^{ab} : \bJ_a (\bz) \bJ_b (\bz) :
\sp c_g = 2 G_{ab} L_g^{ab}
\end{equation}
\begin{equation}
J_a (z) \quad \goto \quad J_{aI} (z) \sp
\bar J_a (\bz) \quad \goto \quad \bar J_{aI}(\bz)
\end{equation}
\begin{equation}
\label{Dgdef}
L_g^{ab} \quad\goto\quad  L_g^{aI,bJ} = \frac{\eta_I^{ab} \delta^{IJ} }
{2 k_I + Q_I} \sp
D_g (T)_\a{}^\be = (L_g^{ab} T_a T_b)_\a{}^\be
\quad \goto \quad
D_g (T)_{\a I}{}^{\be J}
\end{equation}
\begin{equation}
D_g (T)_{\a I}{}^{\be J}  = \left(L_g^{aM,bN} T_{aM} T_{bN}
\right)_{\a I}{}^{\be J} =  \frac{ (\eta_I^{ab}
T_a^I T_b^I)_{\a(I)}{}^{\be (I)}}{2 k_I +Q_I}
 \de_I^J = \Delta_{\sgbn^I } (T^I)
\delta_{\a(I)}{}^{\be (I)} \delta_I^J
\end{equation}
\end{subequations}
where  $: \cdot :$ is operator product normal ordering and
$L^{ab}_g$ is the inverse inertia tensor of the affine-Sugawara
constructions. The quantity $D_g (T)$ is the {\it conformal weight matrix}
corresponding to representation $T$ of $g$ and $\Delta_{\sgbn^I} (T^I) $ is
the conformal weight of representation $T^I$ under  $\gbn^I$.

For permutation invariant systems, these relations take the simple form
\begin{subequations}
\begin{equation}
 \gfrak^I \simeq \gfrak, \quad  a(I)=a , \quad \a(I)=\a
\end{equation}
\begin{equation}
\e^I_{ab} = \e_{ab},\quad \e_I^{ab} = \e^{ab}, \quad f^I_{ab}{}^c = f_{ab}{}^c ,
 \quad T_a^I = T_a,\quad k_I=k
 \end{equation}
 \begin{equation}
  [T_a,T_b] = if_{ab}{}^{c}T_c ,\quad {\rm Tr}(T_a T_b) =y(T)\e_{ab},\quad M(k,T) = \frac{k}{y(T)}
  \one
  \label{eq:MT=1}
\end{equation}
\begin{equation}
\label{eq:D=1} L_g^{aI,bJ} = \frac{\eta^{ab} \delta^{IJ}}{2 k
+Q_{\sgbn} } \sp D_g (T)_{\a I}{}^{\be J} =  \frac{ (\eta^{ab} T_a
T_b){}_{\a}{}^{\be } }{2 k +Q_{\sgbn} } \de_I^J  = \Delta_{\sgbn} (T)
\delta_\a^\be \de_I^J
\end{equation}
\end{subequations}
where $\Delta_{\sgbn} (T)$ is the conformal weight of irrep $T$ under
any copy  $\gbn^I \simeq \gbn$. In this case, the data matrix and the conformal
weight matrix are proportional to $\one$ , and the last relations in
\eqref{eq:MT=1}, \eqref{eq:D=1} hold for simple $g$ as well.

\subsection{WZW Operator Algebra}

For general semisimple $g$, we have the OPEs
\begin{subequations}
\label{calg}
\begin{equation}
\label{curalg}
J_a (z) J_b (w) = \frac{G_{ab}}{(z-w)^2} + \frac{if_{ab}{}^c J_c(w) }{z-w}
+ \Ord (z-w)^0
\end{equation}
\begin{equation}
\label{curalgb}
\bJ_a (\bz) \bJ_b (\bw) = \frac{G_{ab}}{(\bz-\bw)^2} + \frac{if_{ab}{}^c
\bJ_c(\bw) }{\bz-\bw}
+ \Ord (\bz-\bw)^0
\end{equation}
\begin{equation}
\label{curalgubb}
J_a (z) \bJ_b (\bz) = {\rm regular}
\end{equation}
\begin{equation}
\label{utj}
T(z) J_a (w) = \left( \frac{1}{(z-w)^2} + \frac{\partial_w }{z-w} \right) J_a (w)
+\Ord (z-w)^0
\end{equation}
\begin{equation}
\label{utjb}
\bar T(\bz) \bar J_a (\bw) = \left( \frac{1}{(\bz-\bw)^2} + \frac{\partial_{\bw}}{\bz-\bw}
\right) \bar J_a (\bw)
+\Ord (\bz-\bw)^0
\end{equation}
\begin{equation}
\label{utjjb}
T(z) \bar J_a (\bw) = \bar T(\bz) J_a (w)= {\rm regular}
\end{equation}
\end{subequations}
\begin{subequations}
\label{primf}
\begin{equation}
\label{Jgl}
 J_a (z) g(T,\bw,w) = \frac{g(T,\bw,w)}{z-w}  T_a + \Ord (z-w)^0
\end{equation}
\begin{equation}
\label{Jgr}
 \bJ_a (\bz) g(T,\bw,w) = - T_a \frac{g(T,\bw,w)}{\bz-\bw}  +
  \Ord (\bz-\bw)^0
\end{equation}
\begin{equation}
T(z)g(T,\bw,w) = \frac{g(T,\bw,w)D_g(T)}{(z-w)^2} +
\frac{\part_w}{z-w} g(T,\bw,w) + \Ord (z-w)^0
\end{equation}
\begin{equation}
\bar{T}(\bz)g(T,\bw,w) = \left( \frac{D_g (T)}{(\bz-\bw)^2} +
\frac{\part_{\bw}}{\bz-\bw} \right) g(T,\bw,w) + \Ord (\bz-\bw)^0
\end{equation}
\end{subequations}
where $J(z)$, $\bar J (\bz)$ are the currents of affine $(g \oplus g)$ and
 $g(T,\bz,z)$ is the affine primary
field corresponding to matrix representation $T$.
As implied by  \eqref{Jgl} and \eqref{Jgr},
the affine primary field  is also a direct sum
\begin{subequations}
\begin{equation}
 g(T,\bz,z)_\a{}^\be  \goto g(T,\bz,z)_{\a I}{}^{\be J}=g_I(T^I,\bz,z)_{\a(I)}{}^{\be(I)} \d_I{}^J
\end{equation}
\begin{equation}
[ M(k,T), g(T,\bz,z) ] = 0
\end{equation}
\end{subequations}
of the affine primary fields $g_I(T^I,\bz,z)$ of $\gfrak^I$.
In what follows, we often refer to the affine primary fields $g(T,\bz,z)$ as the
WZW vertex operators.

The OPEs \eqref{calg} and \eqref{primf} are equivalent to many familiar commutation
relations, among which we list for reference
\begin{subequations}
\begin{equation}
J_a (z) = \sum_{m \in \sz} J_a (m) z^{-m-1}
\sp \bar J_a (z) = \sum_{m \in \sz} \bar J_a (m) \bz^{-m-1}
\end{equation}
\begin{equation}
T(z) = \sum_{m \in \sz} L(m) z^{-m-2} \sp
\bar T(z) = \sum_{m \in \sz} \bar L(m) z^{-m-2}
\end{equation}
\begin{equation}
\label{jjalg}
[ J_a(m),J_b (n)] = i f_{ab}{}^c J_c (m + n) + m G_{ab} \de_{m+n,0}
\end{equation}
\begin{equation}
\label{jbjbalg}
[ \bar J_a(m), \bar J_b (n)] = i f_{ab}{}^c \bar J_c (m + n) + m G_{ab} \de_{m+n,0}
\end{equation}
\begin{equation}
\label{jjbalg}
[ J_a(m), \bar J_b (n)] = 0
\end{equation}
\begin{equation}
\label{jgco}
\part g (T,\bz,z)  = [ L (-1), g(T,\bz,z) ] \sp
\bar \part g (T,\bz,z) = [ \bar{L} (-1), g(T,\bz,z) ]
\end{equation}
\begin{equation}
\label{Jgut}
[ J_a(m),g(T,\bz,z)] = g(T,\bz,z)\ T_a z^m
\sp [ \bar J_a(m),g(T,\bz,z)] = -\bz^m T_a \ g(T,\bz,z)
\end{equation}
\end{subequations}
where \eqref{jjalg}, \eqref{jbjbalg}, \eqref{jjbalg} is the algebra of affine
$(g \oplus g)$. Then (see for example Ref. \cite{Halpern:1996et}) we may
use \eqref{jgco} and \eqref{Jgut} to obtain the WZW vertex operator
equations
\begin{equation}
\label{difrel}
\part g (T,\bz,z)= 2 L_g^{ab} : J_a \ g (T,\bz,z) T_b :  \sp
\bar\part g (T,\bz,z) = -2 L_g^{ab} : \bJ_a T_b \ g (T,\bz,z) :
\end{equation}
from which the Knizhnik-Zamolodchikov (KZ) equations \cite{Knizhnik:1984nr}
for WZW correlators can be derived \cite{Halpern:1996et}.

\section{The Automorphism Group $H$}

\subsection{Action in Representation $T$}

Let $H \subset {\rm Aut}(g)$ be a symmetry group (automorphism group) of the WZW model
$A_g(H)$ and pick one representative $h_\s \in H$ of each conjugacy class of $H$.
The actions of the $H$ symmetry on the currents $J$, $\bar{J}$ and the affine
primary fields $g(T)$ are
\begin{subequations}
\begin{equation}
J_a (z)' = w (h_\s)_a{}^b J_b (z)\sp
\bJ_a (\bz)' = w (h_\s)_a{}^b \bJ_b (\bz)
\end{equation}
\begin{equation}
g(T,\bz,z)' = W(h_\s;T) g  (T,\bz,z) W^{\dagger} (h_\s;T)
\end{equation}
\begin{equation}
g(T,\bz,z)'_\a{}^\be = W(h_\s;T)_\a{}^\ga g  (T,\bz,z)_\ga{}^\d W^{\dagger}
(h_\s;T)_\d{}^\be
\end{equation}
\begin{equation}
w\hc (h_\s) w (h_\s) = 1 \ , \quad W\hc (h_\s;T) W(h_\s;T) = \one
\end{equation}
\begin{equation}
w(h_\s), W(h_\s;T) \in H \subset {\rm Aut}(g), \quad \srange
\end{equation}
\end{subequations}
where $N_c$ is the number of conjugacy classes of $H$. The matrix $w(h_\s)$
is the action of $h_\s \in H$ in the adjoint of $g$, while
$W(h_\s;T)$ is the action%
\footnote{The explicit form of $W(T)$ is given for permutation
subgroups of ${\rm Aut}(g)$ in Subsec.~\ref{syst} and for
inner-automorphic subgroups of ${\rm Aut}(g)$ in
Subsec.~\ref{innerprop}. To use the formalism for automorphism
groups which include outer automorphisms of simple $\gfrak$, one often needs
to include reducible representations of $\gfrak$.
As an example, we mention the charge conjugation automorphism
$T^{(0)} \leftrightarrow \bar{T}^{(0)} =  - T^{(0)}{}^{\rm T}$
for the 3 and $\bar{3}$ of $SU(3)$ in the standard Cartesian
basis. In this case we may take
$$
J_a' = \w_a{}^b J_b \sp
T_a = \left( \begin{array}{cc}
T_a^{(0)} & 0 \\ 0 & \bar{T}_a^{(0)} \end{array} \right)
\sp a = 1 \ldots {\rm dim}\,\sgbn
$$
$$
W(T) = i \left( \begin{array}{cc} 0 & \one \\ \one & 0  \end{array} \right)
\sp W\hc(T) T_a W(T) = \w_a{}^b T_b =
\left( \begin{array}{cc}
\bar{T}_a^{(0)} & 0 \\ 0 & T_a^{(0)} \end{array} \right)
$$
$$
\w_A{}^A = - \w_I{}^I = 1 \sp \bar{T}_A = T_A \sp \bar{T}_I = -T_I \sp
A = 2,5,7 \sp I = 1,3,4,6,8
$$
as the realization of $W(T)$ and the linkage relation
\eqref{eq:WTWwTa}. We intend to return to this and related examples elsewhere.}
 of $h_\s \in H$ in matrix
representation $T$ of $g$. Automorphic invariance of the
current-current OPEs \eqref{curalg} then requires
\begin{equation}
 w(h_\s)_a{}^c w(h_\s)_b{}^dG_{cd} =G_{ab},\;\; w(h_\s)_a{}^d w(h_\s)_b{}^e
  f_{de}{}^f w\hc(h_\s)_f{}^c  = f_{ab}{}^c,
    \;\; \forall \ h_\s \!\in\! H \!\subset\! {\rm Aut}(g). \label{eq:john2.6c}
\end{equation}
In what follows, these relations are called the $H$ symmetry of $G$ and $f$.
The invariance of $G_{ab}$ also guarantees the $H$ invariance \cite{Halpern:1992gb}
of the inverse inertia tensor and the stress tensors:
\begin{subequations}
\begin{equation}
\label{Lginv}
L_g^{cd}  w(h_\s)_c{}^a w(h_\s)_d{}^b = L_g^{ab}
\end{equation}
\begin{equation}
T(z)' = T(z) \sp \bar T (\bz) ' = \bar T (\bz) \ .
\end{equation}
\end{subequations}
This much is well understood in current-algebraic orbifold theory
\cite{deBoer:1999na,Halpern:2000vj}.

Here we extend this discussion to include the automorphic invariance of the
$Jg$ and $\bar J g$ OPEs, e.g.
\begin{equation}
 J_a (z)' g(T,\bw,w)' = \frac{g(T,\bw,w)'}{z-w}  T_a + \Ord (z-w)^0 \ .
\end{equation}
The invariance of these OPEs requires the {\it linkage relation}
\begin{subequations}
\label{WTWwTa}
\begin{equation}
 W^\hcj(h_\s;T) T_a W(h_\s;T) = w(h_\s)_a{}^b T_b  ,\quad \forall \ h_\s \in H
  \subset {\rm Aut}(g) \label{eq:WTWwTa}
  \end{equation}
  \begin{equation}
 W(h_\s;T)T_aW^\hcj(h_\s;T) = w^\hcj(h_\s)_a{}^b T_b \label{eq:WTWwTb}
 \end{equation}
 \begin{equation}
 w\hc(h_\s)_a{}^b(W\hc(h_\s;T) T_b W(h_\s;T) ) = w(h_\s)_a{}^b(W(h_\s;T)
  T_b W\hc(h_\s;T) ) = T_a \label{eq:WTWwTb-next}
\end{equation}
\end{subequations}
which constrains the action $W$ given the action $w$. Here \eqref{eq:WTWwTb},
\eqref{eq:WTWwTb-next} are equivalent forms of
\eqref{eq:WTWwTa}, and we will refer to the equivalent forms in
\eqref{eq:WTWwTb-next} as the $H$-invariance of the representation matrices
$T$.

The linkage relation tells us that $W$ reduces to $w$
\begin{equation}
 W(h_\s;T^{\rm adj}) = w(h_\s) ,\quad (T^{\rm adj}_a)_b{}^c \equiv -if_{ab}{}^c
\end{equation}
when $T$ is taken to be the adjoint representation. In this case, the linkage
relation is equivalent to the $H$-symmetry of the structure constants $f$ in
\eqref{eq:john2.6c}. One also finds that
\begin{equation}
 T^{\rm adj}(s) \equiv  s T^{\rm adj} s^{-1} \goto W(h_\s;T^{\rm adj}(s)) = sw(h_\s)s^{-1}
\end{equation}
for similarity-transformed versions of the adjoint representation.

The linkage relation also gives us information about the orders $\r(\s)$ and
$R(T,\s)$ of $w(h_\s)$ and $W(h_\s;T)$ respectively. These are the smallest
integers which satisfy
\begin{subequations}
\begin{equation}
 w(h_\s)^{\r(\s)} =1, \quad W(h_\s;T)^{R(T,\s)} =\one
 \end{equation}
 \begin{equation}
  R(T^{\rm adj},\s) = \r(\s) \ .
\end{equation}
\end{subequations}
Using these definitions and the linkage relation, we find in general that
\begin{equation}
 [W(h_\s;T)^{\r(\s)},T_a] =0,\quad w(h_\s)^{R(T,\s)}=1   ,
 \quad \srac{R(T,\s)}{\r(\s)} \in {\mathbb Z}_{>0} \label{eq:[WrhoT]=0}
\end{equation}
so that $R(T,\s)$ can be a positive multiple of $\r(\s)$. In what follows, we
will often abbreviate $R(\s) \equiv R(T,\s)$.

Taken together, the linkage relation, the normalization condition in \eqref{MTA}
and the $H$-invariance \eqref{eq:john2.6c}  of $G_{ab}$ tell us that
the data matrix $M$
\begin{equation}
  W^\hcj(h_\s;T)M(k,T)W(h_\s;T) = M(k,T),\quad \forall \ h_\s \in H
  \subset {\rm Aut}(g) \label{Mtransf}
\end{equation}
is also invariant under transformations  in $H$.

Finally, automorphic invariance of the $T g$ and $\bar T g$ OPEs tells us that
the conformal weight matrix $D_g(T)$ is also invariant
\begin{equation}
\label{cfwinv}
W(h_\s;T) D_g (T) W^{\dagger}  (h_\s;T) = D_g (T) \ .
\end{equation}
According to \eqref{Dgdef}, this invariance also follows from the linkage relation
 and the invariance \eqref{Lginv} of the inverse inertia tensor.

\subsection{The Extended $H$-Eigenvalue Problem \label{EHeig}}

For the action $w(h_\s)$ of $h_\s \in H$ in the adjoint, the {\it $H$-eigenvalue
problem}
\begin{subequations}
\label{HEigen}
\begin{equation}
 w(h_\s)_a{}^b U^\hcj(\s)_b{}^{n(r)\m} = U^\hcj(\s)_a{}^{n(r)\mu}E_{n(r)}(\s),
 \quad \s=0,\ldots,N_c-1
\end{equation}
\begin{equation}
 E_{n(r)}(\s) = e^{-2\pi i\frac{n(r)}{\r(\s)}} ,\quad n(r) \equiv n(r(\s)) \in \z
\end{equation}
\begin{equation}
 E_{n(r)\pm \r(\s)}(\s)=E_{n(r)}(\s),\quad U\hc(\s)_a{}^{n(r)\pm\r(\s),\m} =
 U\hc(\s)_a{}^\nrm
\end{equation}
\begin{equation}
 U\hc(\s)U(\s)=1,\quad U\hc(0)=1,\quad E_{n(r)}(0)=\r(0)=1
\end{equation}
\begin{equation}
 U(\s)_\nrm{}^a U\hc(\s)_a{}^\nsn=\d_\nrm{}^\nsn,
       \quad\d_\nrm{}^\nsn\equiv\d_{n(r)-n(s),0\,\text{mod }\r(\s)}\d_\m{}^\n
\end{equation}
\begin{equation}
\label{nbdef}
 \bar{n}(r) = n(r) - \r(\s)\lfloor\srac{n(r)}{\r(\s)}\rfloor,\quad \bar{n}(r)
 \in \{0,...,\r(\s)-1\}
 \end{equation}
\end{subequations}
was defined and studied in Refs.~\cite{deBoer:1999na}, \cite{Evslin:1999ve} and
\cite{Halpern:2000vj}. Here {$r \equiv r(\s)$} runs over the degenerate
subspaces of the $H$-eigenvalue problem. The unitary eigenvector matrix
$U\hc(\s)$ and the eigenvalues $E(\s)$ are periodic with period $\r(\s)$ in the
spectral index $n(r)$, while $\m = \m(r(\s))$ is a degeneracy index. The
 pull-backs $\bar{n}(r)$ are identified in Ref.~\cite{Halpern:2000vj} as the
 twist classes of the twisted currents of sector $\s$, where $\s=0$ is the
 untwisted sector.

Correspondingly, we define now the {\it extended $H$-eigenvalue problem}
\begin{subequations}
\label{ExtendedHEigen}
\begin{equation}
 W(h_\s;T)_\a{}^\be U^\hcj(T,\s)_\be{}^{N(r)\m} = U^\hcj(T,\s)_\a{}^{N(r)\m}
 E_{N(r)}(T,\s),\quad \s=0,\ldots,N_c-1
\end{equation}
\begin{equation}
 E_{N(r)}(T,\s) = e^{-2\pi i\frac{N(r)}{R(\s)}},\quad N(r)\equiv N(T,r(\s))
 \in \z
\end{equation}
\begin{equation}
 E_{N(r)\pm R(\s)}(T,\s)=E_{N(r)}(T,\s),\quad U\hc(T,\s)_\a{}^{N(r)\pm R(\s),\m}
  = U\hc(T,\s)_\a{}^\Nrm
\end{equation}
\begin{equation}
 U\hc(T,\s)U(T,\s)=\one  ,\quad U\hc(T,0)=\one,\quad E_{N(r)}(T,0)=R(T,0)=1
\end{equation}
\begin{equation}
 U(T,\s)_\Nrm{}^\a U\hc(T,\s)_\a{}^\Nsn=\d_\Nrm{}^\Nsn,
      \;\d_\Nrm{}^\Nsn\equiv\d_{N(r)-N(s),0\,\text{mod }R(\s)}\d_\m{}^\n
\end{equation}
\begin{equation}
 \bar{N}(r) = N(r) - R(\s)\lfloor\srac{N(r)}{R(\s)}\rfloor,\quad \bar{N}(r)
 \in \{0,...,R(\s)-1\}
\end{equation}
\begin{equation}
 W=U\hc EU,\; W\hc = U\hc E^\ast U ,\;  E^\ast E =\one
\end{equation}
\begin{equation}
 WU\hc = U\hc E,\; UW=EU,\; U^\ast W^\ast=E^\ast U^\ast,\; UW\hc=E^\ast U,\;
 W\hc U\hc=U\hc E^\ast
\end{equation}
\begin{equation}
 E(T,\s)_\Nrm{}^{\Nsn} \equiv \d_\Nrm{}^{\Nsn} E_\Nr(T,\s)
\end{equation}
\end{subequations}
for the action $W(h_\s;T)$ of $h_\s \in H$ in matrix representation $T$.
 Here, the order $R(\s)\equiv R(T,\s)$, the spectral index
 $N(r) \equiv  N(T,r(\s))$ and the degeneracy index $\m \equiv \m(T,r(\s))$ can
  all depend on $T$.
 The relations in \eqref{ExtendedHEigen} are alternate forms of the extended
 eigenvalue problem.

Because of the linkage relation \eqref{WTWwTa}, the two eigenvalue problems
\eqref{HEigen} and \eqref{ExtendedHEigen} are not independent. In particular,
 the extended problem includes the original problem as the special case when
 $T=T^{\rm adj}$:
\begin{subequations}
\begin{equation}
 W(h_\s;T^{\rm adj}) = w(h_\s), \quad R(T^{\rm adj},\s)=\r(\s), \quad
 (T_a^{\rm adj})_b{}^{c}  = -if_{ab}{}^{c}
\end{equation}
\begin{equation}
 E(T^{\rm adj},\s)=E(\s)  ,\quad N(r)= n(r) ,\quad U\hc(T^{\rm adj},\s)=U\hc(\s)\ .
\end{equation}
\end{subequations}
Moreover, App.~\ref{braid} remarks on a set of braid relations,
induced by the linkage relation, among the quantities of the two eigenvalue
problems. Solutions of the extended $H$-eigenvalue problem are discussed for
general permutation groups and general groups of inner automorphisms in
Secs.~\ref{permorb} and \ref{innerorb} respectively.

It is known \cite{deBoer:1999na,Halpern:2000vj} that the eigenvectors
$U\hc(\s)$ of the $H$-eigenvalue problem are Fourier basis elements with
 periodicity $\r(\s)$ in the spectral index $n(r)$, and similarly the
 eigenvectors $U\hc(T,\s)$ of the extended problem are Fourier basis elements
 with periodicity $R(\s)=R(T,\s)$ in the spectral index $N(r)$. In what follows,
  all quantities inherit the periodicities
\begin{equation}
 N(r) \goto N(r)\pm R(\s), \quad  n(r) \goto n(r) \pm \r(\s)
\end{equation}
in the spectral indices $\{n(r)\}$, $\{N(r)\}$ of the $H$-eigenvalue problems.
Since $R(T,\s)$ can be a multiple of $\r(\s)$, this means in particular that
 all the orbifold duality transformations of sector $\s$ are discrete Fourier
 transforms with fundamental period $\r(\s)$.

We will also need the behavior of the eigenvalue problems under conjugation in
$H$:
\begin{subequations}
\label{conjugation-in-H}
\begin{equation}
 w(h_\s) \goto v\hc(\s) w(h_\s) v(\s) ,\quad v\hc(\s) v(\s) =1 ,\quad  v(\s)
 \in H \label{evunderconjugation}
\end{equation}
\begin{equation}
 U\hc(\s) \goto v\hc(\s) U\hc(\s)  ,\quad U(\s) \goto U(\s)v(\s)  \label{Uv}
\end{equation}
\begin{equation}
 v(\s)_a{}^{c}  v(\s)_b{}^{d} G_{cd} = G_{ab} ,\quad v(\s)_a{}^{d}  v(\s)_b{}^{e}
  f_{de}{}^{f} v\hc(\s)_f{}^{c} = f_{ab}{}^{c} \label{eq:3.18c}
\end{equation}
\begin{equation}
W(h_\s;T) \goto v\hc(T,\s)W(h_\s;T)v(T,\s) ,\quad v\hc(T,\s) v(T,\s) =\one ,
\quad  v(T,\s) \in H
\end{equation}
\begin{equation}
 U\hc(T,\s) \goto v\hc(T,\s) U\hc(T,\s)  ,\quad  U(T,\s) \goto U(T,\s)v(T,\s)
\end{equation}
\begin{equation}
 v\hc(T,\s) T_a v(T,\s) = v(\s)_a{}^{b} T_b  \ .
\end{equation}
\end{subequations}
The relations in \eqref{eq:3.18c} follow from \eqref{eq:john2.6c} because
$v(\s) \in H$, and the first three relations in \eqref{conjugation-in-H} were
given in Refs.~\cite{deBoer:1999na} and \cite{Halpern:2000vj}.
 Taken together, the conjugation relations \eqref{conjugation-in-H} preserve
 the linkage relation \eqref{WTWwTa}, and will allow us to check that all the
 orbifold duality transformations of this paper are class functions.

\section{Eigenfields \label{Eigf}}

\subsection{Eigencurrents and Eigenprimary Fields}

Following Refs.~\cite{deBoer:1999na} and \cite{Halpern:2000vj}, we define
the eigencurrents $\J (\s)$, $\bar \J (\s)$ and eigenprimary fields
$\sg (T,\s) $ of sector $\s$ as follows
\begin{subequations}
\label{eigencurrents+elements}
\begin{equation}
 \sj_\nrm(z,\s) \equiv \schisig_\nrm U(\s)_\nrm{}^{a}J_a(z) ,\quad
 \bar{\sj}_\nrm(\bz,\s)\equiv \schisig_\nrm U(\s)_\nrm{}^{a}\bar{J}_a(\bz)
 \end{equation}
 \begin{equation}
 \sg(T,\bz,z,\s) \equiv U(T,\s)g(T,\bz,z)U^\hcj(T,\s) \label{eq:scrg}
 \end{equation}
 \begin{equation}
 \sg(T,\bz,z,\s)_\Nrm{}^\Nsn = U(T,\s)_\Nrm{}^\a g(T,\bz,z,\s)_\a{}^\be
 U^\hcj(T,\s)_\be{}^\Nsn
 \end{equation}
 \begin{equation}
 \schisig_{n(r)\pm\r(\s),\m} = \schisig_\nrm ; \quad \schizero_\nrm = 1
 \label{cond-on-schi}
\end{equation}
\end{subequations}
where $U\hc$ and $U\hc(T)$ are the eigenvector matrices respectively of the
$H$-eigenvalue problem \eqref{HEigen} and its extension \eqref{ExtendedHEigen}.
 The quantities $\schisig$,which satisfy the conventions in \eqref{cond-on-schi},
  are otherwise arbitrary normalization constants.

The responses of the eigenfields to elements of the automorphism group are
\begin{subequations}
\label{diag-response-of-JG}
\begin{equation}
    \sj_\nrm(z,\s)' = E_{n(r)}(\s) \sj_\nrm(z,\s)
     = e^{- 2 \pi i \srac{n(r)}{\rho (\s)} } \sj_\nrm(z,\s)
\end{equation}
\begin{equation}
    \sjb_\nrm(\bz,\s)' =     E_{n(r)}(\s) \bar{\sj}_\nrm(\bz,\s)
   = e^{- 2 \pi i \srac{n(r)}{\rho (\s)} } \bar{\sj}_\nrm(z,\s)
\end{equation}
\begin{equation}
\label{sgresp}
\sg(T,\bz,z,\s)' \ = E(T,\s)\sg(T,\bz,z,\s)E(T,\s)^{\ast}
\end{equation}
\begin{equation}
    (\sg(T,\bz,z,\s)')_{N(r)\m}{}^{N(s)\n}
 = e^{-\frac{2\pi i}{R(\s)}(N(r)-N(s))}   \sg(T,\bz,z,\s)_{N(r)\m}{}^{N(s)\n}\ .
\end{equation}
\end{subequations}
The diagonal response of the eigencurrents is well known in the theory of
current-algebraic orbifolds \cite{deBoer:1999na,Halpern:2000vj}, while the
two-sided response of the eigenprimary fields is new.

The next stage in the orbifold program is to rewrite all relations in the
 symmetric theory in terms of the corresponding eigenfields.
This step gives rise to the {\it orbifold duality transformations}
\cite{deBoer:1999na,Halpern:2000vj}, which express
 the twisted tensors in the orbifold as discrete Fourier transforms of
 corresponding $H$-symmetric tensors in the symmetric theory. Moreover, the
 twisted tensors satisfy {\it selection rules} \cite{deBoer:1999na,Halpern:2000vj}
  which are dual to the $H$-symmetry of the original tensors.

The case of the eigencurrent OPEs was worked out in Refs.~\cite{deBoer:1999na}
and \cite{Halpern:2000vj}.
Here one encounters the orbifold duality transformations from the metric $G$ and
structure constants $f$ to the {\it twisted metric} $\sG(\s)$ and the
{\it twisted structure constants} $\scf(\s)$ of sector $\s$
\begin{equation}
 G \dual \sG(G,\s) ,\quad f \dual \scf(f,\s) \ .
\end{equation}
The explicit form of these duality transformations is
\cite{deBoer:1999na,Halpern:2000vj}
\begin{subequations}
\label{sG}
\begin{equation}
\label{sG0}
  \G_{\nrm;\nsn}(\s) =\schisig_\nrm\schisig_\nsn
  U(\s)_\nrm{}^a U(\s)_\nsn{}^b G_{ab}
\end{equation}
\begin{equation}
\label{sel-sG}
=\delta_{n(r)+n(s),0\,\text{mod }\r(\s)}\sG_{\nrm;\mnrn}(\s)
\end{equation}
\begin{equation}
 \sG^{\nrm;\nsn}(\s) =\schisig^{-1}_\nrm\schisig^{-1}_\nsn  G^{ab}
 U\hc(\s)_a{}^\nrm U\hc(\s)_b{}^\nsn
 \end{equation}
 \begin{equation}
 =\d_{n(r)+n(s),0\,\text{mod }\r(\s)}\sG^{\nrm;\mnrn}(\s) \label{sel-sG-inv}
\end{equation}
\begin{equation}
\sG^{\nrm;\ntd}(\s) \sG_{\ntd;\nsn}(\s) =\d_\nsn{}^\nrm
\end{equation}
\begin{equation}
  \F_{\nrm;\nsn}{}^\ntd(\s) =\,\schisig_\nrm\schisig_\nsn
  \schisig^{-1}_\ntd U(\s)_\nrm{}^a U(\s)_\nsn{}^b f_{ab}{}^c
  U\hc(\s)_c{}^\ntd
\end{equation}
\begin{equation}
\label{sel-scf}
= \delta_{n(r)+n(s)-n(t),0\,\text{mod }\r(\s)}
\F_{\nrm;\nsn}{}^{n(r)+n(s),\delta}(\s)
\end{equation}
\begin{equation}
  \scf_{\nrm;\nsn}{}^\ntd(\s)=0 \quad\text{ unless }\quad n(r)+n(s) \in \{n(r)\}
  \ .
\end{equation}
\end{subequations}
The forms \eqref{sel-sG}, \eqref{sel-sG-inv} and \eqref{sel-scf} are solutions to
 selection rules which are dual to the $H$-invariance of $G$ and $f$ in
 \eqref{eq:john2.6c}.
Twisted tensor indices can be lowered and raised with the twisted metric and
its inverse, and the twisted structure constants with all indices down are
totally antisymmetric. The twisted metric and twisted structure constants are
also class functions  \cite{deBoer:1999na,Halpern:2000vj} under conjugation
in $H$
\begin{equation}
 \sG(U(\s)v(\s);\s) = \sG(U(\s);\s) ,\quad \scf(U(\s)v(\s);\s) = \scf(U(\s);\s)
\end{equation}
according to Eqs.~\eqref{eq:john2.6c}, \eqref{conjugation-in-H} and \eqref{sG}.

\subsection{The Twisted Representation Matrices of Sector $\s$ \label{twrepmat} }

We consider next the {\it twisted representation matrices} $\st$, which are
duality transformations
\begin{equation}
 T \dual \st(T,\s)
\end{equation}
from the untwisted representation matrices $T$. The explicit form of the twisted
representation matrices is
\begin{subequations}
\label{explicit-form-of-st}
\begin{equation}
 \st_\nrm(T,\s) \equiv \schisig_\nrm U(\s)_\nrm{}^a \Big{(}\,
 U(T,\s) T_a U\hc(T,\s)\,\Big{)} \label{T-duality-transf}
 \end{equation}
 \begin{equation}
 \st_\nrm(T,\s)_\Nrm{}^\Nsn = \schisig_\nrm U(\s)_\nrm{}^a
 \Big{(}\,U(T,\s) T_a U\hc(T,\s)\,\Big{)}{}_\Nrm{}^\Nsn
\end{equation}
\begin{equation}
 T_a = T_a(\st) =  U\hc(\s)_a{}^\nrm \schisig_\nrm^{-1} \Big{(}\,U\hc(T,\s)
  \st_\nrm(T,\s) U(T,\s)\,\Big{)} \label{T=T(st)}
\end{equation}
\end{subequations}
where we have noted the inverse $T(\st)$ in \eqref{T=T(st)}.
These objects appear for example in the OPEs of the eigencurrents with the
eigenprimary fields
\begin{subequations}
\begin{equation}
\sg(\st,\bz,z,\s) \equiv \sg (T(\T),\bz,z,\s)
\end{equation}
\begin{equation}
\sj_\nrm(z,\s)\sg(\st,\bw,w,\s)  =
\frac{ \sg (\T,\bw,w,\s) }{z-w} \st_{\nrm}(T,\s) + \Ord (z-w)^0
\end{equation}
\begin{equation}
 \sjb\!\!_\nrm(\bz,\s) \sg(\st,\bw,w,\s)  =
-\st_{\nrm}(T,\s) \frac{ \sg (\T,\bw,w,\s) }{\bz-\bw}  + \Ord (\bz-\bw)^0
\end{equation}
\end{subequations}
and will play a central role in the orbifold dynamics below. Examples of $\st$
for specific orbifolds are found in Secs.~\ref{permorb} and \ref{innerorb} and in
App.~\ref{Jperm}.

The twisted representation matrices satisfy many important relations, including
\begin{subequations}
\begin{equation}
 \left[ \st_\nrm(T,\s) ,\st_\nsn(T,\s) \right] =
 i\scf_{\nrm;\nsn}{}^{n(r)+n(s),\d}(\s) \st_{n(r)+n(s),\d}(T,\s)
 \label{eq:reason-for-casimir-of-st}
\end{equation}
\begin{equation}
 [\st_\nrm(T,\s), \st_\nsn(T,\s) \sG^{\nsn;\ntd}(\s)\st_\ntd(T,\s)] = 0
 \label{eq:casimir-of-st}
\end{equation}
\begin{equation}
 \scf_\nrm{}^{\ntd;\nue}(\s) \equiv \scf_{\nrm;\nsn}{}^\nue (\s)
 \sG^{\nsn;\ntd}(\s) = -\scf_\nrm{}^{\nue;\ntd}(\s) \label{eq:antisym-of-scf}
\end{equation}
\end{subequations}
where $\sG(\s)$ and $\scf(\s)$ are the twisted metric and twisted structure
constants of sector $\s$ in \eqref{sG}.
The Casimir property \eqref{eq:casimir-of-st} follows from
\eqref{eq:reason-for-casimir-of-st} because of the antisymmetry of the
twisted structure constants in \eqref{eq:antisym-of-scf}

The normalization condition for the twisted representation matrices
\begin{subequations}
\begin{equation}
\label{eq:trmtt=sg}
  \widehat{\rm Tr}(\sm(\st,\s) \st_\nrm(T,\s)  \st_\nsn(T,\s))  =
  \sG_{\nrm ;\nsn}(\s)
\end{equation}
$$
= \d_{n(r)+n(s),0\, \rmod\,\r(\s)} \, \sG_{\nrm ;-n(r),\n}(\s)
$$
\begin{equation}
\label{trdef}
\widehat{\rm Tr}(AB) \equiv \sum_{r,\m,s,\n} A_\Nrm{}^\Nsn B_\Nsn{}^\Nrm \
\end{equation}
\begin{equation}
  [\sm(\st,\s),\st_\nrm(T,\s)]=0  \label{eq:script-m-T-comm-zero}
\end{equation}
\end{subequations}
follows from \eqref{MTA},\eqref{T-duality-transf} and \eqref{sG0}.
Here we have encountered the {\it twisted data matrix} $\sm (\T,\s)$
\begin{subequations}
\begin{equation}
 M \dual \sm(M,\s)
\end{equation}
\begin{equation}
\label{twisted-data-matrix}
 \sm(\T,\s) \equiv \sm(T(\T),\s) \equiv U(T,\s)M(k,T)U\hc(T,\s)
\end{equation}
\begin{equation}
\label{mtinv}
 \sm^{-1}(\T,\s) = U(T,\s) M^{-1}(k,T) U\hc(T,\s)
\end{equation}
\begin{equation}
 \sm(\T,\s)_{N(r)\m}{}^{N(s)\n} = \sum_I \frac{k_I}{y_I(T^I)}
 \sum_\a U(T,\s)_{N(r)\m}{}^{\a I}  U\hc(T,\s)_{\a I}{}^{N(s)\n}
\end{equation}
\end{subequations}
which is the duality transformation of the data matrix $M$ in \eqref{groupelement}.
Other properties of the twisted data matrix include
\begin{subequations}
\begin{equation}
 [\sm(\T,\s),\sg(\T,\bz,z,\s)] =0 \label{[sm,sg]=0}
\end{equation}
\begin{equation}
\label{eq:msel}
 E(T,\s)\sm(\T,\s)E(T,\s)^\ast = \sm(\T,\s), \quad
 \sm(\T,\s)_{N(r)\m}{}^{N(s)\n}
 \left(1-e^{-\frac{2\pi i (N(r)-N(s))}{R(\s)}}\right) = 0
\end{equation}
\begin{equation}
 \sm(\T,\s)_{N(r)\m}{}^{N(s)\n} = \d_{N(r)-N(s),0\,\text{mod }\r(\s)}
  \sm(\T,\s)_{N(r)\m}{}^{N(r)\n}  \ . \label{eq:sel-rule-for-sm-sol}
\end{equation}
\end{subequations}
The $\sm$-selection rule in \eqref{eq:msel} is dual to the $H$-invariance
\eqref{Mtransf} of the data matrix $M$, and the solution
of this selection rule is given in \eqref{eq:sel-rule-for-sm-sol}.
The twisted data matrix is also a class function
\begin{equation}
 \sm(U(T,\s) v(T,\s);\s) = \sm(U(T,\s);\s)
\end{equation}
according to Eqs.~\eqref{Mtransf}, \eqref{conjugation-in-H} and
\eqref{twisted-data-matrix}.

The twisted representation matrices also satisfy the selection rules
\begin{subequations}
\label{sel-rule-for-st-1}
\begin{equation}
 \st_\nrm(T,\s) = E_{n(r)}(\s)^\ast \big{(}\,E(T,\s)^\ast
 \st_\nrm(T,\s) E(T,\s)\,\big{)}
\end{equation}
\begin{equation}
 \st_\nrm(T,\s)_\Nsn{}^\Ntd ( 1- e^{2\pi i(\frac{n(r)}{\r(\s)}+
 \frac{N(s)-N(t)}{R(\s)}  )}) = 0
\end{equation}
\begin{equation}
 \st_\nrm(T,\s) = E_{n(r)}(\s) \big{(}\,E(T,\s) \st_\nrm(T,\s) E(T,\s)^\ast\,
 \big{)}
\end{equation}
\begin{equation}
\label{selrulet}
 \st_\nrm(T,\s)_\Nsn{}^\Ntd ( 1- e^{-2\pi i(\frac{n(r)}{\r(\s)}+
 \frac{N(s)-N(t)}{R(\s)}  )}) = 0 \ .
\end{equation}
\end{subequations}
These selection rules are dual to the linkage relation \eqref{WTWwTa} and in
 particular to the forms \eqref{eq:WTWwTb-next} of the $H$-invariance of the
 representation matrices $T$. To see this, insert the forms
 \eqref{eq:WTWwTb-next} into the duality transformation \eqref{T-duality-transf}
  and use both eigenvalue problems
 \eqref{ExtendedHEigen} and \eqref{HEigen}. The useful intermediate relations
\begin{subequations}
\begin{equation}
 T_a(U) = w\hc(h_\s)_{a}{}^{b} E(T,\s)^\ast T_b(U) E(T,\s)=
 w(h_\s)_{a}{}^{b} E(T,\s) T_b(U) E(T,\s)^\ast
\end{equation}
\begin{equation}
 T_a(U) \equiv U(T,\s) T_a U\hc(T,\s)
\end{equation}
\end{subequations}
follow from \eqref{eq:WTWwTb-next} and the extended $H$-eigenvalue problem
\eqref{ExtendedHEigen} alone.

The two sets of selection rules in (\ref{sel-rule-for-st-1}) are equivalent, and
the solution of the selection rules is the ``Wigner-Eckart'' relation
\begin{subequations}
\begin{equation}
 \st_\nrm(T,\s)_\Nsn{}^\Ntd = \d_{\frac{R(\s)}{\r(\s)}n(r)+N(s)-N(t),0 \,
 \text{mod}\,R(\s)}
  \st_\nrm(T,\s)_\Nsn{}^{N(s)+\frac{R(\s)}{\r(\s)}n(r),\d} \label{eq:selT}
\end{equation}
\begin{equation}
 \st_\nrm(T,\s)_\Nsn{}^{N(s)+\frac{R(\s)}{\r(\s)}n(r),\d}=0\quad\text{ unless }
 \quad N(s)+\srac{R(\s)}{\r(\s)}n(r)\in\{N(r)\}
\end{equation}
\begin{equation}
 \d_{\frac{R(\s)}{\r(\s)}n(r)+N(s)-N(t),0 \,\text{mod}\,R(\s)}
 =\d_{n(r)+\frac{\r(\s)}{R(\s)}(N(s)-N(t)),0 \,\text{mod}\,\r(\s)}=
 \d_{\frac{n(r)}{\r(\s)}+\frac{N(s)-N(t)}{R(\s)},0 \,\text{mod}\,1} \ .
 \label{3-k-deltas}
\end{equation}
\end{subequations}
The Kronecker delta in \eqref{eq:selT} can be replaced by either of the equivalent forms in \eqref{3-k-deltas}.
We remark that
\begin{equation}
 [\,E(T,\s)^{\r(\s)},\,\st_\nrm(T,\s)\,] = 0 \label{eq:order-Ers}
\end{equation}
follows from (\ref{eq:selT}) or (\ref{eq:[WrhoT]=0}) and the extended
 $H$-eigenvalue problem.

The twisted representation matrices are also class functions under conjugation
in $H$
\begin{equation}
 \st(U(\s)v(\s), U(T,\s)v(T,\s);\s) = \st(U(\s), U(T,\s);\s)
\end{equation}
according to \eqref{conjugation-in-H}, \eqref{explicit-form-of-st} and the
 linkage relation \eqref{WTWwTa}. More generally, all the duality transformations
 of this paper are class functions, although we will not mention this explicitly
  below.

\subsection{Other Duality Transformations}

The affine-Sugawara stress tensors $T(z)$ and $\bar T(\bz)$ can easily be rewritten in
 terms of the eigencurrents
\begin{subequations}
\label{asst}
\begin{eqnarray}
T (z) & = & \sum_{r,\m,\n} {\cL}_{\sgb (\s)}^{\nrm ; \mnrn} (\s)
: \sj_{\nrm} (z,\s) \sj_{\mnrn} (z,\s) : \nn \\
& & \equiv
{\cL}_{\sgb (\s)}^{\nrm ;  \mnrn} (\s)
: \sj_{\nrm} (z,\s) \sj_{\mnrn} (z,\s) :
\end{eqnarray}
\begin{equation}
\bar{T} (\bz) = {\cL}_{\sgb (\s)}^{\nrm ; \mnrn} (\s)
: \bar{\sj}_{\nrm} (\bz,\s) \bar{\sj}_{\mnrn} (\bz,\s) :
\end{equation}
\begin{equation}
L_g \dual {\cL}_{\sgb (\s)}(\s)
\end{equation}
\begin{equation}
\label{twL}
{\cL}_{\sgb (\s)}^{\nrm ; \nsn}(\s)=\schisig^{-1}_\nrm \schisig^{-1}_\nsn
L_g^{ab} U\hc(\s)_a{}^\nrm U\hc(\s)_b{}^\nsn
\end{equation}
\begin{equation}
\label{eigL}
=\delta_{n(r)+n(s),0\,\text{mod }\r(\s)}
{\cL}_{\sgb (\s)}^{\nrm ;\mnrn}(\s)
\end{equation}
\begin{equation}
\label{lgsym}
{\cL}_{\sgb (\s)}^{\nrm ;\nsn}(\s) ={\cL}_{\sgb (\s)}^{\nsn ;\nrm}(\s)
\end{equation}
\end{subequations}
where ${\cL}_{\sgb (\s)}(\s)$ is the duality transformation  called
the {\it twisted inverse inertia tensor}
\cite{Evslin:1999qb,deBoer:1999na,Evslin:1999ve,Halpern:2000vj} of sector $\s$.

We also find the $T \sg$  and $\bar T \sg$ OPEs
\begin{subequations}
\begin{equation}
T (z) \sg (\T,\bw,w,\s) =
  \frac{\sg(\T,\bw,w,\s) \D_{\sgb (\s)} (\T)}{(z-w)^2} +
\frac{\part_w}{z-w} \sg(\T,\bw,w,\s) + \Ord (z-w)^0
\end{equation}
\begin{equation}
\bar{T}(\bz) \sg(\T,\bw,w,\s) =
\left( \frac{\D_{\sgb (\s)} (\T) }{(\bz-\bw)^2} +
\frac{\part_{\bw}}{\bz-\bw} \right) \sg(\T,\bw,w,\s) + \Ord (\bz-\bw)^0 \ .
\end{equation}
\end{subequations}
Here we have encountered the {\it twisted conformal weight matrix}
 $\D_{\sgb (\s)} (\T)$
\begin{subequations}
\label{twcfm0}
\begin{equation}
D_g(T) \dual \D_{\sgb (\s)} (\T)
\end{equation}
\begin{equation}
\label{tcwm} \D_{\sgb (\s)} (\T) \equiv U(T,\s) D_g (T) U\hc
(T,\s) = \lr^{\nrm ;\mnrn} \T_{\nrm}(T,\s) \T_{\mnrn}(T,\s)
\end{equation}
\end{subequations}
which is the duality transformation of the conformal weight matrix in
\eqref{Dgdef}. The twisted conformal weight matrix satisfies the selection rule
\begin{equation}
\D_{\sgb (\s)} (\T)_{N(r)\m}{}^{N(s)\n} =
\delta_{N(r)-N(s),0\,\text{mod }R(\s)} \D_{\sgb (\s)} (\T)_{N(r)\m}{}^{N(r)\n}
\end{equation}
which is dual to the invariance \eqref{cfwinv} of the conformal weight
matrix.

This completes the set of duality transformations that we will need to describe
the WZW orbifolds in this paper. We finally note that all these twisted objects
reduce to their original untwisted counterparts
\begin{subequations}
\begin{equation}
\G (0) = G \sp \F (0) = f \sp {\cL}_{\sgb (0)}(0) = L_g
\end{equation}
\begin{equation}
\T (T,0) = T \sp
\M (T,0) = M (T) \sp \D_{\sgb (0)} (\T)  = D_g (T)
\end{equation}
\end{subequations}
in the untwisted sector $\s = 0$ of each orbifold.

\section{The General WZW Orbifold $A_g(H)/H$ \label{gwzw} }

\subsection{Local Isomorphisms and Monodromies \label{liso} }

To go from the symmetric theory $A_g(H)$ in the eigenfield basis
to sector $\s$ of the WZW orbifold $A_g(H)/H$, we apply the principle of
{\it local isomorphisms}
\cite{Borisov:1997nc,deBoer:1999na,Evslin:1999ve,Halpern:2000vj}.
This principle tells us to replace all eigenfields by corresponding {\it twisted
operators} with definite monodromy,
\begin{subequations}
\label{local-isomorphisms}
\begin{equation}
 \sj_\nrm(z,\s),\bar{\sj}_\nrm(\bz,\s) \dual \hj_\nrm(z,\s),\hjb_\nrm(\bz,\s)
 \end{equation}
 \begin{equation}
 \sg(\st,\bz,z,\s) \dual \hat{g}(\st,\bz,z,\s)
\end{equation}
\end{subequations}
where $\hj (\s)$, $\hjb (\s)$ are the twisted
left and right mover currents and $\hg (\T,\bz,z,\s)$
is the {\it twisted affine primary field} corresponding to the twisted
representation matrix $\T = \T (T,\s)$. By local isomorphisms from \eqref{[sm,sg]=0} the twisted
affine primary fields satisfy
\begin{equation}
\label{Mtgt}
[\sm(\T,\s),\hg(\T,\bz,z,\s)] =0
\end{equation}
and all the twisted operators inherit the periodicities
\begin{subequations}
\begin{equation}
\label{pertc}
\hj_{n(r) \pm \rho (\s),\mu} (z,\s) = \hj_{\nrm} (z,\s)
\sp
\hjb_{n(r) \pm \rho (\s),\mu} (\bz,\s) = \hjb_{\nrm} (\bz,\s)
\end{equation}
\begin{equation}
\label{pertg}
\hg (\T,\bz,z,\s)_{N(r) \pm R(\s),\m}{}^{N(s) \n}
=\hg (\T,\bz,z,\s)_{N(r)\m}{}^{N(s) \pm R(\s), \n}
=\hg (\T,\bz,z,\s)_{N(r)\m}{}^{N(s) \n}
\end{equation}
\end{subequations}
of the spectral indices $\{ n(r) \}$ and $\{ N(r) \}$.

For the twisted currents, the principle of local isomorphisms tells us to
 take the  monodromies  to be the old automorphic responses
\begin{subequations}
\label{jjbmon}
\begin{equation}
\label{jmon}
\hat J_{\nrm} (z e^{2 \pi i}, \s) = E_{n(r)} (\s)
\hat J_{\nrm} (z , \s) = e^{- 2 \pi i \srac{n(r)}{\rho (\s)}} \hat J_{\nrm} (z , \s)
\end{equation}
 \begin{equation}
 \label{jbmon}
\ \hat{\bar{J}}_{\nrm} (\bz e^{-2 \pi i}, \s) = E_{n(r)} (\s)
\hat{\bar{J}}_{\nrm} (\bz , \s) = e^{- 2 \pi i \srac{n(r)}{\rho (\s)}}
\hat{\bar{J}}_{\nrm} (\bz , \s) \ .
\end{equation}
\end{subequations}
The rule \eqref{jmon} was verified for the twisted left mover currents in
Refs.~\cite{deBoer:1999na,Halpern:2000vj} and
the rule \eqref{jbmon} says that the same monodromies are obtained for the
twisted right mover currents when the same path is followed for $z$ and
$\bz = z^\ast$. That this is the correct
rule for the right mover currents can be argued both at
the level of characters (see Subsec.~\ref{quarg}) and at the classical level
(see Subsec.~\ref{clarg}) where it is understood as a consequence of locality.

Using \eqref{asst} and the local isomorphisms, we obtain the orbifold stress
tensors
\begin{subequations}
\label{local-isomorphismsT}
\begin{equation}
 T(z) \dual \hat{T}_\s (z) \ , \quad \bar T(\bz) \dual
 \hat{\bar{T}}_\s (\bz)
 \end{equation}
\begin{equation}
\hat{T}_\s (z) = {\cL}_{\sgb (\s)}^{\nrm ; \mnrn} (\s)
: \hj_{\nrm} (z,\s) \hj_{\mnrn} (z,\s) :
 \end{equation}
\begin{equation}
\hat{\bar{T}}_\s (\bz) = {\cL}_{\sgb (\s)}^{\nrm ; \mnrn} (\s)
: \hjb_{\nrm} (\bz,\s) \hjb_{\mnrn} (\bz,\s) :
\end{equation}
\begin{equation}
\label{ttbmon}
\hat{T}_\s (z e^{2 \pi i})  =\hat{T}_\s (z)
\sp \hat{\bar{T}}_\s (\bz e^{-2 \pi i})  =\hat{\bar{T}}_\s (\bz)
\end{equation}
\end{subequations}
which are the {\it twisted affine-Sugawara constructions}
\cite{Kac:1984mq,Freericks:1988zg,deBoer:1999na,Halpern:2000vj} of sector $\s$.
Here, ${\lr}(\s)$ is the twisted inverse inertia tensor \eqref{twL} of the
twisted affine-Sugawara construction and $:\cdot :$ is OPE normal ordering
\cite{Evslin:1999qb,deBoer:1999na,Evslin:1999ve,Halpern:2000vj} of the twisted
currents.

The  monodromies of the twisted primary fields $\hg (\T,\bz,z,\s)$ are quite
complicated because the monodromies of the untwisted affine primary fields
are complicated  (see e.g. Ref.~\cite{Halpern:1996et}). The monodromies of
$\hg (\T,\bz,z,\s) $ are however determined in principle by the twisted KZ
 equations discussed below. On the other hand, the  monodromies of the
classical limit of the affine primary fields (the WZW group elements)
are trivial, so the {\it classical monodromies} of the classical limit of the
twisted affine primary fields (the group orbifold elements) are easily
deduced from the local isomorphisms (see Subsec.~\ref{clarg}).

\subsection{WZW Orbifold OPEs}

For sector $\s$ of the general WZW orbifold $A_g(H)/H$,
we then obtain the twisted current-current
system
\begin{subequations}
\begin{equation}
 \hj_\nrm(z,\s)\hj_\nsn(w,\s) =\quad \frac{\delta_{n(r)+n(s),0\,\text{mod}\,\r(\s)}
 \G_{\nrm;\mnrn} (\s) }{(z-w)^2} \hskip 3cm
\end{equation}
$$
\hskip 3cm + \frac{i\scf_{\nrm;\nsn}{}^{n(r)+n(s),\delta}(\s)
 \hj_{n(r)+n(s),\delta}(w,\s)}{z-w} + \Ord(z-w)^0
 $$
\begin{equation}
 \hjb_\nrm(\bz,\s)\hjb_\nsn(\bw,\s)=\quad \frac{\delta_{n(r)+n(s),0\,\text{mod}\,\r(\s)}
 \G_{\nrm;\mnrn} (\s) }{(\bz-\bw)^2} \hskip 3cm
\end{equation}
$$
\hskip 3cm + \frac{i\scf_{\nrm;\nsn}{}^{n(r)+n(s),\delta}(\s)
\hjb_{n(r)+n(s),\delta}(\bw,\s)}{\bz-\bw} + \Ord(\bz-\bw)^0
$$
\begin{equation}
\hj_\nrm(z,\s )  \hjb_\nrm(\bw,\s) = \mbox{regular}
\end{equation}
\end{subequations}
by local isomorphisms from the corresponding OPEs of the eigencurrents.
Here $\G (\s)$ and $\F (\s)$ are respectively the twisted metric and the twisted
structure constants given in Eq.~\eqref{sG} .

The twisted affine-Sugawara
constructions $\hat{T}_\s$ and $\hat{\bar{T}}_\s$ satisfy the OPEs of
 Vir$\oplus$Vir, as expected, with central charges
\begin{equation}
\hat{c}(\s) =\hat{\bar{c}} (\s) =
2 \g_{n(r) \mu; -n(r), \nu}(\s) \lr^{n(r) \mu ; -n(r), \nu}(\s)
 =2 G_{ab} L_g^{ab}=c_g \sp \srange
\end{equation}
which are equal to the untwisted affine-Sugawara central charge $c_g$.
Moreover, we find from \eqref{utj}, \eqref{utjb}, \eqref{utjjb} the OPEs
\begin{subequations}
\label{orbTJ}
\begin{equation}
\hat T_\s(z) \hj_{\nrm} (w,\s) = \left( \frac{1}{(z-w)^2} + \frac{\partial_w }{z-w} \right)
\hj_{\nrm} (w,\s)
+\Ord (z-w)^0
\end{equation}
\begin{equation}
\hat{\bar{T}}_\s(\bz) \hjb_{\nrm} (\bw,\s) = \left( \frac{1}{(\bz-\bw)^2} + \frac{\partial_{\bw}}{\bz-\bw}
\right) \hjb_{\nrm} (\bw,\s)
+\Ord (\bz-\bw)^0
\end{equation}
\begin{equation}
\hat T_\s(z) \hjb_{\nrm} (\bw,\s) = \hat{\bar{T}}_\s(\bz)\hj_{\nrm} (w,\s)=
{\rm regular}
\end{equation}
\end{subequations}
of the twisted affine-Sugawara constructions with the twisted currents.

For the OPEs involving the twisted affine primary fields $\hg (\T,\bz,z,\s)$, we obtain
\begin{subequations}
\begin{equation}
\label{hjhgope}
\hj_{\nrm} (z,\s) \hg (\T,\bw,w,\s) = \frac{\hg (\T,\bw,w,\s)}{z-w}
\T_{\nrm}(T,\s) + \Ord (z-w)^0
\end{equation}
\begin{equation}
\hjb_{\nrm} (\bz,\s) \hg (\T,\bw,w,\s) =-\T_{\nrm}(T,\s)
\frac{\hg (\T,\bw,w,\s)}{\bz-\bw} + \Ord (\bz-\bw)^0
\end{equation}
\begin{equation}
\hat{T}_\s (z) \hg (\T,\bw,w,\s) =
  \frac{\hg(\T,\bw,w,\s) \D_{\sgb (\s)} (\T)}{(z-w)^2} +
\frac{\part_w}{z-w} \hg(\T,\bw,w,\s) + \Ord (z-w)^0
\end{equation}
\begin{equation}
\hat{\bar{T}}_\s(\bz)\hg(\T,\bw,w,\s) =
\left( \frac{\D_{\sgb (\s)} (\T) }{(\bz-\bw)^2} +
\frac{\part_{\bw}}{\bz-\bw} \right) \hg(\T,\bw,w,\s) + \Ord (\bz-\bw)^0
\end{equation}
\end{subequations}
where $\T (T,\s)$ and $\D_{\sgb (\s)}(\T) $ are respectively
the twisted representation matrices
and twisted conformal weight matrices of sector $\s$.

The local isomorphisms also give
the {\it twisted vertex operator equations}
\begin{subequations}
\label{tvoe}
\begin{equation}
\label{tvoel}
\partial \hg (\T,\bz,z,\s) = 2 {\cL}_{\sgb (\s)}^{n(r)\m ; - n(r), \n} (\s)
: \hat J_{n(r) \m} (z,\s) \hg (\T,\bz,z,\s) \T_{-n(r), \n}(T,\s) :
\end{equation}
\begin{equation}
\label{tvoer}
\bar \partial \hg (\T,\bz,z,\s) = - 2 {\cL}_{\sgb (\s)}^{n(r)\m ; - n(r),\n} (\s)
: \hat{\bar{J}}_{n(r) \m} (\bz,\s)\T_{-n(r), \n}(T,\s) \hg (\T,\bz,z,\s) :
\end{equation}
\end{subequations}
from the untwisted vertex operator equations \eqref{difrel}. The normal ordering
here is also OPE normal ordering
\cite{Evslin:1999qb,deBoer:1999na,Evslin:1999ve,Halpern:2000vj}. These
world-sheet differential equations for the twisted affine primary fields
are the WZW orbifold operator equations of motion, and they will be
 central in the derivation of the twisted KZ equations below.

\subsection{Mode Form of the General WZW Orbifold Algebra \label{MF} }

Using the monodromies \eqref{jjbmon} and \eqref{ttbmon} we find
the mode relations:
\begin{subequations}
\label{jmod}
\begin{equation}
\label{jmodl}
\hat J_{n(r) \m}(z,\s) = \sum_{m  \in \sz}   \hat J_{n(r) \m} ( m + \srac{n(r)}{\rho (\s)} )
z^{-(m+\srac{n(r)}{\rho (\s)}) -1}
\end{equation}
\begin{equation}
\label{jmodr}
{\hat{\bar{J}}}_{n(r) \m}(\bz,\s) = \sum_{m \in \sz}{\hat{\bar{J}}}_{n(r) \m} ( m + \srac{n(r)}{\rho (\s)} )
\bz^{(m+\srac{n(r)}{\rho (\s)}) -1}
\end{equation}
\begin{equation}
\label{perJ}
 \hat J_{n(r) \pm \rho (\s),\m} ( m \mp 1 + \srac{n(r) \pm \rho (\s) }{\rho (\s)} )
=\hat J_{n(r) \m} ( m + \srac{n(r)}{\rho (\s)} )
\end{equation}
\begin{equation}
\label{perbJ}
 \hat{\bar{J}}_{n(r) \pm \rho(\s), \m} ( m \mp 1 + \srac{n(r) \pm \rho (\s) }{\rho (\s)} )
=\hat{\bar{J}}_{n(r) \m} ( m + \srac{n(r)}{\rho (\s)} )
\end{equation}
\begin{equation}
\hat T_\s (z) = \sum_{m  \in \sz} L_\s(m) z^{-m -2} \sp
{\hat{\bar{T}}}_\s (\bz) = \sum_{m\in \sz} \bar{L}_\s(m) \bz^{-m -2} \ .
\end{equation}
\end{subequations}
The periodicity of the twisted current modes in \eqref{perJ}, \eqref{perbJ}
follows from the periodicity \eqref{pertc} of the twisted currents.

Then the orbifold OPEs give the {\it general twisted current algebra}
\begin{subequations}
\label{mode-current-algebra}
\begin{equation}
\label{mcall}
 [\hj_\nrm(\mnrrs),\hj_\nsn(\nnsrs)]=i\scf_{\nrm;\nsn}{}^{n(r)+n(s),\de}(\s)
 \hj_{n(r)+n(s),\de}(\mnnrnsrs)
 \end{equation}
 $$
 + (\mnrrs)\de_{\mnnrnsrsf,0}\sG_{\nrm;\mnrn}(\s)
$$
\begin{equation}
\label{mcalr}
  [\hjb_\nrm(\mnrrs),\hjb_\nsn(\nnsrs)]=i\scf_{\nrm;\nsn}{}^{n(r)+n(s),\de}(\s)
  \hjb_{n(r)+n(s),\de}(\mnnrnsrs)
\end{equation}
$$
- (\mnrrs)\de_{\mnnrnsrsf,0}\sG_{\nrm;\mnrn}(\s)
$$
\begin{equation}
[\hj_\nrm(\mnrrs),\hjb_\nsn(\nnsrs)] = 0  \sp m,n \in {\mathbb Z}\ ,
 \quad \srange
\end{equation}
\end{subequations}
in each sector $\s$ of all WZW orbifolds $A_g(H)/H$. The general twisted
current algebra \eqref{mode-current-algebra} is a central result of this paper.

We have checked that the orbifold adjoint \cite{Halpern:2000vj} of
the twisted currents
\begin{subequations}
\label{jhat-adj-grp}
\begin{equation}
 \hj_\nrm(\mnrrs)\hc = \sum_\n \sr_\nrm{}^{-n(r),\n}(\s) \hj_{-n(r),\n}
 (\mnrrsminus) \label{jhat-adj}
\end{equation}
\begin{equation}
 \hjb_\nrm(\mnrrs)\hc = \sum_\n \sr_\nrm{}^{-n(r),\n}(\s) \hjb_{-n(r),\n}
 (\mnrrsminus) \label{jhatb-adj}
\end{equation}
\end{subequations}
defines a real form of the general twisted current algebra
\eqref{mode-current-algebra}. Here, $\sr$ is the orbifold conjugation matrix of
Ref.~\cite{Halpern:2000vj}.

The integral affine subalgebra of the general twisted current algebra
\eqref{mode-current-algebra} is
\begin{subequations}
\label{integral-affine-sub-alg}
\begin{equation}
 [\hj_{0\m}(m),\hj_{0\n}(n)] =
 i\scf_{0\m;0\n}{}^{0\d}(\s) \hj_{0\d}(m+n) + m\d_{m+n,0}\sG_{0\m ;0\n}(\s)
\end{equation}
\begin{equation}
 [\hjb_{0\m}(m),\hjb_{0\n}(n)] = i\scf_{0\m;0\n}{}^{0\d}(\s) \hjb_{0\d}(m+n) -
   m\d_{m+n,0}\sG_{0\m ;0\n}(\s)
\end{equation}
\begin{equation}
 [\hj_{0\m}(m),\hjb_{0\n}(0)] = 0 \ .
 \label{integral-affine-sub-algebra-relabel}
\end{equation}
\end{subequations}
 We also note the Lie subalgebra
\begin{subequations}
\label{liesa}
\begin{equation}
 [\hj_{0\m}(0),\hj_{0\n}(0)] = i\scf_{0\m;0\n}{}^{0\d}(\s) \hj_{0\d}(0) ,\quad
 [\hjb_{0\m}(0),\hjb_{0\n}(0)] =    i\scf_{0\m;0\n}{}^{0\d}(\s) \hjb_{0\d}(0)
\end{equation}
\begin{equation}
  [\hj_{0\m}(0),\hjb_{0\n}(0)] = 0
\end{equation}
\end{subequations}
generated by the zero modes of the integral affine subalgebra. It is expected
(see Secs.~\ref{kzsec} and \ref{innerorb}) that these zero modes contribute to the
{\it residual symmetry} in each sector $\s$ of each WZW
orbifold $A_g(H)/H$. Further discussion of the general twisted current algebra is
found in the following subsection, and special cases are worked out in
Secs.~\ref{permorb} and \ref{innerorb}.

The twisted affine-Sugawara generators $L_\s (m)$, $\bar L_\s (m)$ satisfy the algebra of Vir$\oplus$Vir,
and we also find the commutators
\begin{subequations}
\label{modeTcur}
\begin{equation}
[ L_\s (m), \hj_{\nrm} (z,\s) ] = z^m ( z \partial + (m+1) ) \hj_{\nrm} (z,\s)
\end{equation}
\begin{equation}
[ \bar L_\s (m), \hjb_{\nrm} (\bz,\s) ] = \bz^m ( \bz \bar\partial + (m+1) )
\hjb_{\nrm} (\bz,\s)
\end{equation}
\begin{equation}
[ L_\s (m), \hjb_{\nrm} (\bz,\s) ] =[ \bar L_\s (m), \hj_{\nrm} (z,\s) ] = 0
\end{equation}
\begin{equation}
\label{lmco}
[ L_\s (m), \hj_{\nrm} (n + \srac{n(r)}{\rho (\s)}) ] =
- (n + \srac{n(r)}{\rho (\s)})\hj_{\nrm} (n +m+ \srac{n(r)}{\rho (\s)})
\end{equation}
\begin{equation}
\label{lmcor}
[ \bar L_\s (m), \hjb_{\nrm} (n + \srac{n(r)}{\rho (\s)}) ] =
 (n + \srac{n(r)}{\rho (\s)})\hjb_{\nrm} (n -m+ \srac{n(r)}{\rho (\s)})
\end{equation}
\begin{equation}
[ L_\s (m), \hjb_{\nrm} (n + \srac{n(r)}{\rho (\s)}) ] =
[ \bar L_\s (m), \hj_{\nrm} (n + \srac{n(r)}{\rho (\s)}) ] = 0
\end{equation}
\end{subequations}
for the Virasoro generators with the twisted currents.

Finally, we also obtain the commutators with the twisted affine primary fields
\begin{subequations}
\label{mode-current-g}
\begin{equation}
\label{jgct}
[ \hj_\nrm (\mnrrs), \hg (\T,\bz,z,\s) ] = \hg (\T,\bz,z,\s)
\T_\nrm (T,\s)  z^\mnrrs
\end{equation}
\begin{equation}
\label{rjgct}
[ \hjb_\nrm (\mnrrs), \hg (\T,\bz,z,\s) ] = -  \bz^{-(\mnrrs)}
\T_\nrm (T,\s)  \hg (\T,\bz,z,\s)
\end{equation}
\begin{equation}
\label{Lgu}
[L_\s (m), \hg (\T,\bz,z,\s)] = \hg(\T,\bz,z,\s)  (\overleftarrow{\partial} z
+ (m+1)\D_{\sgb (\s)} (\T) )z^m
\end{equation}
\begin{equation}
\label{Lgb}
[\bar L_\s (m), \hg (\T,\bz,z,\s)] = \bz^m (\bz \bar{\partial}
+ (m+1)  \D_{\sgb (\s)} (\T) )\hg (\T,\bz,z,\s)
\end{equation}
\end{subequations}
and we have checked all Jacobi identities among the commutators
\eqref{mode-current-algebra}, \eqref{modeTcur} and \eqref{mode-current-g}.

\subsection{Identification of the General Twisted Current Algebra \label{quarg}}

In this subsection, we discuss the form of the general twisted current
algebra \eqref{mode-current-algebra}.

The general twisted left mover current algebra \eqref{mcall} was called
$\gfrakh(\s)=\gfrakh(H\subset {\rm Aut}(g);\s)$ in Refs.~\cite{deBoer:1999na} and
\cite{Halpern:2000vj}. The twisted right mover algebra in \eqref{mcalr} appears
for the first time in this paper. The twisted right mover algebra has the same
 form as the twisted left mover algebra, but with
the opposite sign of the central term. Both algebras are consistent with the
Jacobi identity \cite{Halpern:2000vj}, since the $\scf$ and $\sG$ terms satisfy
the identity separately. We will find it convenient to refer to the general
twisted current algebra \eqref{mode-current-algebra} as
\begin{subequations}
\begin{equation}
 {\gfrakh}(h_\s) \oplus {\bar{\gfrakh}}(h_\s)
\end{equation}
\begin{equation}
 \gfrakh(h_\s)  \equiv \gfrakh ( H \subset {\rm Aut}(g); h_\s )
 \equiv \gfrakh(H \subset {\rm Aut}(g);\s)
\end{equation}
\end{subequations}
and more generally to follow the convention $\s \rightarrow h_\s$ in the
discussion below.

In fact the twisted right mover algebra of sector $\s$ is not a new algebra.
To understand this, consider first the mode-number reversed form of this algebra
\begin{subequations}
\label{double-bar-alg}
$$
[\hjb^{\;{\rm R}}_\nrm(m\+\srac{n(r)}{\r(\s)}), \hjb^{\; \rm R}_\nsn(n\+\srac{n(s)}{\r(\s)}) ] =
 i\scf_{-n(r),\m;-n(s),\n}{}^{\!-n(r)-n(s),\d}(\s)\hjb^{\; \rm R}_{n(r)+n(s),\d}
 (m\+n\+\srac{n(r)+n(s)}{\r(\s)})
$$
\begin{equation}
+ (m+\srac{n(r)}{\r(\s)})\,\d_{m+n+\frac{n(r)+n(s)}{\r(\s)},\,0}\;
\sG_{-n(r),\m;n(r),\n}(\s)
\end{equation}
\begin{equation}
 \hjb^{\; \rm R}_\nrm(\mnrrs) \equiv \hjb_{-n(r),\m} (-m-\srac{n(r)}{\r(\s)})
\end{equation}
\end{subequations}
which exhibits the correct sign of the central term.

We claim that the algebra \eqref{double-bar-alg} is isomorphic to the twisted
left mover current algebra of the sector corresponding to $h_\s^{-1} \in H$
\begin{equation}
 \bar{\gfrakh}(h_\s) \simeq \gfrakh(h_\s^{-1}) \ .
\end{equation}
To see this, one needs to compare the corresponding forms of the $H$-eigenvalue problem
\begin{subequations}
\begin{equation}
 w(h_\s^{-1}) = w\hc(h_\s), \quad \r(h_\s^{-1}) = \r(h_\s)
\end{equation}
\begin{equation}
 w(h_\s)U\hc(h_\s) = U\hc(h_\s) E_{n(r)}(h_\s)
\end{equation}
\begin{equation}
 w(h_\s^{-1}) U\hc(h_\s) = w\hc(h_\s) U\hc(h_\s) = U\hc(h_\s) E_{n(r)}(h_\s)^\ast
 \label{wUhc=whcUhc=UhcEstar}
\end{equation}
\end{subequations}
where we used \eqref{ExtendedHEigen} to obtain \eqref{wUhc=whcUhc=UhcEstar}.
 This allows us to choose
\begin{equation}
 U\hc(h_\s^{-1})_{a}{}^{\nrm} = U\hc(h_\s)_{a}{}^{-n(r),\m}
\end{equation}
and then the duality transformations for $\sG(\s)$ and $\scf(\s)$ in \eqref{sG}
imply that
\begin{subequations}
\begin{equation}
 \sG_{\nrm;\nsn}(h_\s^{-1})\!=\!\sG_{-n(r),\m;-n(s),\n}(h_\s) ,\;
    \scf_{\nrm;\nsn}{}^{\!\ntd}(h_\s^{-1}) \!= \!\scf_{-n(r),\m;-n(s),\n}
    {}^{\!-n(t),\d}(h_\s)
\end{equation}
$$
  \{\hjb^{\;\rm R}_\nrm(\mnrrs),\hjb^{\; \rm R}_\nsn(\nnsrs) \} =
  i\scf_{\nrm;\nsn}{}^{n(r)+n(s),\d}(h_\s^{-1})
  \hjb^{\; \rm R}_{n(r)+n(s),\d}(m+n+\srac{n(r)+n(s)}{\r(\s)})
$$
\begin{equation}
     \+(\mnrrs)\d_{m+n+\frac{n(r)+n(s)}{\r(\s)},0} \sG_{\nrm;\mnrn}(h_\s^{-1})
\end{equation}
\end{subequations}
which establishes the claim. This argument identifies the general twisted
current algebra \eqref{mode-current-algebra} as
\begin{equation}
 \rm{affine}(g\oplus g) \dual \gfrakh(h_\s) \oplus \bar{\gfrakh}(h_\s) \simeq
  \gfrakh(h_\s) \oplus \gfrakh(h_\s^{-1}) \label{gfrakhbar=gfrakhinv}
\end{equation}
in sector $\s$ of $A_g(H)/H$.

Although the form \eqref{mode-current-algebra} of the general
 twisted current algebra is new, it is important to note that this form
 is in fact consistent with what is known about
  partition functions and characters of general orbifolds \cite{Dijkgraaf:1989hb}.
   The point is that when one sees
\begin{equation}
 Z= \sum_{\s,i} | \chi_i(q,\s) |^2
\end{equation}
in the literature, it is supposed to be read as
\begin{equation}
 Z= \sum_{\s,i}  \chi_{i}(q,\s) \chi_{i^\vee}(\bar{q},\s) \label{Z=chichivee}
\end{equation}
where $i^\vee$ denotes the charge conjugate field of $i$. Moreover, it is known
that when $i$ is a representation in the sector twisted by $h_\s$ then $i^\vee$
is a representation in the sector twisted by $h_\s^{-1}$ (so that the two sectors
 can combine to the identity). Thus, in our notation, what is required by
 \eqref{Z=chichivee}
\begin{equation}
 Z = \sum_{\s,i}  \chi_{i}(q,h_\s) \chi_{i^\vee}(\bar{q},h_\s^{-1})
\end{equation}
is in precise agreement with the identification $\gfrakh(h_\s) \oplus \gfrakh(h_\s^{-1})$
of the general twisted current algebra \eqref{mode-current-algebra}.

We also point out that the general twisted current algebra
   \eqref{mode-current-algebra} guarantees
the consistency of the equations for general twisted boundary states
\begin{equation}
\left(\hj_\nrm(\mnrrs) +  \hjb_\nrm(\mnrrs) \right) | B \rangle_\s = 0 \sp
\s = 0, \ldots ,N_c-1
\end{equation}
because the sum of the twisted left and right mover currents
 has no central term.

\subsection{The Rectification Problem  \label{recpr} }

The identification \eqref{gfrakhbar=gfrakhinv} of the general twisted current
algebra leaves open the question of when
the twisted right and left mover current algebras are copies
\begin{equation}
\gfrakh(h_\s^{-1}) \simeq \gfrakh(h_\s)
\end{equation}
which we call {\it the rectification
problem}.

In a simpler example, we find for all constants $\theta (\s)$ that
the further redefinition
\begin{subequations}
\begin{equation}
\label{rmcot}
 \hjbb_\nrm(\mnrrs) \equiv \theta_\nrm(\s) \hjb_{-n(r),\m}
 (-m-\srac{n(r)}{\r(\s)})
\end{equation}
\begin{equation}
\label{rmco}
[ \bar L_\s (m), \hjbb_{\nrm} (n + \srac{n(r)}{\rho (\s)} )]
= - (n + \srac{n(r)}{\rho (\s)} ) \hjbb_{\nrm} (n + m + \srac{n(r)}{\rho (\s)})
\end{equation}
\begin{equation}
 \hjb_\nrm(\bz,\s) = \theta^{-1}_{-n(r),\m}(\s) \sum_{m \in \sz} \hjbb_{-n(r),\m}
 (m-\srac{n(r)}{\r(\s)}) \bz^{-(m-\frac{n(r)}{\r(\s)}) -1}
\end{equation}
\end{subequations}
rectifies the right mover commutator \eqref{lmcor} into a copy
\eqref{rmco} of the left mover commutator \eqref{lmco}. In the language of
Refs.~\cite{Evslin:1999ve,Halpern:1996et}, the commutators \eqref{lmco} and
\eqref{rmco} identify the currents $\hj$ and $\hjbb$
\begin{subequations}
\begin{eqnarray}
\hj_\nrm  (z,\s) & = &  \sum_{m \in \sz} \hj_\nrm (m + \srac{n(r)}{\rho (\s)})
z^{- (m + \srac{n(r)}{\rho (\s)}) -1} \\
\hjbb_\nrm  (\bz,\s) & \equiv & \sum_{m \in \sz} \hjbb_\nrm (m + \srac{n(r)}{\rho (\s)})
\bz^{- (m + \srac{n(r)}{\rho (\s)}) -1} = \theta_\nrm (\s) \hjb_{\mnrm} (\bz,\s)
\;\;\; \;\;\; \\
\hat{\bar{T}}_\s (\bz)  \hjbb_\nrm (\bw,\s) & = &
\left( \frac{1}{(\bz -\bw)^2} + \frac{\partial_{\bw}}{\bz-\bw} \right)
\hjbb_\nrm (\bw,\s) +\Ord (\bz-\bw)^0
\end{eqnarray}
\begin{equation}
\hj_\nrm (z e^{2 \pi i} ) = e^{- 2 \pi i \srac{n(r)}{\rho (\s)} }
\hj_\nrm (z) \sp
\hjbb_\nrm (\bz e^{-2 \pi i} ) = e^{2 \pi i \srac{n(r)}{\rho (\s)} }
\hjbb_\nrm (\bz)
\end{equation}
\end{subequations}
as twisted (1,0) and twisted (0,1) operators respectively under the left and
right mover twisted affine-Sugawara constructions.

For the current algebra itself, the rectification problem may be trivially solved by
$\theta (0) =1$ in the untwisted sector $\s = 0$ of any orbifold
\begin{subequations}
\begin{equation}
U(0) =1 \sp \G (0) = G \sp \scf (0) = f \sp
J (m) \equiv \hat J (m) \sp   \bar J^\sharp (m) \equiv \hjb (- m)
\end{equation}
\begin{equation}
[J_a (m),J_b(n) ] = if_{ab}{}^c J_c (m+n) + m G_{ab} \de_{m+n,0}
\end{equation}
\begin{equation}
[\jbb_a (m),\jbb_b(n) ] = if_{ab}{}^c \jbb_c (m+n) + m G_{ab} \de_{m+n,0}
\end{equation}
\begin{equation}
[J_a (m),\jbb_b(n) ] = 0
\end{equation}
\end{subequations}
because $h_0 = h_0^{-1} = 1$.

So far as the general twisted current algebra is concerned, the situation does
not seem so simple: We shall find for the permutation orbifolds
 (see Subsecs.~\ref{cyclorb}, \ref{snorb} and \ref{unrec})
and the inner-automorphic orbifolds (see Subsec.~\ref{innerprop}) that it is
possible for all $\s$ to choose the constants $\theta (\s)$ to
rectify the right mover algebra  $\gfrakh(h_\s^{-1})$ into a
copy $\gfrakh(h_\s)$ of the left mover algebra.
However (see also App.~\ref{recpro}), we do not have a proof  that such
rectification can be extended across all twisted right mover algebras
$\gfrakh(h_\s^{-1})$.

\subsection{Orbifold Correlators and $L_\s (0)$, $\bar L_\s(0)$ Ward Identities}

For all WZW orbifolds $A_g(H)/H$, the correlators of the twisted primary fields
are
\begin{equation}
\label{cortp}
\hat A (\s) \equiv
\hat A_\s(\T,\bz,z) \equiv {}_\s \langle 0| \hg (\T^{(1)},\bz_1,z_1,\s) \cdots
\hg (\T^{(N)},\bz_N,z_N,\s) | 0 \rangle_\s \hskip 1cm
\end{equation}
$$
= \langle  \hg (\T^{(1)},\bz_1,z_1,\s) \cdots
\hg (\T^{(N)},\bz_N,z_N,\s)  \rangle_\s
$$
and in what follows, we use the notation\footnote{The distinction between these
Greek letters and those of the degeneracy indices will be clear in
context.}
\begin{equation}
T^{(\m)}, \bz_\m ,z_\m \sp \m = 1 \ldots N \quad : \quad
[\T^{(\m)},\T^{(\n)}] = 0 \quad \mbox{when}\;\,\mu \neq \n
\end{equation}
for the coordinates and twisted representation matrices.
In \eqref{cortp}, the ground state $|0 \rangle_\s$ (or twist field state%
\footnote{As seen in \eqref{virprim}, twist field states $|0\rangle_\s = \tau_\s (0) | 0 \rangle$ are not
in general $Sl(2)\oplus Sl(2)$ invariant so that translation invariance
is lost \cite{Borisov:1997nc} in the twisted sectors.}) of
sector $\s$ is primary  under Vir$\oplus$Vir
\begin{subequations}
\label{virprim}
\begin{equation}
L_\s (m \geq 0) | 0 \rangle_\s = \de_{m,0} \hat \Delta_0 (\s) | 0 \rangle_\s
\ , \quad
\bar L_\s (m \geq 0) | 0 \rangle_\s = \de_{m,0} \hat{\bar{\Delta}}_0 (\s) | 0 \rangle_\s
\end{equation}
\begin{equation}
{}_\s \langle 0 | L_\s (m \leq 0) = {}_\s \langle 0 | \hat \Delta_0 (\s)\de_{m,0}
\ , \quad
{}_\s \langle 0 | \bar L_\s (m \leq 0) =
{}_\s \langle 0 | \hat{\bar{\Delta}}_0 (\s)\de_{m,0} \ .
\end{equation}
\end{subequations}
Using \eqref{Lgu}, \eqref{Lgb}, this gives the $L_\s (0)$, $\bar L_\s(0)$
Ward identities
\begin{equation}\label{L0wi}
 \hat A (\s)  \sum_{\m=1}^N \left(\overleftarrow{\partial_\m} z_\m
  + \D_{\sgb (\s)} (\T^{(\m)} ) \right) = 0 \sp
 \sum_{\m=1}^N \left( \bz_\m \bar \partial_\m + \D_{\sgb (\s)} (\T^{(\m)} )
 \right) \hat A (\s)  =0
\end{equation}
where $\D_{\sgb (\s)} (\T^{(\m)}) $ is the twisted conformal weight matrix
\eqref{twcfm0} of
twisted representation $\T^{(\m )}$. Other Ward identities exist which are associated
to the Lie subalgebra \eqref{liesa} and other subalgebras
of the general twisted current algebra,  and we discuss
these on a case by case basis below. We also note that many orbifolds have other
Ward identities associated to extended Virasoro algebras
(see Ref.\cite{Borisov:1997nc}) but we will not study these here.

\subsection{The Classical Theory of WZW Orbifolds  \label{clarg} }

We finally consider the corresponding classical theory of all WZW orbifolds, where we find that
locality dictates the monodromies \eqref{jjbmon} of the twisted right
and left mover currents.

In the classical WZW model on semisimple $g$, we know that the matrix currents have
the local form
\begin{subequations}
\begin{equation}
J(T,z) = J_a (z) G^{ab} T_b \propto g^{-1} (T,\bz,z) \bar
\partial g(T,\bz,z)
\end{equation}
\begin{equation}
\bJ(T,\bz) = \bJ_a (z) G^{ab} T_b \propto g (T,\bz,z)
\partial g^{-1} (T,\bz,z)
\end{equation}
\begin{equation}
\label{jeom}
g\hc (T,\bz,z) g(T,\bz,z) = 1 \sp
\bar \partial J(T,z) = \partial \bJ (T,\bz) = 0
\end{equation}
\end{subequations}
where $g(T,\bz,z)$ is the {\it group element} in matrix irrep $T$. We note in particular
that the linkage relation \eqref{WTWwTa} guarantees that an $H$-transformation
of a group element is still a group element
\begin{subequations}
\label{gclas0}
\begin{equation}
\label{gclas}
 g(T,\bz,z) = e^{i\be^a(\bz,z) T_a }
 \end{equation}
 \begin{equation}
  g(T,\bz,z)' = W(h_\s;T)g(T,\bz,z)W\hc(h_\s;T) = e^{i\be^a(\bz,z)
  W(h_\s;T)T_a W\hc(h_\s;T)} = e^{i\be^a(\bz,z)' T_a }
\end{equation}
\begin{equation}
  \be^a(\bz,z)' \equiv \be^b(\bz,z)w\hc(h_\s)_b{}^a
\end{equation}
\end{subequations}
where the linkage relation in the form \eqref{eq:WTWwTb}
 was used to obtain the final form.

The group elements $g(T,\bz,z)$ are the classical limit \cite{Halpern:1996et}
of the affine primary fields, so, following \eqref{eq:scrg}, we define the
{\it eigengroup elements} in terms of the group elements as
\begin{subequations}
\label{resp0}
\begin{equation}
  \sg(\st,\bz,z,\s) \equiv U(T,\s)g(T,\bz,z) U\hc(T,\s)
         = e^{i\bfrak^\nrm (\bz,z,\s) \T_\nrm (T,\s) }
\end{equation}
\begin{equation}
\sg\hc(\st,\bz,z,\s) \sg(\st,\bz,z,\s) = \one \sp
\bfrak^\nrm (\bz,z,\s) = \schisig^{-1}_\nrm \, \be^a (\bz,z) \, U\hc(\s)_a{}^\nrm
\end{equation}
\begin{equation}
\sg (\T,\bz,z,\s)' = E(T,\s) \sg (\T,\bz,z,\s) E (T,\s)^\ast \sp
\bfrak^\nrm(\bz,z,\s)' =  \bfrak^\nrm(\bz,z,\s) E_{n(r)}(\s)^\ast
 \label{response-of-sb}
\end{equation}
\end{subequations}
where $\{ \bfrak^{\nrm} \}$ are the tangent space eigencoordinates. With the
selection rules \eqref{sel-rule-for-st-1} of the twisted representation
matrices $\T$, the equivalence of the automorphic responses in
\eqref{response-of-sb} is easily checked to all orders in $\bfrak$.

Then by local isomorphisms we obtain the classical twisted matrix currents
 and the classical {\it group orbifold elements} $\hg (\T,\bz,z,\s)$
\begin{subequations}
\label{locrel}
\begin{equation}
{\cal{J}}_\nrm (z,\s) \dual \hat J_\nrm(z,\s) \sp
{\bar{\cal{J}}}_\nrm (\bz,\s) \dual \hat{\bar{J}}_\nrm (\bz,\s)
\end{equation}
\begin{equation}
 \sg (\T,\bz,z,\s) \dual \hg (\T,\bz,z,\s) \sp
 \sg^{-1} (\T,\bz,z,\s) \dual \hg^{-1} (\T,\bz,z,\s)
 \end{equation}
\begin{equation}
\label{clrell}
\hat{J} (\T,z,\s) = \hat{J}_{\nrm} (z,\s) \G^{\nrm ;  \mnrn} (\s)
\T_{\mnrn} (T,\s) \propto \hg^{-1} (\T,\bz,z,\s) \bar
\partial \hg(\T,\bz,z,\s)
\end{equation}
\begin{equation}
\label{clrelr}
\hat{\bar{J}} (\T,\bz,\s) = \hat{\bar{J}}_{\nrm} (\bz,\s) \G^{\nrm ;
\mnrn} (\s) \T_{\mnrn} (T,\s) \propto \hg (\T,\bz,z,\s) \partial
\hg^{-1}(\T,\bz,z,\s) \hskip 1.2cm
\end{equation}
\begin{equation}
\hg\hc(\T,\bz,z,\s) \hg (\T,\bz,z,\s) = \one \sp
\bar{\partial} \hj (\T,z,\s) = \partial \hjb (\T,\bz,\s) = 0 \ .
\end{equation}
\end{subequations}
The group orbifold elements are the classical limit of the twisted affine
primary fields. Because the classical group elements have trivial monodromy, the
 local isomorphisms also give the {\it classical monodromies} of the group
 orbifold elements
\begin{subequations}
\label{clmon}
\begin{equation}
\hg (\T,\bz e^{-2\pi i},z e^{ 2 \pi i},\s) = E(T,\s) \hg
(\T,\bz,z,\s) E(T,\s)^\ast
\end{equation}
\begin{equation}
\label{clmona}
\hg (\T,\bz e^{-2 \pi i}, z e^{2 \pi i},\s )_{N(r) \m}{}^{N(s) \n} =
e^{- 2 \pi i \frac{N(r) -N(s)}{R (\s)}} \hg (\T,\bz,z,\s)_{N(r)\m}{}^{N(s) \n}
\end{equation}
\begin{equation}
\hg^{-1} (\T,\bz e^{-2 \pi i}, z e^{2 \pi i},\s )_{N(r) \m}{}^{N(s)
\n} = e^{- 2 \pi i \frac{N(r) -N(s)}{R (\s)}} \hg^{-1}
(\T,\bz,z,\s)_{N(r) \m}{}^{N(s) \n}
\end{equation}
\end{subequations}
as the old automorphic responses \eqref{response-of-sb} of the eigengroup elements.

Using local isomorphisms also for the tangent space eigencoordinates, the
group orbifold elements can be considered as the exponentiation of the Lie
algebra \eqref{eq:reason-for-casimir-of-st} of the twisted representation matrices
\begin{subequations}
\begin{equation}
\bfrak^\nrm \dual \hat \be^\nrm
\end{equation}
\begin{equation}
\hg (\T,\bz,z,\s) = e^{i \hat{\be}^{\nrm} (\bz,z,\s) \T_{\nrm} (T,\s)}
\end{equation}
\begin{equation}
\label{betah}
\hat{\be}^{\nrm} (\bz e^{-2 \pi i},z e^{2 \pi i},\s) =
\hat{\be}^{\nrm} (\bz,z,\s) e^{2 \pi i \srac{n(r)}{\rho (\s)}}
\end{equation}
\end{subequations}
where the quantities $\{\hat \be^\nrm \}$ are called the twisted tangent space coordinates.
The selection rules \eqref{sel-rule-for-st-1} for $\T$ are again necessary to
check the consistency of the monodromies
\eqref{betah} and \eqref{clmona}. We also note that the multiplication law
\begin{subequations}
\begin{equation}
\hg_3 (\T,\bz,z,\s) \equiv \hg_1 (\T,\bz,z,\s) \hg_2(\T,\bz,z,\s)
\end{equation}
\begin{equation}
\hat{\be}_3 (\bz,z,\s) = \hat{\be}_1 (\bz,z,\s) + \hat{\be}_2 (\bz,z,\s)
+ \Ord (\hat{\be}^2)
\end{equation}
\begin{equation}
\hg_i (\T,\bz e^{-2\pi i},z e^{2 \pi i},\s) = E(T,\s) \hg_i(\T,\bz,z,\s)
E(T,\s)^\ast \sp i = 1,2,3
\end{equation}
\end{subequations}
preserves the monodromies of the group orbifold elements.

We turn now to the monodromies of the twisted currents. These are easily
obtained as above by local isomorphisms, but we wish to study these
monodromies as they are dictated by the {\it local} relations
\eqref{clrell}, \eqref{clrelr} in the classical theory. Using these relations
and the monodromies \eqref{clmon} of the group orbifold elements, we find that
the monodromies of the twisted matrix currents are the same
\begin{subequations}
\label{clmon0}
\begin{equation}
\label{clmonc}
\hat{J}(\T,z e^{2 \pi i},\s )_{N(r) \m}{}^{N(s) \n} = e^{- 2 \pi i
\frac{N(r) -N(s)}{R (\s)}}  \hat{J}(\T,z,\s)_{N(r) \m}{}^{N(s) \n}
\end{equation}
\begin{equation}
\label{clmond}
\hat{\bar{J}}(\T,\bz e^{-2 \pi i},\s )_{N(r) \m}{}^{N(s) \n} = e^{- 2
\pi i \frac{N(r) -N(s)}{R (\s)}}  \hat{\bar{J}}(\T,z,\s)_{N(r)
\m}{}^{N(s) \n}
\end{equation}
\end{subequations}
when the same path is  followed for $z$ and $\bz = z^\ast$. Moreover, the
selection rule \eqref{selrulet} for the twisted representation matrices
\begin{equation}
e^{- 2 \pi i \frac{n(r)}{\rho (\s)}}
\T_{\mnrm}(T,\s)_{N(s)\n}{}^{N(t)\de} = \T_{\mnrm}(T,\s)_{N(s)\n}{}^{N(t)\de}
e^{-2  \pi i \frac{N(s)-N(t)}{R (\s)}}
\end{equation}
shows that the monodromies \eqref{clmon0}  are equivalent
to the monodromies
\begin{equation}
\label{jjbmon0}
\hj_\nrm (z e^{2 \pi i},\s) = e^{ -2 \pi i \srac{n(r)}{\rho (\s)}}
\hj_\nrm (z,\s) \sp
\hjb_\nrm (\bz e^{-2 \pi i},\s) = e^{ -2 \pi i \srac{n(r)}{\rho (\s)}}
\hjb_\nrm (\bz,\s)
\end{equation}
of the twisted currents $\hat J_{\nrm}$, ${\hat{\bar{J}}}_\nrm$.
As promised in Subsec.~\ref{liso}: The classical monodromies \eqref{jjbmon0},
which followed from the local classical relations \eqref{clrell} and
\eqref{clrelr}, are in agreement with the quantum
monodromies \eqref{jjbmon} of the twisted currents.

We can also obtain the corresponding set of WZW orbifold actions as follows.
The general WZW action \cite{Novikov:1982ei,Witten:1984ar} for semisimple $g$
on the cylinder $(t,\xi)$ with $0 \leq \xi \leq 2 \pi$ is
\begin{subequations}
\begin{equation}
 S_{WZW}[M,g] =-\frac{1}{8\pi}\int d^2\xi\0b {\rm Tr}\Big{(}M(k,T)\0b\ginv (T)\pl_+
 g(T)\0b\ginv(T)\pl_-g(T)\Big{)}
 \end{equation}
 $$
    \quad\quad\quad\quad\quad-\frac{1}{12\pi}\int_{\Gamma} {\rm Tr}\Big{(}M(k,T)\0b
    (\ginv(T)dg(T))^3\0b\Big{)} ,\quad d^2\xi\equiv dt\,d\xi \label{action-a}
$$
\begin{equation}
 (\ginv dg)^3 = dt\,d\xi\,d\r\;\epsilon^{ABC}(\ginv \pl_Ag)(\ginv \pl_Bg)
 (\ginv \pl_Cg) ,\quad \{A,B,C\}=\{t,\xi,\r\}
\end{equation}
\begin{equation}
g (T) \equiv g(T,\xi,t) \sp \partial_{\pm} \equiv \partial_t \pm \partial_\xi
\end{equation}
\end{subequations}
where $\epsilon^{t \xi  \rho } =1$, $M (k,T)$ is the data matrix in
\eqref{groupelement} and $g(T)$ in the solid cylinder  $\Gamma$ is also a function
of $\rho$. This form of the action provides a convenient packaging for the
expected sum over the simple components of $g$,
\begin{subequations}
\begin{equation}
\be^a \quad \rightarrow \quad \be^{aI}
\end{equation}
\begin{equation}
g(T)_\a{}^\be \quad \rightarrow \quad g(T)_{\a I}{}^{\be J}
= g_I (T^I)_{\a (I)}{}^{\be (I)} \de_I^J
\sp g_I (T^I) = e^{i \sum_a \be^{a I} T_a^I}
\end{equation}
\begin{equation}
S_{\rm WZW}[ M,g]
=-\frac{1}{8\pi}\sum_I \frac{k_I}{y_I(T^I)}\int d^2\xi \ {\rm Tr}\Big{(}
g_{I}^{-1} (T^I)\pl_+g_I(T^I)\0b g_{I}^{-1}(T^I)\pl_-g_I(T^I)\Big{)} +  \ldots
\end{equation}
\end{subequations}
where we have used \eqref{TaI}, \eqref{2.2f} and \eqref{gclas}. The automorphic
responses of the group elements on the cylinder are given by \eqref{gclas0}
with $(\bz,z) \rightarrow (\xi,t)$, so we find that the $H$-invariance of the
general WZW action
\begin{equation}
S_{\rm WZW}[ M,g'] = S_{\rm WZW}[ M,g] \sp
g(T,\xi,t)' = W(h_\s;T) g(T,\xi,t) W\hc (h_\s;T)
\end{equation}
is guaranteed by the $H$-invariance \eqref{Mtransf} of the data matrix.

Eigengroup elements on the cylinder are also defined as in \eqref{resp0} with
$(\bz,z) \rightarrow (\xi,t)$.
Then local isomorphisms give the {\it general WZW orbifold
action}
\begin{subequations}
\label{gorbact}
\begin{equation}
S_{\rm WZW} [\M, \sg] \equiv S_{\rm WZW} [M, g(\sg)] \dual \hat S_\s
[\M,\hg ]
\end{equation}
\begin{equation}
\label{orbact}
  \hat{S}_\s [{\cal{M}},\hg ] = -\frac{1}{8\pi}\int d^2\xi
 \0b \widehat{\rm Tr}\big{(}\;\sm(\st,\s)\0b\hat{g}^{-1} (\st,\s)
\pl_+\hat{g}(\st,\s)\0b\hat{g}^{-1}
(\st,\s)\pl_-\hat{g}(\st,\s)\;\big{)}
\end{equation}
$$
  -\frac{1}{12\pi}\int_{\Gamma} \widehat{\rm Tr}\big{(}\;\sm(\st,\s)\0b(\;\hat{g}^{-1}
  (\st,\s) d\hat{g}(\st,\s)\;)^3\,\big{)} \sp \s = 0, \ldots , N_c -1
$$
\begin{equation}
    (\hat{g}^{-1} d\hat{g})^3 = dt\,d\xi\,d\r\;\epsilon^{ABC}(\hat{g}^{-1} \pl_A\hat{g})(\hat{g}^{-1} \pl_B\hat{g})(\hat{g}^{-1} \pl_C\hat{g}),\quad
        \{A,B,C\}=\{t,\xi,\r\}%
\end{equation}
\begin{equation}
\hg\hc (\T,\xi,t,\s) \hg (\T,\xi,t,\s) = 1
\end{equation}
\begin{eqnarray}
\label{Glmon}
\hg (\T,\xi + 2\pi,t,\s)_{\Nrm}{}^{\Nsn} & = &  e^{-2 \pi i
\frac{N(r)-N(s)}{R(\s)}} \hg (\T,\xi,t,\s)_{\Nrm}{}^{\Nsn} \\
\label{Grmon}
\hg^{-1} (\T,\xi + 2\pi,t,\s)_{\Nrm}{}^{\Nsn} & = &  e^{-2 \pi i
\frac{N(r)-N(s)}{R(\s)}} \hg^{-1} (\T,\xi,t,\s)_{\Nrm}{}^{\Nsn} \\
\label{smon}
 \hat{S}_\s[\sm(\st,\s),\hg(\st,\xi+2\pi,t,\s)] & = &
 \hat{S}_\s[\sm(\st,\s),\hg(\st,\xi,t,\s)]
\end{eqnarray}
\end{subequations}
for sector $\s$ of each WZW orbifold $A_g(H)/H$. Here,
$\hg (\T,\s)\equiv \hg(\T,\xi,t,\s)$ is the
group orbifold element on the cylinder%
\footnote{The automorphic response of the group element $g$ and hence the
$\xi$-monodromy of $\hg$ is the same when the group orbifold elements are
extended to the solid cylinder $\Gamma$.}, the orbifold trace
$\widehat{\rm Tr}$ is defined
in \eqref{trdef} and the quantity ${\cal{M}}(\T,\s)$
is the twisted data matrix in \eqref{twisted-data-matrix}. The trivial
monodromy \eqref{smon} of the WZW orbifold action follows from the monodromies
\eqref{Glmon}, \eqref{Grmon} of the group orbifold elements and the selection rule
\eqref{eq:msel} of the twisted data matrix.

Because the twisted data matrix is also
invertible and commutes with $\hg (\T,\s)$
(see Eqs.~\eqref{mtinv} and \eqref{Mtgt}), the equations of motion of this action
\begin{subequations}
\begin{equation}
  \hj(\st,\xi,t,\s) \equiv -\frac{i}{2}\hat{g}^{-1}(\st,\xi,t,\s)\pl_+
  \hat{g}(\st,\xi,t,\s)
   = \hat{J}_\nrm(\xi,t,\s)\sG^{n(r)\m;-n(r),\n}(\s)\st_{-n(r),\n}(T,\s)
   \label{twistedEOMJmatrix-start}
   \end{equation}
   \begin{equation}
 \hjb(\st,\xi,t,\s) \equiv -\frac{i}{2}\hat{g}(\st,\xi,t,\s)\pl_-\hg^{-1}
 (\st,\xi,t,\s) = \hjb_\nrm(\xi,t,\s)\sG^{n(r)\m;-n(r),\n}(\s)\st_{-n(r),\n}(T,\s)
        \label{jbh=g-pl-ginv}
  \end{equation}
   \begin{equation}
 \pl_-\hj(\st,\xi,t,\s) = \pl_+\hjb(\st,\xi,t,\s) = 0
  \label{twistedEOMJmatrix}
 \end{equation}
 \end{subequations}
are equivalent to those found on the sphere in \eqref{jeom}. Moreover, following
the development on the sphere, one finds the additional monodromies on the
cylinder
\begin{subequations}
\begin{equation}
\label{admon}
\hg (\T, \xi, t,\s) = e^{i \hat \be^\nrm (\xi,t,\s) \T_\nrm (T,\s)}
\sp \hat \be^\nrm (\xi + 2 \pi,t,\s) = \hat \be^\nrm (\xi,t,\s)
e^{2 \pi i \srac{n(r)}{\rho (\s)}}
\end{equation}
\begin{equation}
\hj_\nrm (\xi + 2 \pi ,t,\s) =e^{-2 \pi i \srac{n(r)}{\rho (\s)}}
\hj_\nrm (\xi  ,t,\s) \sp
\hjb_\nrm (\xi + 2 \pi ,t,\s) =e^{-2 \pi i \srac{n(r)}{\rho (\s)}}
\hjb_\nrm (\xi ,t,\s)
\end{equation}
\end{subequations}
which are also equivalent to those found on the sphere.

Further remarks on the classical theory of WZW orbifolds are found in
Subsecs.~\ref{unrec} and \ref{RUS}

\section{Mode Normal Ordering}
\subsection{Exact Operator Products}

In this subsection, we introduce a mode normal ordering \cite{Halpern:2000vj}
to discuss exact operator products of the twisted currents with themselves and with the
twisted affine primary fields. Moreover, we will follow the usual
\cite{Halpern:1996et} right-left mover factorization
\begin{subequations}
\begin{equation}
\hg (\T,\bz,z,\s) = \hg_- (\T,\bz,\s) \hg_+ (\T,z,\s)
\end{equation}
\begin{equation}
[ {\hat{\bar{J}}}_\nrm ( m + \srac{n(r)}{\rho (\s)} ),\hg_+ (\T,z,\s) ] = 0
\end{equation}
\begin{equation}
[ \hat J_\nrm ( m + \srac{n(r)}{\rho (\s)}),\hg_- (\T,\bz,\s) ] = 0
\end{equation}
\begin{equation}
\label{tvoel1}
\partial \hgp(\T,z,\s) = 2 \lr^{n(r) \mu; -n(r), \nu}(\s)
 : \hat{J}_{n(r) \mu}(z,\s) \hgp (\T,z,\s) \T_{-n(r), \nu} :
\end{equation}
\begin{equation}
\label{tvoer1}
\bar \partial \hgm(\T,\bz,\s) = -2 \lr^{n(r) \mu; -n(r), \nu}(\s)
: \hjb_{n(r) \mu}(\bz,\s)\T_{-n(r), \nu}  \hgm (\T,\bz,\s) :
\end{equation}
\begin{equation}
\hat A(\s) = \hat A_- (\s) \hat A_+ (\s)
\end{equation}
\begin{equation}
\hat A_+ (\s) \equiv \langle \hgp (\T^{(1)},z_1,\s) \ldots
\hgp (\T^{(N)},z_N,\s) \rangle_\s
\end{equation}
\begin{equation}
\hat A_- (\s) \equiv \langle \hgm (\T^{(1)},\bz_1,\s) \ldots
\hgm (\T^{(N)},\bz_N,\s) \rangle_\s
\end{equation}
\end{subequations}
of the twisted primary fields, and we will generally abbreviate
\begin{subequations}
\begin{equation}
\hj_\nrm(z) \equiv \hj_\nrm (z,\s) \sp
\hjb_\nrm(\bz) \equiv \hjb_\nrm (\bz,\s) \sp
\T_\nrm \equiv \T_\nrm (T,\s)
  \end{equation}
   \begin{equation}
\hgp (\T,z) \equiv \hgp (\T,z,\s) \sp \hgm (\T,\bz) \equiv \hgm (\T,\bz,\s)
\end{equation}
 \end{subequations}
by suppressing the sector label $\s$.

We begin with the exact operator products of the currents
\begin{subequations}
\label{exopejj}
$$
 \hj_{n(r) \mu}(z) \hj_{n(s)\nu}(w)=
 \frac{\G_{n(r) \mu;n(s) \nu}(\s)}{(z-w)^2}
+\frac{i \F_{n(r) \mu;n(s) \nu}{}^{ n(r)+n(s), \delta}(\s)
\hj_{n(r)+n(s), \delta}(w)}{z-w}
$$
\begin{equation}
 + : \hj_{n(r) \mu}(z) \hj_{n(s)\nu}(w) :
\end{equation}
$$
 \hjb_{n(r) \mu}(\bz) \hjb_{n(s)\nu}(\bw)=
\frac{\G_{n(r) \mu;n(s) \nu}(\s)}{(\bz-\bw)^2}
 +\frac{i \F_{n(r) \mu;n(s) \nu}{}^{ n(r)+n(s),
\delta}(\s) \hjb_{n(r)+n(s), \delta}(\bw)}{\bz-\bw}
$$
\begin{equation}
+ : \hjb_{n(r) \mu}(\bz) \hjb_{n(s)\nu}(\bw) :
\end{equation}
\end{subequations}
where $: \cdot :$ is operator product normal ordering. Then the mode normal
orderings $M$ and $\bM$ are defined as
\begin{subequations}
\label{mmbno}
\begin{eqnarray}
:\hj_{n(r) \mu}(m+\srac{n(r)}{\r(\s)}) \hj_{n(s) \nu}
(n+\srac{n(s)}{\r(\s)}):_{M} \!\!\!\!\!\!\!\!\!
&  & \equiv
 \theta(m+\srac{n(r)}{\r(\s)} \geq 0 ) \ \hj_{n(s)
\nu}(n+\srac{n(s)}{\r(\s)}) \hj_{n(r) \mu}(m+\srac{n(r)}{\r(\s)}) \nn \\
 & & +   \theta(m+\srac{n(r)}{\r(\s)} < 0) \
\hj_{n(r) \mu}(m+\srac{n(r)}{ \r(\s)})
\hj_{n(s) \nu}(n+\srac{n(s)}{ \r(\s)}) \;\;\;\;\;\;\;\;\;\;
\label{mnbnol}
\end{eqnarray}
\begin{eqnarray}
:\hjb_{n(r) \mu}(m+\srac{n(r)}{\r(\s)}) \hjb_{n(s) \nu}
(n+\srac{n(s)}{\r(\s)}):_{\bM} \!\!\!\!\!\!\!\!\!
& & \equiv
 \theta(m+\srac{n(r)}{\r(\s)} \leq 0 ) \ \hjb_{n(s)\nu}(n+\srac{n(s)}{\r(\s)})
 \hjb_{n(r) \mu}(m+\srac{n(r)}{\r(\s)}) \nn \\
 & & + \theta(m+\srac{n(r)}{\r(\s)} > 0) \
\hjb_{n(r) \mu}(m+\srac{n(r)}{ \r(\s)})\hjb_{n(s) \nu}(n+\srac{n(s)}{ \r(\s)})
\;\;\;\;\;\;\;\;\;\;
\label{mnbnor}
\end{eqnarray}
\end{subequations}
where $M$ normal ordering was studied for the twisted left movers in
Ref.~\cite{Halpern:2000vj}.
The $M$ and $\bM$  orderings are truly normal only for the semisimple
orbifold affine algebras
\cite{Borisov:1997nc,Evslin:1999qb,deBoer:1999na,Evslin:1999ve,Halpern:2000vj}
of the permutation orbifolds (see Sec.~\ref{permorb}), but we can also use
them to obtain exact relations for any WZW orbifold $A_g(H)/H$.

Using \eqref{mmbno} and the general twisted current algebra
\eqref{mode-current-algebra}, we find the exact operator products
\begin{subequations}
\label{opejjmno}
$$
 \hj_{n(r) \mu}(z) \hj_{n(s)\nu}(w)=
 (\frac{w}{z})^{\frac{\nb}{\r(\s)}} \Big\{
[\frac{1}{(z-w)^2} +
\frac{ \nb/\r(\s)}{w(z-w)}] \G_{n(r) \mu;n(s)\nu}(\s)  \hskip 4cm
$$
\begin{equation}
\label{opejjl} \hskip 3cm   +\frac{i \F_{n(r) \mu;n(s) \nu}{}^{
n(r)+n(s), \delta}(\s) \hj_{n(r)+n(s), \delta}(w)}{z-w} \Big\} +
:\hj_{n(r) \mu}(z) \hj_{n(s) \nu}(w):_{M}
\end{equation}
$$
 \hjb_{n(r) \mu}(\bz) \hjb_{n(s)\nu}(\bw)=
 (\frac{\bw}{\bz})^{\frac{\mnb}{\r(\s)}} \Big\{
[\frac{1}{(\bz-\bw)^2} +
\frac{ \mnb/\r(\s)}{\bw(\bz-\bw)}] \G_{n(r) \mu;n(s)\nu}(\s)  \hskip 4cm
$$
\begin{equation}
\hskip 3cm  +\frac{i \F_{n(r) \mu;n(s) \nu}{}^{ n(r)+n(s),
\delta}(\s) \hjb_{n(r)+n(s), \delta}(\bw)}{\bz-\bw} \Big\} +
:\hjb_{n(r) \mu}(\bz) \hjb_{n(s) \nu}(\bw):_{\bM}
\end{equation}
\end{subequations}
in terms of $M$ and $\bM$ normal ordered products. To obtain these results,
we have used the definition of $\nb$  in \eqref{nbdef},
 the summation identities in App.~\ref{sumid}, and the fact that
\cite{Halpern:2000vj}
\begin{equation}
\label{mnbdef}
\overline{-n(r)}=\overline{-\bar{n}(r)}=\left\{
\begin{array}{cc}
\r(\s)-\bar{n}(r) \textup{ when } \bar{n}(r)\neq 0 \\
0 \comment{\hsp{.4} when $\bar{n}(r)= 0$}
\end{array} \right.
\end{equation}
which follows from the definition of $\nb$. The left mover result in
\eqref{opejjl}  was given in Ref.~\cite{Halpern:2000vj}.

Comparing \eqref{opejjmno} with
\eqref{exopejj}, we find the relations among the normal ordered products
\begin{subequations}
$$
:\hj_{n(r) \mu}(z) \hj_{n(s) \nu}(z): \  =  \ :\hj_{n(r) \mu}(z)
\hj_{n(s) \nu}(z):_{M} \hskip 8cm
$$
 \begin{equation}
 -\frac{i}{z} \frac{ \nb }{\r(\s) }
 \F_{n(r) \mu;n(s) \nu}{}^{ n(r)+n(s), \delta}(\s)
 \hj_{n(r)+n(s), \delta}(z)
+ \frac{1}{z^2}  \frac{\nb}{2 \r(\s)}\left(1-\frac{\nb}{\r(\s)}\right)
 \G_{n(r) \mu;n(s) \nu}(\s)
\end{equation}
$$
:\hjb_{n(r) \mu}(\bz) \hjb_{n(s) \nu}(\bz): \  =  \
:\hjb_{n(r) \mu}(\bz) \hjb_{n(s) \nu}(\bz):_{\bM} \hskip 8cm
$$
 \begin{equation}
 -\frac{i}{\bz} \frac{ \mnb }{\r(\s) }
 \F_{n(r) \mu;n(s) \nu}{}^{ n(r)+n(s), \delta}(\s)
 \hjb_{n(r)+n(s), \delta}(\bz)
+ \frac{1}{\bz^2}  \frac{\mnb}{2 \r(\s)}\left(1-\frac{\mnb}{\r(\s)}\right)
 \G_{n(r) \mu;n(s) \nu}(\s) \ .
\end{equation}
\end{subequations}
These relations allow us to express the stress tensors of the twisted
affine-Sugawara constructions in terms of
$M$ and $\bM$ normal ordered products (see App.~\ref{gcaob} with
${\cL} \rightarrow \lr$).
 The corresponding Virasoro generators of the twisted affine-Sugawara
 constructions are
\begin{subequations}
\label{Virgent}
\begin{equation}
L_\s (m) =  \lr^{n(r) \mu; -n(r), \nu}(\s) \
\sum_{p \in \sz} :\hat{J}_{n(r) \mu}(p+\srac{n(r)}{\r(\s)}) \hat{J}_{-n(r),
 \nu}(m-p-\srac{n(r)}{\r(\s)}):
\end{equation}
 $$
= \lr^{n(r) \mu; -n(r), \nu}(\s) \
\{ \sum_{p \in \sz} :\hat{J}_{n(r) \mu}(p+\srac{n(r)}{\r(\s)})
\hat{J}_{-n(r), \nu}(m-p-\srac{n(r)}{\r(\s)}):_M
$$
\begin{equation}
\label{lmnol}
 -i\frac{\bar{n}(r)}{\r(\s)} \F_{n(r) \mu; -n(r), \nu}{}^{0 \delta}(\s)
 \hat{J}_{0 \delta}(m)
+ \delta_{m,0} \frac{\bar{n}(r)}{2 \r(\s)}\left(1-\frac{\bar{n}(r)}{\r(\s)}
\right)
\g_{n(r) \mu;-n(r), \nu}(\s) \}
\end{equation}
\begin{equation}
\bar L_\s (m) =  \lr^{n(r) \mu; -n(r), \nu}(\s) \
\sum_{p \in \sz} :\hjb_{n(r) \mu}(p+\srac{n(r)}{\r(\s)})
\hjb_{-n(r),
 \nu}(-m-p-\srac{n(r)}{\r(\s)}):
\end{equation}
 $$
=   \lr^{n(r) \mu; -n(r), \nu}(\s) \ \{ \sum_{p \in
\sz} :\hjb_{n(r) \mu}(p+\srac{n(r)}{\r(\s)}) \hjb_{-n(r),
\nu}(-m-p-\srac{n(r)}{\r(\s)}):_{\bM} $$
 \begin{equation}
 \label{lmnor}
 -i\frac{\mnb }{\r(\s)} \F_{n(r) \mu; -n(r), \nu}{}^{0 \delta}(\s)
 \hjb_{0 \delta}(-m)
+ \delta_{m,0}\frac{\mnb}{2 \r(\s)}\left(1-\frac{\mnb}{\r(\s)}\right)
\g_{n(r) \mu;-n(r), \nu}(\s) \} \ .
\end{equation}
\begin{equation}
\label{viriden}
 \frac{\mnb}{2 \r(\s)}\left(1-\frac{\mnb}{\r(\s)}\right)
= \frac{\bar{n}(r)}{2 \r(\s)}\left(1-\frac{\bar{n}(r)}{\r(\s)} \right)
\end{equation}
\end{subequations}
in sector $\s$ of each WZW orbifold $A_g(H)/H$.
The identity in \eqref{viriden} follows from \eqref{mnbdef}.

We turn next to the exact operator products of the twisted currents with the
twisted primary fields, starting with the operator products
\begin{subequations}
\label{jgopeb}
\begin{equation}
\label{jgopet}
 \hat J_{n(r) \mu }  (z) \hgp(\T,w) = \frac{1}{z-w} \hgp(\T,w) \T_{n(r) \mu}
  + : \hat J_{n(r) \mu }
(z) \hgp(\T,w)
:
\end{equation}
\begin{equation}
\label{rjgopet}
 \hjb_{n(r) \mu } (\bz) \hgm(\T,\bw) = -\frac{1}{\bz-\bw}
\T_{n(r) \mu} \hgm(\T,\bw)
+ : \hjb_{n(r) \mu }(\bz) \hgm(\T,\bw) :
\end{equation}
\end{subequations}
in terms of operator product normal ordering. For these operators, we
introduce the corresponding $ M$ and $\bM $ normal  ordering as follows
\begin{subequations}
\label{MNOing}
\begin{equation}
\label{lMNOing}
:\hat{J}_{n(r) \mu}(m+\srac{n(r)}{\r(\s)}) \hgp (\T,z) :_M \
\equiv \theta(m+\srac{n(r)}{\r(\s)} \geq 0) \  \hgp (\T,z)
\hat{J}_{n(r) \mu}(m+\srac{n(r)}{\r(\s)})
\end{equation}
$$
\hskip 4cm + \theta(m+\srac{n(r)}{\r(\s)} < 0) \
 \hat{J}_{n(r) \mu}(m+\srac{n(r)}{\r(\s)})
\hgp (\T,z)
$$
\begin{equation}
\label{rMNOing}
:\hjb_{n(r) \mu}(m+\srac{n(r)}{\r(\s)}) \hgm (\T,\bz) :_{\bM} \
\equiv
 \theta(m+\srac{n(r)}{\r(\s)} \leq 0) \
\hgm (\T,\bz)
 \hjb_{n(r) \mu}(m+\srac{n(r)}{\r(\s)})
\end{equation}
$$
\hskip 4cm + \theta(m+\srac{n(r)}{\r(\s)} > 0) \
\hjb_{n(r) \mu}(m+\srac{n(r)}{\r(\s)})
\hgm (\T,\bz) \ .
$$
\end{subequations}
Again, these mode orderings are truly normal only for permutation orbifolds
(see Sec.~\ref{permorb}).
Then using the commutators in \eqref{jgct}, \eqref{rjgct} we find the
exact operator products
\begin{subequations}
\label{jgoperl}
\begin{equation}
\label{jgopeM}
 \hat J_{n(r) \mu }  (z) \hgp(\T,w) = \left( \frac{w}{z}
 \right)^{ \srac{\bar n(r)}{\r(\s)} }
 \frac{1}{z-w} \hgp(\T,w) \T_{n(r) \mu}
  + : \hat J_{n(r) \mu }
(z) \hgp(\T,w)
:_M
\end{equation}
\begin{equation}
\label{rjgopeM}
 \hjb_{n(r) \mu }  (\bz) \hgm(\T,\bw) = -\left( \frac{\bw}{\bz}
 \right)^{ \srac{\mnb}{\r(\s)} }
 \frac{1}{\bz-\bw}\T_{n(r) \mu} \hgm(\T,\bw)
  + : \hjb_{n(r) \mu }
(\bz) \hgm(\T,\bw)
:_{\bM}
\end{equation}
\end{subequations}
in terms of the $M$ and $\bM$ normal ordered products.

Comparing \eqref{jgoperl}  with
\eqref{jgopeb}, we also obtain the relations among the normal ordered products
\begin{subequations}
\label{relmno}
\begin{equation}
: \hat J_{n(r) \mu }(z) \hgp(\T,z) : =
: \hat J_{n(r) \mu }(z) \hgp(\T,z) :_M -\srac{\bar n(r)}{\r(\s)}
\frac{1}{z}
\hgp (\T,z) \T_{n(r) \mu }
\end{equation}
\begin{equation}
: \hjb_{n(r) \mu }(\bz) \hgm(\T,\bz) : \ = \
: \hjb_{n(r) \mu }(\bz) \hgm(\T,\bz) :_{\bM} + \srac{\mnb}{\r(\s)}
\frac{1}{\bz} \T_{n(r) \mu }\hgm (\T,\bz)
\end{equation}
\end{subequations}
in each sector $\s$ of each $A_g(H)/H$.

\subsection{Mode-Ordered Form of the Vertex Operator Equations}

As an application of \eqref{relmno}, we may reexpress the twisted vertex operator
equations \eqref{tvoel1}, \eqref{tvoer1}
\begin{subequations}
\label{difrelM}
\begin{equation}
\label{difreltM}
\partial \hgp(\T,z) = 2 \lr^{n(r) \mu; -n(r), \nu}(\s)
\left( : \hat{J}_{n(r) \mu}(z) \hgp (\T,z) :_M -
\srac{\bar n(r)}{\r(\s)} \frac{1}{z} \hgp (\T,z)\T_{n(r) \mu} \right) \T_{-n(r), \nu}
\end{equation}
\begin{equation}
\label{rdifreltM}
\bar \partial \hgm(\T,\bz) = -2 \lr^{n(r) \mu; -n(r), \nu}(\s)
\T_{-n(r), \nu}
\left( : \hjb_{n(r) \mu}(\bz) \hgm (\T,\bz) :_{\bM} +
\srac{\mnb}{\r(\s)}  \frac{1}{\bz} \T_{n(r) \mu} \hgm (\T,\bz) \right)
\end{equation}
\end{subequations}
in terms of $M$ and $\bM$ normal ordered products.

Using Eq.~\eqref{MNOing},
we give the explicit form of these mode ordered products
\begin{subequations}
\label{lmexp}
\begin{equation}
\label{jgMno}
 : \hat{J}_{n(r) \mu}(z) \hgp (\T,z) :_M \  =
\hat J_{n(r) \mu}^-(z) \hgp (\T,z) +  \hgp (\T,z)  \hat J_{n(r) \mu}^+(z)
\end{equation}
\begin{eqnarray}
\label{jm0} \hat J_{n(r) \mu}^-(z) & \equiv &  \sum_m \theta
(m+\srac{n(r)}{\r(\s)} <0) \hat J_{n(r) \mu} (m+\srac{
n(r)}{\r(\s)}) z^{-(m+\srac{ n(r)}{\r(\s)})-1} \\
\label{jm}
 & = &  \sum_{m \leq -1}
\hat J_{n(r) \mu} (m+\srac{\bar n(r)}{\r(\s)})
z^{-(m+\srac{\bar n(r)}{\r(\s)})-1}
\end{eqnarray}
\begin{eqnarray}
\label{jp0} \hat J_{n(r) \mu}^+(z) & \equiv &  \sum_{m} \theta
(m+\srac{n(r)}{\r(\s)} \geq 0) \hat J_{n(r) \mu}
(m+\srac{n(r)}{\r(\s)}) z^{-(m+\srac{n(r)}{\r(\s)})-1} \\
\label{jp}   & =  & \sum_{m \geq 0} \hat J_{n(r) \mu}
(m+\srac{\bar n(r)}{\r(\s)}) z^{-(m+\srac{\bar n(r)}{\r(\s)})-1}
\end{eqnarray}
\begin{equation}
\hat J_{n(r) \mu}^-(z) + \hat J_{n(r) \mu}^+(z) =\hat J_{n(r) \mu}(z)
\end{equation}
\end{subequations}
\begin{subequations}
\label{rmexp}
\begin{equation}
\label{rjgMno}
 : \hjb_{n(r) \mu}(\bz) \hgm (\T,\bz) :_{\bM} \  =
\hjb_{n(r) \mu}^+(\bz) \hgm (\T,\bz) +  \hgm (\T,\bz)  \hjb_{n(r) \mu}^-(\bz)
\end{equation}
\begin{eqnarray}
\label{rjm0} \hjb_{n(r) \mu}^+(\bz) & \equiv & \sum_m \theta
(m+\srac{n(r)}{\r(\s)} > 0) \hjb_{n(r) \mu} (m+\srac{
n(r)}{\r(\s)}) \bz^{(m+\srac{ n(r)}{\r(\s)})-1} \\
\label{rjm} & =&  \sum_{m \geq 0} \hjb_{n(r) \mu} (m+\srac{\bar
n(r)}{\r(\s)}) \bz^{(m+\srac{\bar n(r)}{\r(\s)})-1}
-\delta_{\bar n(r),0} \hjb_{0 \mu} (0) \bz^{-1} \\
\label{rjm1} & = & \sum_{m > 0} \hjb_{n(r) \mu} (m -
\srac{\mnb}{\r(\s)}) \bz^{(m-\srac{\mnb}{\r(\s)})-1}
\end{eqnarray}
\begin{eqnarray}
\label{rjp0} \hjb_{n(r) \mu}^-(\bz) & \equiv &  \sum_{m} \theta
(m+\srac{n(r)}{\r(\s)} \leq 0) \hjb_{n(r) \mu}
(m+\srac{n(r)}{\r(\s)}) \bz^{(m+\srac{n(r)}{\r(\s)})-1} \\
\label{rjp} & = &  \sum_{m < 0} \hjb_{n(r) \mu}
(m+\srac{\bar n(r)}{\r(\s)}) \bz^{(m+\srac{\bar n(r)}{\r(\s)})-1}
+\delta_{\bar n(r),0} \hjb_{0 \mu} (0) \bz^{-1} \\
\label{rjp1} &  = &  \sum_{m \leq 0} \hjb_{n(r) \mu} (m -
\srac{\mnb}{\r(\s)}) \bz^{(m-\srac{\mnb}{\r(\s)})-1}
\end{eqnarray}
\begin{equation}
\hjb_{n(r) \mu}^-(\bz) + \hjb_{n(r) \mu}^+(\bz) =\hjb_{n(r) \mu}(\bz)
\end{equation}
\end{subequations}
where $\hj^\pm$, $\hjb^\pm$ are called the twisted partial currents.
In obtaining the final  form for the twisted right mover partial currents, we
also used the identity
\begin{equation}
 \hjb_{n(r) \mu} (m - \srac{\mnb}{\r(\s)})
 = \hjb_{n(r) \mu}  (m-1+\srac{\bar n(r)}{\r(\s)})
\sp \bar n(r) \neq 0
\end{equation}
which follows from \eqref{mnbdef}.

Finally, it will be useful to know the commutation relations of the partial
currents
\begin{subequations}
\label{jpmg}
\begin{equation}
[\hat J_{n(r) \mu}^+(z), \hgp (\T,w) ] = \frac{1}{z-w} \left(
\frac{w}{z} \right)^{\srac{\bar n(r)}{\r(\s)} } \hgp (\T,z) \T_{n(r) \mu}
\sp |z| > | w|
\end{equation}
\begin{equation}
[\hat J_{n(r) \mu}^-(z), \hgp (\T,w) ] = -\frac{1}{z-w} \left(
\frac{w}{z} \right)^{\srac{\bar n(r)}{\r(\s)} } \hgp (\T,z) \T_{n(r) \mu}
\sp |w| > | z|
\end{equation}
\begin{equation}
[\hjb_{n(r) \mu}^-(\bz), \hgm (\T,\bw) ] = -\frac{1}{\bz-\bw} \left(
\frac{\bw}{\bz} \right)^{\srac{\mnb}{\r(\s)} }
\T_{n(r) \mu} \hgm (\T,\bz) \sp |\bz| > | \bw|
\end{equation}
\begin{equation}
[\hjb_{n(r) \mu}^+(\bz), \hgm (\T,\bw) ] = \frac{1}{\bz-\bw} \left(
\frac{\bw}{\bz} \right)^{\srac{\mnb}{\r(\s)} }
\T_{n(r) \mu} \hgm (\T,\bz) \sp |\bw| > | \bz|
\end{equation}
\end{subequations}
with the twisted primary fields $\hg (\T)$.

\subsection{A Consistency Check on the Twisted Vertex Operator Equations \label{consvoe}}

The forms \eqref{tvoe} (and hence the forms \eqref{difrelM}) of the twisted vertex
operator equations followed from local isomorphisms. In this subsection,
we provide a consistency check on these results, starting with the
differential equations
\begin{equation}
\label{lmg}
\part \hgp (\T,z) =[ L_\s (-1), \hgp (\T,z) ]  \sp
\bar \part \hgm (\T,\bz) = [\bar L_\s (-1), \hgm (\T,\bz) ]
\end{equation}
which are special cases of the more general commutation relations in \eqref{Lgu},
\eqref{Lgb}.

Then using \eqref{lmnol}, \eqref{lmnor} and
\eqref{jgct}, \eqref{rjgct} one finds after some algebra
a somewhat different form of the twisted
vertex operator equations
\begin{subequations}
\label{altform}
\begin{equation}
\label{difreltMd}
\partial \hgp(\T,z) =  \lr^{n(r) \mu; -n(r), \nu}(\s)
\Big( 2 : \hat{J}_{n(r) \mu}(z) \hgp (\T,z) :_M  \T_{-n(r), \nu} \hskip 4cm
\end{equation}
$$\hskip 1cm  -\hgp (\T,z) \frac{1}{z} \T_{n(r) \mu}  \T_{-n(r), \nu}
(1 - \delta_{\bar n(r),0} ) - i \srac{\bar n(r)}{\r(\s)} \frac{1}{z}\hgp (\T,z)
 \F_{n(r) \mu;-n(r), \nu}{}^{0,\delta}  \T_{0, \delta} \Big)
$$
\begin{equation}
\label{rdifreltMd}
\bar \partial \hgm(\T,\bz) =  -\lr^{n(r) \mu; -n(r), \nu}(\s)
\Big( 2  \T_{-n(r), \nu} : \hjb_{n(r) \mu}(\bz) \hgm (\T,\bz) :_{\bM} \hskip 4cm
\end{equation}
$$\hskip 1cm + \frac{1}{\bz}  \T_{-n(r), \nu}\T_{n(r) \mu} \hgm (\T,\bz) (1
- \delta_{\bar n(r),0} ) - \frac{i}{\bz} \srac{\mnb}{\r(\s)}
 \F_{n(r) \mu;-n(r), \nu}{}^{0,\delta}  \T_{0, \delta}\hgm (\T,\bz) \Big) \ .
$$
\end{subequations}
These forms agree with those given in \eqref{difrelM} iff
\begin{subequations}
\label{ident}
\begin{equation}
\label{id1}
 2 \lr^{n(r) \mu; -n(r), \nu}(\s)
\srac{\bar n(r)}{\r(\s)} \T_{n(r) \mu}  \T_{-n(r), \nu} \hskip 6cm
\end{equation}
$$
\hskip 2cm =\lr^{n(r) \mu; -n(r), \nu}(\s)  \left( \T_{n(r) \mu}  \T_{-n(r), \nu}
(1 - \delta_{\bar n(r),0} )  + i \srac{\bar n(r)}{\r(\s)}
 \F_{n(r) \mu;-n(r), \nu}{}^{0,\delta}  \T_{0, \delta} \right)
$$
\begin{equation}
\label{rid1}
 2 \lr^{n(r) \mu; -n(r), \nu}(\s)
\srac{\mnb }{\r(\s)} \T_{-n(r), \nu} \T_{n(r) \mu} \hskip 6cm
\end{equation}
$$
\hskip 2cm =\lr^{n(r) \mu; -n(r), \nu}(\s)  \left( \T_{-n(r), \nu}\T_{n(r) \mu}
 (1 - \delta_{\bar n(r),0} ) - i \srac{\mnb }{\r(\s)}
 \F_{n(r) \mu;-n(r), \nu}{}^{0,\delta}  \T_{0, \delta} \right) \ .
$$
\end{subequations}
These consistency relations are in fact identities. To check the first identity,
replace the $i\F_{n(r) \mu;-n(r), \nu}{}^{0,\delta}  \T_{0, \delta} $
term by the commutator
 $[\T_{n(r) \mu}, \T_{-n(r), \nu}]$
and note that the identity is trivially satisfied for the $\bar n(r)
=0$ terms. For the $\bar n \neq 0$ terms, use the symmetry \eqref{lgsym} of
$\lr$ and the change
of dummy variables $n \rightarrow -n$, which induces $\bar n/\rho
\rightarrow 1- \bar n/\rho$ (see Eq.~\eqref{mnbdef}).
The second identity then follows from the first identity  by
 $n(r) \rightarrow -n(r)$ and the symmetry \eqref{lgsym} of $\lr$.

\section{The WZW Permutation Orbifolds \label{permorb} }

\subsection{Systematics \label{syst} }

In this and the following two sections, we apply our general
development above to the special case of the WZW permutation
orbifolds
\begin{equation}
\frac{A_g(H)}{H} \sp H (\textup{permutation}) \subset S_N \ .
\end{equation}
We begin by reprising the salient features of  the
permutation-invariant theory $A_g(H)$
\begin{subequations}
\label{salf}
\begin{equation}
 g=\oplus_{I=0}^{K-1} \gfrak^I, \quad K \leq N  ,
 \quad \gfrak^I \simeq \gfrak ,\quad a \goto aI, \quad \a \goto \a I,
\end{equation}
\begin{equation}
 G_{aI,bJ} = k\e_{ab}\d_{IJ}, \quad f_{aI,bJ}{}^{cK} = f_{ab}{}^c \d_{IJ}
 \d_J{}^K ,
    \quad (T_{aI})_{\a J}{}^{\be K} = (T_a)_\a{}^\be \d_{IJ} \d_J{}^{K}
\end{equation}
\begin{equation}
 [T_a, T_b] = if_{ab}{}^c T_c \sp
 {\rm Tr}(T_a T_b) = y(T)\e_{ab}, \quad M(k,T)_{\a I}{}^{\be J} = \frac{k}{y(T)}
 \d_\a{}^{\be} \d_I{}^{J}
\end{equation}
\begin{equation}
\label{deltaeq}
L_g^{aI,bJ} = \frac{\eta^{ab} \delta^{IJ} }{2 k +Q} \sp
D_g (T)_{\a I}{}^{\be J} = \Delta_{\sgbn} (T) \delta_\a^\be \de_I^J
\end{equation}
\begin{equation}
 I=0, \ldots ,K-1, \quad a=1 \ldots \text{dim}\,\gfrak,
 \quad \a = 1 \ldots \text{dim}\,T
\end{equation}
\end{subequations}
where $I$ is the semisimplicity index. Here each copy $\gfrak^I$ is taken at level $k$ with Killing metric $\eta_{ab}$,
structure constants $f_{ab}{}^c$ and irrep $T_a$. In Eq.~\eqref{deltaeq},
$\Delta_{\sgbn} (T)$ is the conformal weight of irrep $T$ under any copy
$\gfrak^I \simeq \gfrak$.

We next discuss the form of the symmetry transformations $w$ and $W$
when $H$ is any subgroup of $S_N$. The action in the adjoint is
\cite{Halpern:2000vj}
\begin{equation}
 w(h_\s)_{aI}{}^{bJ} = \d_a{}^b w(h_\s)_I{}^J, \quad \forall \ h_\s \in
 H(\mbox{permutation}) \subset {\rm Aut}(g) \ . \label{semisimple-w}
\end{equation}
Then we must find
$W(h_\s;T)$, the corresponding action of $h_\s$ in matrix representation $T$.
It is straightforward to check that
\begin{equation}
 W(h_\s;T)_{\a I}{}^{\be J} = \d_\a{}^\be w(h_\s)_I{}^J
 \sp \forall \ h_\s (\mbox{permutation}) \subset {\rm Aut}(g)
 \label{semisimple-W}
\end{equation}
solves the linkage relation \eqref{eq:WTWwTa} so long as
\vspace{-.05in}
\begin{equation}
  w\hc(h_\s)_I{}^J w(h_\s)_J{}^K = w(h_\s)_J{}^I\d_I{}^K ,\quad \forall \ I,J,K
\end{equation}
and these relations indeed hold for all $h_\s \in H$ because each
$w(h_\s)_I{}^{J}$ is a permutation matrix.

The simple forms \eqref{semisimple-w} and \eqref{semisimple-W} tell us that, for
 permutation orbifolds, the extended $H$-eigenvalue problem
 \eqref{ExtendedHEigen} and the $H$-eigenvalue problem \eqref{HEigen} are both
  simply related to the solution of the  reduced eigenvalue problem for
  $w(h_\s)_I{}^{J}$
\begin{subequations}
\begin{equation}
 \nrm \goto n(r)aj ,\quad \Nrm \goto N(r)\a j
\end{equation}
\begin{equation}
 R(T,\s)=\r(\s),\quad  E(T,\s)= E(\s),\quad N(r)=n(r)
\end{equation}
\begin{equation}
 U\hc(T,\s)_{\a I}{}^{n(r)\be j} = \d_\a{}^\be U\hc(\s)_I{}^{n(r)j} ,
 \quad U\hc(\s)_{aI}{}^{n(r)bj} =\d_a{}^b U\hc(\s)_I^{n(r)j}
\end{equation}
\begin{equation}
 w(h_\s)_I{}^J U\hc(\s)_J{}^{n(r)j} = U\hc(\s)_I{}^{n(r)j} E_{n(r)}(\s),
 \quad E_{n(r)}(\s)=e^{-2\pi i \frac{n(r)}{\r(\s)}} \label{perm:WUdagger=UdaggerE}
\end{equation}
\end{subequations}
which is given in \eqref{perm:WUdagger=UdaggerE}.
This gives the simplified forms of the duality transformations
\begin{subequations}
\label{permo}
\begin{eqnarray}
 \sG_{n(r)aj;n(s)bl}(\s) &  =  &  k\e_{ab} \schisig_{n(r)aj} \schisig_{n(s)bl}
 \sum_I U(\s)_{n(r)j}{}^I U(\s)_{n(s)l}{}^I \\
  & = & \d_{n(r)+n(s),0\,\text{mod}\,\r(\s)} \sG_{n(r)aj;-n(r),bl}(\s)
\end{eqnarray}
\begin{eqnarray}
\label{permg}
\G^{n(r)aj;n(s)bl} (\s) & = &  \frac{\eta^{ab} }{k}
\schisig^{-1}_{n(r)aj} \schisig^{-1}_{n(s)bl}
\sum_I U(\s)_I{}^{n(r)j} U(\s)_I{}^{n(s)l} \\
& = & \d_{n(r)+n(s),0\,\text{mod}\,\r(\s)} \sG^{n(r)aj;-n(r),bl}(\s)
\end{eqnarray}
\begin{equation}
\label{scperm}
\scf_{n(r)aj;n(s)bl}{}^{n(t)cm}(\s) = f_{ab}{}^c \schisig_{n(r)aj}
\schisig_{n(s)bl}  \schisig_{n(t)cm}^{-1} \sum_I
U(\s)_{n(r)j}{}^I U(\s)_{n(s)l}{}^I U\hc(\s)_I{}^{n(t)m}
\end{equation}
\begin{equation}
= \d_{n(r)+n(s)-n(t),0\,\text{mod}\,\r(\s)}
\scf_{n(r)aj;n(s)bl}{}^{n(r)+n(s),cm}(\s)
\end{equation}
\begin{equation}
\label{perml}
{\cL}_{\sgb (\s)}^{n(r)aj;n(s)bl} (\s) = \frac{k}{2k + Q_{\sgbn}} \G^{n(r)aj;n(s)bl} (\s)
\end{equation}
\begin{equation}
\label{repmatperm}
(\T_{n(r)aj})_{n(s) \a l}{}^{n(t) \be m} = (T_a)_\a{}^\be \chi (\s)_{n(r) a j}
\sum_I U(\s)_{n(r) j}{}^I U(\s)_{n(s) l}{}^I U\hc (\s)_I{}^{n(t) m}
\end{equation}
\begin{equation}
= \delta_{n(r)+n(s)-n(t),0 \,\rmod \rho(\s)} (\T_{n(r) aj })_{n(s) \a l}
 {}^{n(r) + n(s), \be m}
\end{equation}
\begin{equation}
\label{twmperm}
 \sm(\T,\s)_{n(r) \a j}{}^{n(s) \beta l} = \frac{k}{y(T)}
 (\one)_{n(r) \a j}{}^{n(s) \beta l}
\end{equation}
\begin{equation}
\label{twcfm}
\D_{\sgb (\s)}(\T) = {\cL}_{\sgb (\s)}^{n(r) a j;-n(r), b l} (\s)
\T_{n(r) aj} \T_{ -n(r), b l} = \Delta_{\sgbn} (T) \one
\end{equation}
\begin{equation}
(\one )_{n(r) \a j}{}^{n(s) \be l} = \delta_{n(r)-n(s), 0 \,\rmod \rho(\s)}
\delta_{\a}^{\be } \de_j^l
\end{equation}
\end{subequations}
in sector $\s$ of any WZW orbifold $A_g(H)/H$. More explicit forms of
$\G$, $\F$, $\lr$ and $\T$ are given in Subsecs.~\ref{cyclorb},
\ref{snorb}, \ref{unrec} and App.~\ref{Jperm}.

Because  the twisted conformal weight matrix \eqref{twcfm} is proportional
 to unity, we also find the simpler OPEs
\begin{subequations}
\begin{equation}
\hg (\T,\bz,z)_{n(r) \a j}{}^{n(s) \be l}
= (\hgm (\T,z) \hgp(\T,\bz))_{n(r) \a j}{}^{n(s) \be l}
 \sp \a, \be = 1 \ldots {\rm dim}\,T
\end{equation}
\begin{equation}
{\hat{T}}_\s (z) \hgp (\T,w) = \left(
\frac{\Delta_{\sgbn} (T)}{(z-w)^2} + \frac{\partial_w}{z-w} \right)
\hgp (\T,w) + {\cal{O}} (z-w)^0
\end{equation}
\begin{equation}
{\hat{\bar{T}}}_\s (\bz) \hgm (\T,\bw) = \left(
\frac{\Delta_{\sgbn} (T)}{(\bz-\bw)^2} + \frac{\partial_{\bw}}{\bz-\bw} \right)
\hgm (\T,\bw) + {\cal{O}} (\bz-\bw)^0
\end{equation}
\end{subequations}
for the WZW permutation orbifolds and the WZW orbifolds on simple $g$.

\subsection{The Ground State of Sector $\s$ and Simple Results \label{srperm} }

For the permutation orbifolds, the ground state (twist field state)
of sector $\s$ satisfies
\begin{subequations}
\label{vacc}
\begin{equation}
\label{vacl}
 0 = \hat J_{n(r)aj} (m +\srac{ n(r)}{\r(\s)} \geq 0) | 0
 \rangle_\s =
 {}_\s \langle 0 | \hat J_{n(r)aj} (m +\srac{ n(r)}{\r(\s)} \leq 0)
\end{equation}
\begin{equation}
\label{vacr}
 0 = \hjb_{n(r)aj} (m +\srac{ n(r)}{\r(\s)} \leq 0) | 0
 \rangle_\s =
 {}_\s \langle 0 | \hjb_{n(r)aj} (m +\srac{ n(r)}{\r(\s)} \geq 0)
\end{equation}
\begin{equation}
\langle \hat{J}_{n(r) aj} (z) \rangle_\s =
\langle \hat{\bJ}_{n(r) aj} (\bz) \rangle_\s = 0
\end{equation}
\begin{equation}
\langle  : \hat{J}_{n(r) aj} (z) \hat{J}_{n(s) bl} (w):_M  \rangle_\s
=
\langle  : \hat{\bJ}_{n(r) aj} (\bz) \hat{\bJ}_{n(s) bl}
(\bw):_{\bM}
  \rangle_\s = 0
\end{equation}
\end{subequations}
The left mover conditions \eqref{vacl} were given in Ref.~\cite{Halpern:2000vj}
and the right mover conditions \eqref{vacr} can be understood (see also
Subsec.~\ref{unrec}) from the mode-number reversed form \eqref{double-bar-alg} of the
twisted right mover current algebra.

Using  the exact operator products in \eqref{opejjmno}
we find that the current-current correlators
\begin{subequations}
\label{grpo}
\begin{eqnarray}
 \langle \hat{J}_{n(r) aj}(z) \hat{J}_{n(s) bl}(w) \rangle_\s
 & = & \delta_{n(r)+n(s),0 \rmod \r(\s)}
\left(\frac{w}{z}\right)^{\frac{\bar{n}(r)}{\r(\s)}} \nn \\
 && \times \left[\frac{1}{(z-w)^2} +
\frac{\bar{n}(r)/\r(\s)}{w(z-w)} \right] \g_{n(r) aj;-n(r), bl}(\s)\label{grpol} \\
\langle \hat{\bJ}_{n(r) aj} (\bz) \hat{\bJ}_{n(s)bl} (\bw)
\rangle_\s & =  &\delta_{n(r) + n(s), 0 \rmod \rho (\s) }
\left( \frac{\bw}{\bz} \right)^{\frac{\mnb}{\rho (\s)}} \nn \\
 & & \times \left[\frac{1}{(\bz-\bw)^2} +  \frac{\mnb/\rho (\s)}{\bw(\bz-\bw)}
\right]
\G_{n(r)aj ; -n(r),bl} (\s)
\end{eqnarray}
\end{subequations}
follow for all permutation orbifolds from the ground state conditions \eqref{vacc}.
The left mover result \eqref{grpol} was given in Ref.~\cite{Halpern:2000vj}.

Using \eqref{permg}, \eqref{scperm}, \eqref{perml} and
\eqref{Virgent}, we find that there are no $\hj$ or $\hjb$ linear terms in the
Virasoro generators of the WZW permutation orbifolds
\begin{subequations}
\begin{eqnarray}
L_\s(m) =   \lr^{n(r) aj; -n(r),bl}(\s) \!\!\!\!\!\!\!\!\!  & &
\Big\{ \sum_{p \in \sz} :\hat{J}_{n(r)aj}(p+\srac{n(r)}{\r(\s)})
\hat{J}_{-n(r),bl} (m-p-\srac{n(r)}{\r(\s)}):_M \nn \\
 & & + \delta_{m,0} \frac{\bar{n}(r)}{2 \r(\s)}\left(1-\frac{\bar{n}(r)}{\r(\s)}
\right)
\g_{n(r)aj;-n(r),bl}(\s) \Big\} \label{lnll}
\end{eqnarray}
\begin{eqnarray}
\bar L_\s (m)  =   \lr^{n(r) aj; -n(r),bl}(\s) \!\!\!\!\!\!\!\!\! &
& \Big\{ \sum_{p \in \sz} :\hjb_{n(r)aj}(p+\srac{n(r)}{\r(\s)})
\hjb_{-n(r),bl}(-m-p-\srac{n(r)}{\r(\s)}):_{\bM} \nn \\ & & +
\delta_{m,0}\frac{\mnb}{2
\r(\s)}\left(1-\frac{\mnb}{\r(\s)}\right) \g_{n(r)aj;-n(r),bl}(\s)
\Big\} \ . \label{lnlr}
\end{eqnarray}
\end{subequations}
Then the ground state conditions \eqref{vacl}, \eqref{vacr}  allow us to compute
the ground state conformal weights
(conformal weights of the twist fields)
\begin{subequations}
\begin{equation}
L_\s (m \geq 0) | 0 \rangle_\s = \delta_{m,0} \hat{\Delta}_0 (\s)
| 0 \rangle_\s \sp
\bar{L}_\s (m \geq 0) | 0 \rangle_\s = \delta_{m,0} \hat{\bar{\Delta}}_0
(\s) | 0 \rangle_\s
\end{equation}
\begin{equation}
\hat{\Delta}_0 (\s) =
\frac{ c_{\sgbn}}{2}
\sum_r \frac{\nb}{2 \rho (\s)} \left( 1 - \frac{\nb}{\rho (\s)} \right)
{\rm dim} [ \nb ] \hskip 4cm
\end{equation}
$$
\hskip 1.5cm = \frac{ c_{\sgbn}}{2}
\sum_r \frac{\mnb}{2 \rho (\s)} \left( 1 - \frac{\mnb}{\rho (\s)} \right)
{\rm dim} [ \nb ] =
\hat{\bar{\Delta}}_0 (\s)
 \sp c_{\sgbn} = \frac{2 k\ {\rm dim}\, \gfrak}{2k + Q_{\sgbn}} \ .
$$
\end{subequations}
The left mover result $\hat \Delta_0 (\s)$ was given in
Ref.~\cite{Halpern:2000vj}. Here ${\rm dim} [ \nb ]$ is the
dimension of the degenerate subspace labeled by the spectral index
$\nb $. The equality of the left and right mover ground state
conformal weights follows from \eqref{viriden}, and this equality provides  a
consistency check on the form \eqref{mode-current-algebra} of the
general twisted current algebra. For the cyclic permutation
orbifolds and the $S_N$ permutation orbifolds, more explicit forms
of these conformal weights are given in
Ref.~\cite{Halpern:2000vj}. Similarly, one finds equality of the
ground state conformal weights in each sector of the general
permutation orbifold (see  App.~\ref{gcaob}).

\subsection{The Cyclic Permutation Orbifolds \label{cyclorb} }

In this and the following subsection we discuss the cyclic permutation orbibolds
$A_g(\z_\l)/\z_\l$ and the $S_N$ permutation orbifolds $A_g(S_N)/S_N$ in further
detail, concentrating on the {\it rectification problem} and the twisted
representation matrices for these cases.

For the cyclic permutation group $H={\mathbb Z}_\l$, it is known
\cite{deBoer:1999na,Halpern:2000vj} that
\begin{subequations}
\label{datacycl}
\begin{equation}
 n(r) =r , \quad \bar{n}(r) =\bar{r}, \quad E_r(\s) = e^{-2\pi i\frac{r}{\r(\s)}}
\end{equation}
\begin{equation}
  U\hc(\s)_I{}^{rj} = \srac{1}{\sqrt{\r(\s)}} e^{2\pi i
  \frac{N(\s)r(j-I)}{\l}}\d_{j,I\,\text{mod}\,\frac{\l}{\r(\s)} }
            ,\quad \schisig_{raj}=\sqrt{\r(\s)}
\end{equation}
\begin{equation}
  \sG_{raj;sbl}(\s)=\r(\s)k\e_{ab}\d_{jl}\d_{r+s,0 \,\text{mod}\,\r(\s)} ,\quad
  \sG^{raj;sbl}(\s)=\frac{1}{\r(\s)k}\e^{ab}\d^{jl}\d_{r+s,0\,\text{mod}\,\r(\s)},
\end{equation}
\begin{equation}
  \scf_{raj;sbl}{}^{tcm}(\s)=f_{ab}{}^c\d_{jl}\d_l{}^m
  \d_{r+s-t,0\;\text{mod}\r(\s)}
\end{equation}
\begin{equation}
\label{Lcycl}
\lr^{raj; sbl}(\s)=\frac{k}{2k + Q_{\sgbn} }
\g^{raj; sbl}(\s)
\end{equation}
\begin{equation}
 a=1 \ldots \text{dim }\gfrak,\quad \bar{r}=0,\ldots,\r(\s)-1,
 \quad j=0,\ldots,\srac{\l}{\r(\s)}-1,\quad \s=0,\ldots,\l-1
\end{equation}
\end{subequations}
where $I= 0,\ldots,\l-1$ and the integers $N(\s)$ are defined in
Ref.~\cite{deBoer:1999na}. In the standard notation
$\hj_{aj}^{(r)} \equiv \hj_{raj}$, $\hjb_{aj}^{(r)} \equiv \hjb_{raj} $, this
 gives the twisted current algebra of the cyclic permutation orbifolds:
\begin{subequations}
\label{twisted{J,J}Z-lambda}
\begin{eqnarray}
  [\hj^{(r)}_{aj}(m+\srac{r}{\r(\s)}),\hj^{(s)}_{bl}(n+\srac{s}{\r(\s)})]
 & = &\d_{jl} \big{(} if_{ab}{}^c\hat{J}^{(r+s)}_{cj}(m+n+\srac{r+s}{\r(\s)})
 \nn \\ &&
+ \r(\s)k \e_{ab}(m+\srac{r}{\r(\s)})\d_{m+n+\srac{r+s}{\r(\s)},0}\,\!\big{)}
\label{twisted{J,J}Z-lambda-first}
\end{eqnarray}
\begin{eqnarray}
[ \hjbb_{aj} {}^{(r)}(m+\srac{r}{\r(\s)}), \hjbb_{bl} {}^{(s)}(n+\srac{s}{\r(\s)})]
&=&\d_{jl} \big{(}\, if_{ab}{}^c\hjbb_{cj} {}^{(r+s)}(m+n+\srac{r+s}{\r(\s)})
\nn \\  & & +
\r(\s)k\e_{ab}(m+\srac{r}{\r(\s)})\d_{m+n+\srac{r+s}{\r(\s)},0}\,\!\big{)}
\label{right-mover-alg-Z-l}
\end{eqnarray}
\begin{equation}
[ \hj^{(r)}_{aj}(m+\srac{r}{\r(\s)}), \hjbb_{bl} {}^{(s)}(n+\srac{s}{\r(\s)})] = 0
\end{equation}
\begin{equation}
 \hjbb_{aj} {}^{(r)}(m+\srac{r}{\r(\s)}) \equiv
 \hjb_{aj}^{(-r)}(-m-\srac{r}{\r(\s)}) \ . \label{hjbb}
\end{equation}
\end{subequations}
Here the twisted left mover algebra \cite{deBoer:1999na,Halpern:2000vj}
is a semisimple orbifold affine algebra
\cite{Borisov:1997nc}, with commuting orbifold affine subalgebras labeled by $j$,
 at orbifold affine level $\hat{k}=\r(\s)k$.
Moreover, we have used the mode-number reversed relabelling
$\hjbb$ in \eqref{hjbb} to
{\it rectify} the twisted right mover algebra into a copy
\eqref{right-mover-alg-Z-l} of the twisted left mover algebra.

We also find that the left and right mover Virasoro generators of
$A_g(\z_\l)/\z_\l$ are copies of each other
\begin{subequations}
\begin{eqnarray}
L_\s(m) =   \frac{1}{2k + Q_{\sgbn}}  \!\!\!\!\!\!\!\!\!  & &
\Big\{ \frac{\eta^{ab}}{\rho (\s)}
\sum_{r=0}^{\rho (\s) -1} \sum_{j=0}^{\srac{\l}{\rho (\s)} -1} \sum_{p \in \sz}
: \hj_{aj}^{(r)} ( p + \srac{r}{\rho (\s)}) \hj_{bj}^{(-r)}
(m-p-\srac{r}{\r(\s)}):_M \nn   \\
 & &  + \delta_{m,0} \frac{k \l {\rm dim}\,\gfrak}{12} \Big( 1 - \srac{1}{\r(\s)}
 \Big) \Big\} \label{Vircyc}
\end{eqnarray}
\begin{eqnarray}
\bar L_\s(m) =   \frac{1}{2k + Q_{\sgbn}}  \!\!\!\!\!\!\!\!\!  & &
\Big\{ \frac{\eta^{ab}}{\rho (\s)}
\sum_{r=0}^{\rho (\s) -1} \sum_{j=0}^{\srac{\l}{\rho (\s)} -1} \sum_{p \in \sz}
: \hjbb_{aj}{}^{(r)} ( p + \srac{r}{\rho (\s)}) \hjbb_{bj}{}^{(-r)}
(m-p-\srac{r}{\r(\s)}):_M \nn   \\
 & &  + \delta_{m,0} \frac{k \l {\rm dim}\,\gfrak}{12} \Big( 1 - \srac{1}{\r(\s)}
 \Big) \Big\}
\end{eqnarray}
\end{subequations}
when the right mover Virasoro generators are expressed in terms of the
rectified current modes $\hjbb$. The left mover result in \eqref{Vircyc} was given
in Ref.~\cite{deBoer:1999na}.

The twisted representation matrices of $A_g(\z_\l)/\z_\l$
\begin{subequations}
\label{st-algebra-Z-lambda}
\begin{equation}
R(\s) = \rho (\s) \sp N(r) = n(r) = r \sp \bar{r} = 0 , \ldots , \rho (\s) -1
\end{equation}
\begin{equation}
  \st^{(r)}_{aj}(T,\s) \equiv \st_{raj}(T,\s) \sp
  \st^{(r \pm \rho (\s))}_{aj}(T,\s) =\st^{(r)}_{aj}(T,\s)
\end{equation}
\begin{equation}
  \st^{(r)}_{aj}(T,\s)_{s\a l}{}^{t\be m}=(T_a)_\a{}^\be \d_{jl}\d_l{}^m
  \d_{r+s-t, \,0 \;mod\,\r(\s)} \label{st-in-AZl}
\end{equation}
\begin{equation}
  [\st^{(r)}_{aj}(T,\s), \st^{(s)}_{bl}(T,\s)]=\d_{jl}
  if_{ab}{}^c\st^{(r+s)}_{cj}(T,\s) \label{st-algebra-Z-lambda-c}
\end{equation}
\end{subequations}
also follow from the simplified formula \eqref{repmatperm} and the data
\eqref{datacycl}. The algebra \eqref{st-algebra-Z-lambda-c} of the twisted
representation matrices shows the same semisimplicity
 as the twisted left and right mover current algebras in
 \eqref{twisted{J,J}Z-lambda}. The example
\begin{equation}
[ \hj_{aj}^{(r)} (n + \srac{r}{\rho (\s)}), \hgp(\T,z,\s)_{s \a l}{}^{t \be m} ]
=\hgp(\T,z,\s)_{s \a l}{}^{s' \a' l'} \T_{aj}^{(r)} (T,\s)_{s' \a' l'}{}^{t \be m}
z^{ n + \srac{r}{\rho (\s)} }
\end{equation}
illustrates the action of the twisted representation matrices in this case

Further study of the WZW cyclic permutation orbifolds is found in
Subsec.~\ref{cyclfl}.

\subsection{The $S_N$ Permutation Orbifolds \label{snorb} }

For the permutation group $H=S_N$, it is known \cite{Halpern:2000vj} that
\begin{subequations}
\label{datasn}
\begin{equation}
 I \goto \hjs j, \quad n(r)l \goto \hls l
\end{equation}
\begin{equation}
\label{moeig}
 U\hc(\vec{\s})_{\hat{j} j}{}^{\hat{l}l} = \frac{\d_j{}^l}{\sqrt{\s_j}}
 e^{-2\pi i\frac{\hjss\hat{l}}{\s_j}},
     \quad E_{\hjs}^j (\vec{\s}) = e^{-2\pi i \frac{\hjs}{\s_j}} ,
     \quad  \frac{n(r)}{\r(\s)} = \frac{\hjs}{\s_j}
\end{equation}
\begin{equation}
 \sum_{j=0}^{n(\vec{\s})-1} \s_j = N,\quad\quad \s_{j+1} \leqslant \s_j
\end{equation}
\begin{equation}
 \hj_{n(r)aj} \goto  J_{aj}^{(\hjs)}\equiv \hj_{\hjs aj} , \quad \hjb_{n(r)aj}
  \goto \bar{J}_{aj}^{(\hjs)}  \equiv \hjb_{\hjs aj} ,\quad \schi_{\hjs aj}(\vec{\s})
   = \sqrt{\s_j}
\end{equation}
\begin{equation}
 \sG_{\hjs aj; \hls bl} (\vec{\s}) = \d_{jl} \s_j k \e_{ab} \d_{\hjs+\hls ,0\,
 \text{mod}\,\s_j } ,\quad  \sG^{\hjs aj; \hls bl} (\vec{\s}) =
 \frac{\d^{jl}}{\s_j} \frac{\e^{ab}}{k} \d_{\hjs+\hls ,0\,\text{mod}\,\s_j }
\end{equation}
\begin{equation}
 \scf_{\hjs aj; \hls bl}{}^{\hms cm} (\vec{\s}) = f_{ab}{}^{c} \d_{jl}\d_l{}^{m}
  \d_{\hjs + \hls -\hmss,0 \,\text{mod}\,\s_j }
\end{equation}
\begin{equation}
a= 1 \ldots {\rm dim} \,  \gfrak \sp
 j,l=0,\ldots,n(\vec{\s})-1,\quad \bhj=0,\ldots,\s_j-1,\quad \bhl=0,\ldots,\s_l-1 \ .
\end{equation}
\end{subequations}
Here each sector $\vec{\s} = \{\s_j\}$ is a partition of $N$, which also labels
a representative element $h_{\vec{\s}}$ of each conjugacy class of $S_N$:
The integer $\hjs$ runs within each disjoint cycle $\s_j$ (an element of
$\z_{\s_j}$ with size and order $\s_j$), and the integer $j$ runs over the set of all disjoint
cycles of $h_{\vec{\s}} \in S_N$. Note that the matrix of eigenvectors
$U\hc(\vec{\s})$ in \eqref{moeig} is a direct sum in the space of disjoint cycles.

This leads to the twisted current algebra of  the permutation orbifolds
$A_g(S_N)/S_N$
\begin{subequations}
\label{SN-CA}
\begin{equation}
 [\hj_{aj}^{(\hjs)}(m\+\srac{\hjs}{\s_j}),\hj_{bl}^{(\hls)}(n\+\srac{\hls}{\s_l})]
= \d_{jl} \big{(} if_{ab}{}^c\hj_{cj}^{(\hjs+\hls)}\!
(m\+n\+\srac{\hjs+\hls}{\s_j})
    + (\s_jk)\e_{ab}(m\+\srac{\hjs}{\s_j})\d_{m+n+\frac{\hjs+\hls}{\s_j},0}
     \big{)} \label{S_N-{J,J}-begin}
\end{equation}
\begin{equation}
[\hjbb_{aj} {}^{(\hjs)}(m\+\srac{\hjs}{\s_j}),
\hjbb_{bl} {}^{(\hls)}(n\+\srac{\hls}{\s_l})]
    = \d_{jl} \big{(} if_{ab}{}^c\hjbb_{cj} {}^{(\hjs+\hls)}\!
    (m\+n\+\srac{\hjs+\hls}{\s_j})
+ (\s_jk)\e_{ab}(m\+\srac{\hjs}{\s_j})\d_{m+n+\frac{\hjs+\hls}{\s_j},0} \big{)}
\label{SN-CA-barred}
\end{equation}
\begin{equation}
 [\hj_{aj}^{(\hjs)}(m\+\srac{\hjs}{\s_j}),
 \hjbb_{bl} {}^{(\hls)}(n\+\srac{\hls}{\s_l})] = 0 \label{S_N-{J,J}-end}
\end{equation}
\begin{equation}
 \hjbb_{aj} {}^{(\hjs)}(m+\srac{\hjs}{\s_j}) \equiv \hjb_{aj}^{(-\hjs)}
 (-m-\srac{\hjs}{\s_j}) \label{S_N-rectify} \ .
\end{equation}
\end{subequations}
Here the twisted left mover algebra \cite{Halpern:2000vj} in
\eqref{S_N-{J,J}-begin} is a semisimple orbifold affine algebra
\cite{Borisov:1997nc}, with commuting subalgebras
 labeled by $j$, at orbifold affine level $\hat{k}_j=\s_j k$. Moreover, we have
 used the mode-number reversed relabelling \eqref{S_N-rectify} to rectify the
 twisted right mover algebra into a copy \eqref{SN-CA-barred} of the twisted left
  mover algebra. The left and right mover Virasoro generators are also found
  to be copies (see Subsec.~\ref{unrec}) when the right mover generators are
  expressed in terms of the rectified current modes $\hjbb$.

Finally, we give the twisted representation matrices of $A_g(S_N)/S_N$
\begin{subequations}
\begin{equation}
\frac{N(r)}{R(\s)} = \frac{n(r)}{\r(\s)} = \frac{\hjs}{\s_j} \sp
\hat{\bar{j}} = 0 , \ldots , \s_j-1
\end{equation}
\begin{equation}
 \st_{aj}^{(\hjs)}(T,\vec{\s}) \equiv \st_{\hjs aj}(T,\vec{\s}) \sp
 \st_{aj}^{(\hjs \pm \s_j )}(T,\vec{\s}) =  \st_{aj}^{(\hjs)}(T,\vec{\s})
\end{equation}
\begin{equation}
 \st_{aj}^{(\hjs)}(T,\vec{\s})_{\hls \a l}{}^{\hms \be m} = (T_a)_\a{}^{\be} \d_{jl}
 \d_l{}^{m} \d_{\hjs+\hls-\hms ,0\,\text{mod}\,\s_j } \label{st-in-SN}
\end{equation}
\begin{equation}
\label{twralg}
 [\st_{aj}^{(\hjs)}(T, \vec{\s}), \st_{bl}^{(\hls)}(T, \vec{\s})] =
 \d_{jl} i f_{ab}{}^c \st_{cj}^{(\hjs+\hls)}(T, \vec{\s})
\end{equation}
\end{subequations}
which also follow from the data \eqref{datasn} and Eq.~\eqref{repmatperm}.
The algebra \eqref{twralg} of the twisted representation matrices shows
the same semisimplicity as the twisted current algebra \eqref{SN-CA}.
The example
\begin{equation}
[ \hj_{aj}^{(\hjs)} (n + \srac{\hjs}{\s_j}), \hgp(\T,z,\s)_{\hls \a l}
{}^{\hms \be m} ]
=\hgp(\T,z,\s)_{\hls \a l}{}^{\hls' \a' l'} \T_{aj}^{(\hjs)} (T,\s)_{ \hls' \a' l'}{}^{\hms \be m}
z^{ n + \srac{\hjs}{\s_j} }
\end{equation}
illustrates the action of the twisted representation matrices in this case.

Further discussion of the $S_N$ permutation orbifolds can be found in
Ref.~\cite{Halpern:2000vj}.

\subsection{Universal Forms and Rectification of the Permutation Orbifolds \label{unrec}}

For all permutation groups $H \subset S_N$ it is well known that a representative element
$h_\s$ of each conjugacy class of $H$ can be expressed in terms of disjoint
cycles (see e.g. Subsec.~\ref{snorb}). The matrix of eigenvectors $U\hc (\s)$
can always be taken as a direct sum in the space of disjoint cycles, so we
find%
\footnote{A qualitative version of this argument was noted in
Refs.~\cite{deBoer:1999na} and \cite{Halpern:2000vj}. See also
Ref.~\cite{Bantay:1998ek}.}
{\it for all permutation orbifolds} that
the twisted left mover current algebra $ \gfrakh(h_\s)$ of sector $\s$ is
isomorphic%
\footnote{That is, the normalization constants $\chi_{n(r) a j}
(\s)$ can be chosen to obtain the form \eqref{coalg}.} to the
following commuting set of orbifold affine algebras:
\begin{subequations}
\label{csoal}
\begin{equation}
\gfrakh (h_\s)  = \gfrakh (H(\mbox{permutation})\subset {\rm Aut}(g) ;\s)
\end{equation}
\begin{eqnarray}
   [ \hj_{n(r) aj } (m + \srac{n(r)}{\rho (\s)}),
 \hj_{n(r) bl} (n + \srac{n(s)}{\rho (\s)})]  & = &
 \de_{jl} \Big( if_{ab}{}^c \hj_{n(r)+n(s),cj }
 (m +n+ \srac{n(r)+n(s)}{\rho (\s)}) \nn \\
\label{coalg}
 & &
+  \hat k_j(\s) \eta_{ab} (m + \srac{n(r)}{\rho (\s)}) \
\de_{\mnnrnsrsf,0}\ \Big) \;\;\;\;\;\;\;\;\; \\
 \hat k_j(\s) \equiv k f_j (\s)  \sp a,b,c &  \!\!\!\!\!\! = \!\!\!\!\!\! &
 1 \ldots {\rm dim}\,  \gfrak \sp \s = 0 , \ldots , N_c-1 \ .
\end{eqnarray}
\end{subequations}
Here $\{ \hat k_j (\s)\}$ is a set of orbifold affine levels, where $f_j(\s)$
is the size and order%
\footnote{We already know that $f_j(\s)= \r (\s)$ for $H =\z_\l$ and
$f_j(\s) = \s_j$ for $S_N$.}
 of each disjoint cycle $j$ in $h_\s \in H(\mbox{permutation})$. The ratios
\begin{equation}
\frac{N(r)}{R(\s)} = \frac{n(r)}{\r(\s)} = \frac{\hjs}{f_j(\s)} \sp
\hat{\bar{j}} = 0 , \ldots , f_j (\s)-1
\end{equation}
are also determined as in Subsec.~\ref{snorb}.
The twisted right mover current algebra $\gfrakh(h_\s^{-1}) $ commutes with
 \eqref{csoal} and has the same form with $k \rightarrow - k $.

Looking back, we see that
the mode-number reversed relabellings \eqref{hjbb}
and \eqref{S_N-rectify} rectify each commuting right mover subalgebra separately,
and indeed it is easily checked as above that the mode-number reversed relabelling
\begin{subequations}
\label{recjperm0}
\begin{eqnarray}
   [ \hjbb_{n(r) aj } (m + \srac{n(r)}{\rho (\s)}),
 \hjbb_{n(r) bl} (n + \srac{n(s)}{\rho (\s)})]  & = &
 \de_{jl} \Big( if_{ab}{}^c \hjbb_{n(r)+n(s),cj }
 (m +n+ \srac{n(r)+n(s)}{\rho (\s)}) \nn \\
 & &
+  \hat k_j(\s) \eta_{ab} (m + \srac{n(r)}{\rho (\s)}) \
\de_{\mnnrnsrsf,0}\ \Big) \;\;\;\;\;\; \\
\label{recjperm} \hjbb_{n(r)aj}(m+\srac{n(r)}{\r(\s)})  & \equiv &
 \hjb_{-n(r),aj}(-m-\srac{n(r)}{\r(\s)})
\end{eqnarray}
\end{subequations}
provides a {\it universal rectification} of the twisted right mover algebra
$(\gfrakh(h_\s^{-1}) \simeq \gfrak(h_\s))$ for each sector of all permutation
 orbifolds. (In the language of Eq.~\eqref{rmcot}, this rectification chooses
 $\theta (\s) =1$ for all $\s, n(r),aj$).
  In terms of the rectified right mover modes the ground state
 conditions \eqref{vacr} take the form
 \begin{equation}
 0 = \hjbb_{n(r) aj } (m + \srac{n(r)}{\rho (\s)} \geq  0) | 0 \rangle_\s
 = {}_\s \langle 0 |  \hjbb_{n(r) aj } (m + \srac{n(r)}{\rho (\s)} \leq  0)
\end{equation}
which are a copy of the twisted left mover conditions in \eqref{vacl}.

In the basis \eqref{coalg}, we note that the twisted structure constants and
twisted metric take the universal form
\begin{subequations}
\label{fgperm}
\begin{equation}
\label{fgperm0}
\F_{n(r) aj ;n(s) b l } {}^{n(t)cm} (\s)  = f_{ab}{}^c \de_{jl} \de_l^m
\de_{n(r) + n(s) - n(t), 0 \rmod \rho(\s) }
\end{equation}
\begin{equation}
\label{fgperm2}
\G_{n(r)a j;n(s)b l } (\s) = \hat k_j (\s) \eta_{ab} \de_{jl}
\de_{n(r) + n(s), 0 \rmod \rho(\s) }
\end{equation}
\begin{equation}
\label{symprop}
\F_{-n(r),aj ;-n(s), b l } {}^{-n(t),cm} (\s)
= \F_{n(r) aj ;n(s) b l } {}^{n(t)cm} (\s) \sp
\G_{-n(r),aj;-n(s),b l } (\s)= \G_{n(r)a j;n(s)b l } (\s)
\end{equation}
\end{subequations}
for all sectors of all permutation orbifolds.
Moreover, as required by the discussion of Subsec.~\ref{quarg}, the universal
rectification \eqref{recjperm0} of the permutation orbifolds follows from  the
symmetries \eqref{symprop} alone.

>From \eqref{eq:reason-for-casimir-of-st} and \eqref{fgperm0} we find the
universal form of the algebra of the twisted representation matrices
\begin{equation}
[\T_{n(r)aj} (T,\s), \T_{n(s)bl}(T,\s) ] = \de_{jl} i f_{ab}{}^c
\T_{n(r) + n(s),cj} (T,\s) \sp a,b,c = 1 \ldots {\rm dim}\,\gfrak
\end{equation}
in this basis. The {\it factorized form} $\T = T t$ of the twisted representation
matrices
\begin{subequations}
\label{Tcycl0}
\begin{equation}
\label{Tcycl}
\T_{n(r) aj } (T,\s) = T_a \otimes t_{n(r)j} (\s) \equiv T_a t_{n(r) j} (\s)
\end{equation}
\begin{equation}
[T_a , T_b] = i f_{ab}{}^c T_c \sp
t_{n(r)j} (\s) t_{n(s) l}  (\s) = \de_{jl} t_{n(r)+n(s),j} (\s) \sp
t_{n(r) \pm \rho (\s),j} (\s) = t_{n(r)j} (\s)
\end{equation}
\begin{equation}
t_{n(r)j} (\s)_{n(s)l}{}^{n(t)m} = \de_{jl} \de_l^m
\de_{n(r)+n(s)-n(t) ,0 \,\rmod \rho(\s)} \sp \srange
\end{equation}
\end{subequations}
is discussed in Subsec.~\ref{cyclfl} and App.~\ref{Jperm}.

 Moreover, using \eqref{perml} and \eqref{fgperm2}, we find the universal
 forms
\begin{subequations}
\label{lgs00}
\begin{eqnarray}
\G^{n(r)a j;n(s)b l } (\s) & = & \hat k_j^{-1} (\s) \eta^{ab} \de^{jl}
\de_{n(r) + n(s), 0 \rmod \rho(\s) } \\
\lr^{n(r) aj;n(s)bl} (\s) & =  & \frac{k/\hat k_j(\s)}{2k + Q_{\sgbn}}
\eta^{ab} \delta^{jl}
\delta_{n(r) + n(s), 0 \rmod \rho(\s) } \label{lgs0} \\
\G^{-n(r), aj;-n(s),b l} (\s)  =   \G^{n(r)a j;n(s)b l } (\s)  &, &
\lr^{-n(r), aj;-n(s),b l} (\s)  =   \lr^{n(r)aj;n(s)b l } (\s) \label{lgs1} \\
\lr^{n(r) bl;-n(r),aj} (\s) & = &  \lr^{n(r) aj;-n(r),bl} (\s) \ .  \label{lgs2}
\end{eqnarray}
\end{subequations}
The final symmetry in \eqref{lgs2} is sufficient to verify the rectification identity
$$
 \lr^{n(r) aj;-n(r),bl} (\s)  \sum_{p \in \sz}
: \hjb_{n(r)aj}( p +\srac{n(r)}{\rho (\s)}) \hjb_{-n(r), bl}
( -m -p +\srac{n(r)}{\rho (\s)}) :_{\bM}
 $$
 \begin{equation} \label{E11}
  =
\lr^{n(r) aj;-n(r),bl} (\s)  \sum_{p \in \sz}
: \hjbb_{n(r)aj}( p +\srac{n(r)}{\rho (\s)}) \hjbb_{-n(r), bl}
( m -p +\srac{n(r)}{\rho (\s)}) :_{M}
\end{equation}
where $\hjbb$ are the rectified right mover modes in \eqref{recjperm} and
$M$ and $\bM$ ordering are defined in \eqref{mnbnol} and \eqref{mnbnor}.

Moreover, we may use \eqref{viriden}, \eqref{recjperm} and
 \eqref{E11} to express the right mover Virasoro generators
\eqref{lnlr} as a copy
\begin{eqnarray}
\bar L_\s (m)  =   \lr^{n(r) aj; -n(r),bl}(\s) \!\!\!\!\!\!\!\!\! &
& \Big\{ \sum_{p \in \sz} :\hjbb_{n(r)aj}(p+\srac{n(r)}{\r(\s)})
\hjbb_{-n(r),bl}(m-p-\srac{n(r)}{\r(\s)}):_{M} \nn \\ & & +
\delta_{m,0}\frac{\nb}{2
\r(\s)}\left(1-\frac{\nb}{\r(\s)}\right) \g_{n(r)aj;-n(r),bl}(\s)
\Big\} \label{lnlr2}
\end{eqnarray}
of the left mover generators in \eqref{lnll}.
It is reasonable to expect more generally (see also Subsec.~\ref{amno}) that this
situation will occur whenever the twisted right mover currents are
rectifiable.

We finally give the universal form of the classical WZW orbifold action
(see Subsec.~\ref{clarg}) in the case of WZW permutation orbifolds:
\begin{subequations}
\begin{equation}
\hg (\T,\xi,t,\s) = e^{i \hat \be^{n(r)aj}(\xi,t,\s) T_a t_{n(r)j}(\s)}
\sp
\hg (\T,\xi,t,\s)_{n(r)\a j}{}^{n(s)\be l}
 =\de_j^l \ \hg_j(\T,\xi,t,\s)_{n(r)\a }{}^{n(s)\be }
\end{equation}
\begin{equation}
 \hg_j(\T,\xi,t,\s) = e^{i \sum_{n(r)a}\hat \be^{n(r)aj}(\xi,t,\s) T_a \tau_{n(r)}(\s)}
 \sp \tau_{n(r)}(\s)_{n(s)}{}^{n(t)} \equiv \de_{n(r) + n(s)-n(t), \, 0 \rmod \rho(\s)}
\end{equation}
\begin{eqnarray}
\hat \be^{n(r)aj} (\xi+2\pi,t,\s) & = &
 \hat \be^{n(r) aj } (\xi,t,\s)
   e^{2\pi i \frac{n(r)}{\rho (\s)}} \\
\hg(\st,\xi+2\pi,t,\s)_{n(r)\a j}{}^{n(s)\be l} & = &
   e^{-2\pi i \frac{n(r)-n(s)}{\rho (\s)}}
   \hg(\st,\xi,t,\s)_{n(r)\a j}{}^{n(s) \be l} \\
\hg_j(\st,\xi+2\pi,t,\s)_{n(r)\a}{}^{n(s)\be} & = &
   e^{-2\pi i \frac{n(r)-n(s)}{\rho (\s)}}
   \hg_j(\st,\xi,t,\s)_{n(r)\a}{}^{n(s) \be}
\end{eqnarray}
\begin{eqnarray}
\hat S_\s [ \M, \hg] =-\frac{k}{y(T)}\sum_{j} \BIG{(}
\frac{1}{8\pi} \!\!\!\!\!\! & & \int\! d^2\xi\,\widehat{\hat{\rm Tr}}\big{(}\,\hat{g}^{-1}_j(\st,\s)
\pl_+\hat{g}_j(\st,\s)\0b\hat{g}^{-1}_j(\st,\s)\pl_-\hat{g}_j(\st,\s)\;\big{)}
\nn \\
    & & +\frac{1}{12\pi}\int_{\Gamma}
    \widehat{\hat{\rm Tr}}\big{(}\,(\;\hat{g}^{-1}_j(\st,\s) d\hat{g}_j(\st,\s)\;)^3\,
    \big{)} \BIG{)}
        \label{action-Z-l}
\end{eqnarray}
\begin{equation}
 \widehat{\hat{\rm Tr}}(AB) \equiv \sum_{n(r),n(s)}  \sum_{\a,\be}
 A_{n(r) \a}{}^{n(s) \be} B_{n(s)\be}{}^{n(r)\a}\ .
\end{equation}
\end{subequations}
In obtaining these results, we used Eqs.~\eqref{gorbact}, \eqref{admon},
\eqref{twmperm}
and \eqref{Tcycl0}. Note that, as it should be, the orbifold action is separable
in the semisimplicity index $j$ of the universal orbifold affine algebra
\eqref{csoal}. The required integers  and the range of $j$ are given
explicitly for $A_g(\z_\l)/\z_\l$ and $A_g(S_N)/S_N$ in
Subsecs.~\ref{cyclorb} and \ref{snorb}.

\section{The Twisted KZ Equations of WZW Permutation Orbifolds \label{kzsec}}

\subsection{Preliminaries}

For the WZW  permutation orbifolds the vertex operator equations
\eqref{difrelM} take the form
\begin{subequations}
\begin{equation}
\label{difreltMp}
\partial \hgp(\T,z) = 2 \lr^{n(r)aj;-n(r),bl}(\s)
\left( : \hat{J}_{n(r)aj}(z) \hgp (\T,z) :_M -
\srac{\bar n(r)}{\r(\s)} \frac{1}{z} \hgp (\T,z)\T_{n(r)aj} \right)
\T_{-n(r),bl}
\end{equation}
\begin{equation}
\label{rdifreltMp}
\bar \partial \hgm(\T,\bz) = - 2 \lr^{n(r)aj;-n(r),bl}(\s) \T_{-n(r),bl}
\left( : \hjb_{n(r)aj}(z) \hgm (\T,\bz) :_{\bM} +
\srac{\mnb}{\r(\s)} \frac{1}{\bz} \T_{n(r)aj} \hgm (\T,\bz) \right) \ .
\end{equation}
\end{subequations}
Similarly, we obtain the alternate form of
the vertex operator equations
\begin{subequations}
\label{difrelpp}
\begin{eqnarray}
\partial \hgp(\T,z) & = &  \lr^{n(r)aj;-n(r),bl}(\s)
\left( 2 : \hat{J}_{n(r)aj}(z) \hgp (\T,z) :_M  \T_{-n(r),bl}
\right. \nn \\
\label{difreltM2p}
 &  & \left. -
 \frac{1}{z} \hgp (\T,z)\T_{n(r) aj}\T_{-n(r),bl} (1 - \delta_{\bar n(r),0} )
 \right)
\end{eqnarray}
\begin{eqnarray}
\bar \partial \hgm(\T,\bz) & = &  - \lr^{n(r)aj;-n(r),bl}(\s)
\left( 2  \T_{-n(r),bl} : \hjb_{n(r)aj}(z) \hgm (\T,\bz) :_{\bM}
\right. \nn \\
\label{rdifreltM2p}
& &  \left. +
 \frac{1}{\bz} \T_{-n(r),bl}\T_{n(r)aj} \hgm (\T,\bz) (1 - \delta_{\bar n(r),0} )
 \right)
\end{eqnarray}
\end{subequations}
where we have used \eqref{scperm} to show that the $\F$-terms of
\eqref{altform} are zero for
WZW permutation orbifolds. Correspondingly,   the
identities \eqref{ident} take the form
\begin{subequations}
\label{idp}
\begin{equation}
2 \lr^{n(r)aj ; -n(r),bl}(\s) \srac{\bar n(r)}{\r(\s)}
\T_{n(r)aj} \T_{-n(r),bl}
=\lr^{n(r)aj ; -n(r),bl} (\s) \T_{n(r)aj} \T_{-n(r),bl} (1 - \delta_{\bar n(r),0} )
\end{equation}
\begin{eqnarray}
2 \lr^{n(r)aj ; -n(r),bl} (\s) \srac{\mnb }{\r(\s)} \T_{-n(r),bl} \T_{n(r)aj}
& = & \lr^{n(r)aj ;-n(r),bl} (\s)  \T_{-n(r),bl} \T_{n(r)aj}  (1 -
\delta_{\bar n(r),0} ) \label{idp2} \\
& =&   \lr^{n(r)aj; -n(r),bl} (\s)  \T_{n(r)bl} \T_{-n(r),aj}  (1 -
\delta_{\bar n(r),0} ) \hskip 1cm \label{idp3}
\end{eqnarray}
\end{subequations}
for the WZW permutation orbifolds. The last form in \eqref{idp3} follows
by $n \rightarrow -n$ from \eqref{idp2}.

Using the ground state conditions \eqref{vacc}, we find
that the twisted partial currents obey
\begin{subequations}
\begin{equation}
\label{vacp}
0 =\hat J_{n(r)aj}^+ (z)| 0\rangle_\s ={}_\s \langle 0 |
\hat J_{n(r)aj}^- (z) ={}_\s \langle 0 | :\hgp (\T,z)
\hat J_{n(r)aj} (z) :_M | 0\rangle_\s
\end{equation}
\begin{equation}
\label{rvacp} 0 =\hjb_{n(r)aj}^- (\bz)| 0\rangle_\s ={}_\s \langle
0 | \hjb_{n(r)aj}^+ (\bz) ={}_\s \langle 0 | :\hgm (\T,\bz)
\hjb_{n(r)aj} (\bz) :_{\bM} | 0\rangle_\s
\end{equation}
\end{subequations}
where we used \eqref{lmexp} and \eqref{rmexp} for the left and right movers
respectively.

\subsection{The Twisted Left Mover KZ Equations \label{twlmKZ} }

Using the twisted vertex operator equation \eqref{difreltMp},
the ground state conditions \eqref{vacp} and the commutation relations
\eqref{jpmg}, we obtain after some algebra the {\it twisted Knizhnik-Zamolodchikov
 equations}
\begin{subequations}
\label{tlmkzeq}
\begin{equation}
\frac{A_g (H)}{H} \sp H(\mbox{permutation}) \subset {\rm Aut}(g)
\end{equation}
\begin{equation}
\hat A_+ (\T,z,\s) \equiv \langle  \hgp(\T^{(1)},z_1,\s) \hgp(\T^{(2)},z_2,\s)
\cdots \hgp(\T^{(N)},z_N,\s)  \rangle_\s
\end{equation}
\begin{equation}
\part_\mu \hat A_+(\T,z,\s) = \hat A_+ (\T,z,\s) \hat W_\mu (\T,z,\s) \sp
\s = 0, \ldots ,N_c -1
\end{equation}
\begin{equation}
\label{kzct}
\hat W_{\mu}(\T,z,\s) = 2 \lr^{n(r)aj;-n(r),bl} (\s)
 \left[ \sum_{\nu \neq \mu}\left( \frac{z_{\nu}}{z_{\mu}}
\right)^{ \srac{\bar n(r)}{\r(\s)}} \frac{1}{z_{\mu \nu}}
\T_{n(r)aj}^{(\nu)} \T_{-n(r),bl}^{(\mu)}
- \srac{\bar n(r)}{\r(\s)} \frac{1}{z_{\mu}}
\T_{n(r)aj}^{(\mu)} \T_{-n(r),bl}^{(\mu)} \right]
\end{equation}
\begin{equation}
\T^{(\n)} \T^{(\m)} \equiv
\T^{(\n)} \otimes \T^{(\m)} \sp z_{\mu \nu } \equiv z_\mu - z_\nu
\sp \sum_{\nu \neq \mu} \equiv \sum_{{\nu =1 \atop \nu \neq \mu}}^N
\end{equation}
\end{subequations}
for each twisted left mover sector $\s$ of all WZW permutation orbifolds.
These equations are a central result of this paper.
We remind the reader of the data in Subsecs.~\ref{cyclorb} and \ref{snorb} and
the universal forms of $\T = T t$ and $\lr (\s)$ in Eqs.~\eqref{Tcycl0} and
\eqref{lgs00} respectively (see also Subsec.~\ref{cyclfl}).

By construction the twisted connection \eqref{kzct} is flat
\begin{equation}
\part_\mu \hat W_\nu (\s)- \part_\nu \hat W_\mu(\s) +
[\hat W_\mu (\s) ,\hat W_\nu (\s) ] = 0 \sp
\hat W_\mu (\s) \equiv \hat W_\mu (\T,z,\s)
\end{equation}
and in fact this twisted connection is abelian flat
\begin{subequations}
\begin{equation}
\label{abfl1}
\part_\mu \hat W_\nu (\s) - \part_\nu \hat W_\mu (\s) = 0
\end{equation}
\begin{equation}
\label{abfl2}
[\hat W_\mu (\s),\hat W_\nu (\s) ] = 0 \ .
\end{equation}
\end{subequations}
To check \eqref{abfl1}, compute the derivatives explicitly. For $\bar n(r)=0$ the
identity is immediate. For $\bar n(r) \neq 0$ use in
the second term the symmetry \eqref{lgsym} of $\lr$ and relabel $ n \rightarrow -n $ so
that $\srac{\bar n}{\rho} \rightarrow
1-\srac{\bar n}{\rho }$. Then the two terms sum to zero.
The condition \eqref{abfl2}  is checked explicitly for the
cyclic permutation orbifolds in Subsec.~\ref{cyclfl}. The $z_\m^{-1}$ terms
in \eqref{tlmkzeq} can be traced back to the fact that the twist field states
$| 0\rangle_\s = \tau_\s (0) | 0 \rangle$ are not invariant under $Sl(2) \oplus
Sl(2)$.

For completeness, we also give the alternate form of the twisted connection
\begin{equation}
\label{altkzcon}
\hat W_{\mu}(\s) =  \lr^{n(r)aj;-n(r),bl} (\s)
 \left[ \sum_{\nu \neq \mu}\left( \frac{z_{\nu}}{z_{\mu}}
\right)^{ \srac{\bar n(r)}{\r(\s)}} \frac{2}{z_{\mu \nu}}
\T_{n(r)aj}^{(\nu)} \T_{-n(r),bl}^{(\mu)}
- (1 - \delta_{\bar n(r),0} )  \frac{1}{z_{\mu}}
\T_{n(r)aj}^{(\mu)} \T_{-n(r),bl}^{(\mu)} \right]
\end{equation}
that follows from the alternate form \eqref{difreltM2p}
of the twisted vertex operator equation.

We also find global Ward identities corresponding to the
{\it residual symmetry} generated by the elements of the
 Lie subalgebra \eqref{liesa} of sector $\s$.
For the left mover sectors of the WZW permutation orbifolds, these
Ward identities read
\begin{subequations}
\label{gidt}
\begin{equation}
\langle [ \hat J_{0aj  }(0), \hgp(\T^{(1)},z_1,\s) \cdots
\cdots \hgp(\T^{(N)},z_N,\s)  ] \rangle_\sigma = 0
\quad \Rightarrow \quad
\hat A_+(\T,z,\s) \hQ_{aj} (\s) =0 \qquad
\end{equation}
\begin{equation}
\hQ_{aj} (\s) \equiv \sum_{\mu =1}^N \T_{0aj}^{(\mu)} \sp
a = 1 \ldots {\rm dim}\,\gbn \sp \forall \, j
\end{equation}
\begin{equation}
[ \hQ_{aj} (\s) , \hQ_{bl} (\s) ] = i \scf_{0aj;0bl}{}^{0cm} \hQ_{cm} (\s)
\end{equation}
\end{subequations}
in terms of the twisted representation matrices $\T$ with $n(r)=0$.
Using properties of the twisted representation matrices (see
App.~\ref{Jperm} and in particular Eq.~\eqref{idlast}), we find that
\begin{equation}
\label{wTcom}
[ \hat W_\mu (\s) ,  \hQ_{aj} (\s) ] = 0
\end{equation}
so that the global Ward identities \eqref{gidt}
are consistent with the twisted KZ equations.

Finally, we note the simple form of the $L_\s (0)$ Ward identities
\begin{equation}
\label{L0wi2}
 \hat A_+  (\T,z,\s) \sum_{\m=1}^N \left(\overleftarrow{\partial_\m} z_\m
  + \Delta_{\sgbn}  (T^{(\m)} ) \right) = 0
\end{equation}
which follows from  \eqref{L0wi} and \eqref{twcfm} for the
WZW permutation orbifolds. In fact (as in the untwisted case) we find that the
$L_\s (0)$ Ward identities are automatically satisfied when both
the twisted KZ equations \eqref{tlmkzeq} and the global Ward identities
\eqref{gidt} are satisfied.
To see this, the reader may find helpful the following intermediate steps:
\begin{subequations}
\label{lsumid1}
\begin{equation}
\lr^{n(r)a j;-n(r),bl} (\s) \sum_{{\m, \n \atop \m \neq \n}}
\frac{z_\m}{z_{\m \n}} \left( \frac{z_\n}{z_{\m}} \right)^{\srac{\nb}{\rho(\s)}}
\T_{n(r)aj}^{(\nu)} \T_{-n(r),bl}^{(\mu)}(1 - \delta_{\bar n(r),0} ) = 0
\end{equation}
\begin{equation}
 \lr^{0a j;0bl} (\s) \sum_{{\m, \n \atop \m \neq \n}} \frac{z_\n}{z_{\m \n}}
\T_{0aj}^{(\nu)} \T_{0bl}^{(\mu)}
= - \lr^{0a j;0bl} (\s) \sum_{{\m, \n \atop \m \neq \n}} \frac{z_\m}{z_{\m \n}}
\T_{0aj}^{(\nu)} \T_{0bl}^{(\mu)}
\end{equation}
\begin{equation}
\sum_\m z_\m \hat W_\m (\s) =\lr^{0a j;0bl} (\s)
\sum_{{\m, \n \atop \m \neq \n}}
\T_{0aj}^{(\nu)} \T_{0bl}^{(\mu)} -
\lr^{n(r)a j;-n(r),bl} (\s)\sum_{\m} \T_{n(r)aj}^{(\mu)} \T_{-n(r),bl}^{(\mu)}
(1 - \delta_{\bar n(r),0} )
\end{equation}
\begin{equation}
\hat A_+ (\s)  \ \lr^{0a j;0bl}(\s)
\sum_{{\m, \n \atop \m \neq \n}}
\T_{0aj}^{(\nu)} \T_{0bl}^{(\mu)}
=-\hat A_+ (\s) \ \lr^{0a j;0bl} (\s) \sum_\m \T_{0aj}^{(\mu)} \T_{0bl}^{(\mu)}
\end{equation}
\begin{equation}
\hat A_+ (\s)  \sum_\m z_\m \hat W_\m  (\s) =
- \hat A_+ (\s) \ \lr ^{n(r)a j;-n(r),bl} (\s) \sum_\m \T_{n(r)aj}^{(\mu)}
\T_{-n(r),bl}^{(\mu)} =-\hat A_+ (\s)  \sum_\m \Delta_{\sgbn} (T^{(\m)} ) \ .
\end{equation}
\end{subequations}
The first identity in \eqref{lsumid1} holds for each $(\m \n) + (\n \m)$ pair of
terms separately, without further summation. To see this use $n\rightarrow -n$
in the second term. The second identity follows on $\m \leftrightarrow \n$
exchange, as in the untwisted case. The third identity then follows immediately
from the alternate form \eqref{altkzcon} of the twisted KZ connection.
The final identities follow from the global Ward identities and \eqref{twcfm}.

We finally recall from \eqref{salf} that for $\s =0$
\begin{equation}
\bar  n (r) =0 \sp j ,l \rightarrow J,L = 0 \, \ldots , K-1 \sp
{\cL}_{\sgbn (0)} (0) \rightarrow L_g^{aJ,bL} \sp
\T (0) \rightarrow T_{aJ}
\end{equation}
so our twisted KZ system reduces to the appropriate untwisted KZ system
in the untwisted sector of each permutation orbifold.

\subsection{The Twisted Right Mover KZ Equations \label{twrmKZ} }

For the twisted right mover sectors, the derivation proceeds along the same lines.
Using the twisted vertex operator equation \eqref{rdifreltMp},
the ground state conditions \eqref{rvacp} and the commutation relations \eqref{jpmg},
we obtain the twisted KZ equations for the right movers
\begin{subequations}
\label{rkzct0}
\begin{equation}
\hat A_- (\s) \equiv \hat A_- (\T,\bz,\s)= \langle  \hgm(\T^{(1)},\bz_1,\s) \hgm(\T^{(2)},\bz_2,\s) \cdots
\hgm(\T^{(N)},\bz_N,\s)  \rangle_\s
\end{equation}
\begin{equation}
 \bar \part_\mu \hat A_- (\s) = {\hat{\bar W}}_\mu (\s)\hat A_-  (\s) \sp
\hat{\bar{W}}_\m (\s) \equiv \hat{\bar{W}}_\m (\T,\bz,\s)
\end{equation}
\begin{equation}
\label{rkzct}  {\hat{\bar W}}_{\mu}(\s)= 2 \lr^{n(r)aj;-n(r),bl} (\s)
 \left[ \sum_{\nu \neq \mu}\left( \frac{\bz_{\nu}}{\bz_{\mu}}
\right)^{ \srac{\mnb}{\r(\s)}} \frac{1}{\bz_{\mu \nu}}
\T_{-n(r),bl}^{(\mu)} \T_{n(r)aj}^{(\nu)}  - \srac{\mnb}{\r(\s)}
 \frac{1}{\bz_{\mu}}\T_{-n(r),bl}^{(\mu)} \T_{n(r)aj}^{(\mu)}
 \right]
\end{equation}
\begin{equation}
\bar z_{\mu \nu} \equiv \bar z_\mu - \bar z_\nu \sp
\s = 0, \ldots ,N_c -1  \ .
\end{equation}
\end{subequations}
By construction the twisted right mover connection
$\hat{\bar{W}} (\s)$ is also flat
\begin{equation}
\bar \partial_\mu {\hat{\bar W}}_{\nu}(\s)-
\bar \partial_\nu {\hat{\bar W}}_{\mu}(\s)
- [  {\hat{\bar W}}_{\mu}(\s),{\hat{\bar W}}_{\nu}(\s) ] = 0
\end{equation}
and it is easily  checked
as above that this connection is also abelian flat.
Note also the alternate forms of the right mover connection
\begin{subequations}
\begin{eqnarray}
 {\hat{\bar W}}_{\mu} (\s) & =  & \lr^{n(r)aj;-n(r),bl} (\s)
 \left[ \sum_{\nu \neq \mu}\left( \frac{\bz_{\nu}}{\bz_{\mu}}
\right)^{ \srac{\mnb }{\r(\s)}} \frac{2}{\bz_{\mu \nu}}
 \T_{-n(r),bl}^{(\mu)} \T_{n(r)aj}^{(\nu)}
- (1 - \delta_{\bar
n(r),0} )  \frac{1}{\bz_{\mu}}
\T_{-n(r),bl}^{(\mu)} \T_{n(r)aj}^{(\mu)} \right] \nn \\ \\
& = &  2 \lr^{n(r)aj;-n(r),bl} (\s)
 \left[ \sum_{\nu \neq \mu}\left( \frac{\bz_{\nu}}{\bz_{\mu}}
\right)^{ \srac{\bar n(r) }{\r(\s)}} \frac{1}{\bz_{\mu \nu}}
\T_{n(r)aj}^{(\mu)} \T_{-n(r),bl}^{(\nu)}   - \srac{\bar n(r) }{\r(\s)}
 \frac{1}{\bz_{\mu}}
\T_{n(r)aj}^{(\mu)}
\T_{-n(r),bl}^{(\mu)} \right] \nn \\ \label{rkzctm}  \\
& = &   \lr^{n(r)aj;-n(r),bl} (\s)
 \left[ \sum_{\nu \neq \mu}\left( \frac{\bz_{\nu}}{\bz_{\mu}}
\right)^{ \srac{\bar n(r) }{\r(\s)}} \frac{2}{\bz_{\mu \nu}}
\T_{n(r)aj}^{(\mu)} \T_{-n(r),bl}^{(\nu)}   -(1 - \delta_{\bar
n(r),0} )  \frac{1}{\bz_{\mu}} \T_{n(r)aj}^{(\mu)}
\T_{-n(r),bl}^{(\mu)} \right] \nn \\
\end{eqnarray}
\end{subequations}
where the first form follows from the vertex operator equation
\eqref{rdifreltM2p}, while the second form follows from \eqref{rkzct} by
relabelling $n\rightarrow -n$. The third form then follows from \eqref{idp3}. This
last form shows clearly that the twisted right mover connection differs from the
twisted left mover connection \eqref{altkzcon} only by $z_\m \rightarrow \bz_\m$
and reversal of the order of the twisted representation matrices $\T$.

Finally, the  Ward identities for the twisted right mover correlators
\begin{subequations}
\begin{equation}
\label{rgidt} \langle [ \hat{\bar{J}}_{0aj} (0),
\hgm(\T^{(1)},\bz_1,\s)  \cdots
\hgm(\T^{(N)},\bz_N,\s) ]
\rangle_\sigma = 0 \quad \Rightarrow \quad
\hQ_{aj} (\s) \hat A_- (\s) = 0
\end{equation}
\begin{equation}
\label{L0wi2r}
\sum_{\m=1}^N \left(  \bz_\m \bar \partial_\m
  + \Delta_{\sgbn}  (T^{(\m)} ) \right) \hat A_- (\s)   = 0
\end{equation}
\end{subequations}
are obtained from \eqref{liesa}, \eqref{rjgct} and \eqref{L0wi}. Moreover,
one finds again that the $\bar L_\s (0)$ Ward identity \eqref{L0wi2r} is
satisfied given the twisted right mover KZ equation \eqref{rkzct0}
and the global Ward identities in \eqref{rgidt}.

\section{Examples in the WZW Permutation Orbifolds \label{correl}}

\subsection{General One-Point Correlators \label{gopc} }

The twisted KZ equation for the left mover one-point correlators
\begin{subequations}
\begin{equation}
\part \langle \hgp (\T,z,\s) \rangle_\s = -
\frac{2}{z}
\langle \hgp (\T,z,\s) \rangle_\s
\lr^{n(r)aj;-n(r),bl} (\s) \srac{\bar n(r)}{\r(\s)} \T_{n(r)aj} \T_{-n(r),bl}
\end{equation}
\begin{equation}
\label{glwiop0}
\langle \hgp (\T,z,\s) \rangle_\s \T_{0 aj} = 0 \sp a = 1 \ldots {\rm dim}\,\gbn
\sp \forall \, j
\end{equation}
\end{subequations}
follows from Eq.~\eqref{tlmkzeq}, and we have included the global Ward identity
in \eqref{glwiop0}. The solution of this system is
\begin{subequations}
\label{opf1}
\begin{equation}
\langle \hgp (\T,z,\s) \rangle_\s  = C_+ (\T,\s) z^{-2
\lr^{n(r)aj;-n(r),bl} (\s)
\srac{\bar n(r)}{\r(\s)}
\T_{n(r)aj} \T_{-n(r),bl} } \hskip 3cm
\end{equation}
\begin{equation}
\label{for2}
=C_+ (\T,\s) z^{- {\cL}_{\sgb (\s)}^{n(r) aj ;-n(r), b l} (\s)
\T_{n(r) aj } \T_{-n(r), b l} ( 1- \de_{\bar n(r),0} ) }
\end{equation}
\begin{equation}
\label{for3}
= C_+ (\T,\s) z^{- \Delta_{\sgbn} (T) +{\cL}_{\sgb (\s)}^{0 aj;0 b l} (\s)
\T_{0 aj } \T_{0 b l}}
\end{equation}
\begin{equation}
\label{glwiop}
C_+ (\T,\s) \T_{0 aj} = 0 \sp a =1 \ldots {\rm dim}\,\gbn \;,\forall \;j
\end{equation}
\end{subequations}
where we have used \eqref{id1} to obtain \eqref{for2} and
\eqref{twcfm} to obtain \eqref{for3}.

On the other hand, the $L_\s(0)$ Ward identities \eqref{L0wi2} read in this case
\begin{equation}
\label{opf2}
\langle \hgp (\T,z,\s) \rangle_\s ( \overleftarrow{\partial_z}  z +
\Delta_{\sgbn} (T) ) =0
\quad \Rightarrow \quad \langle \hgp (\T,z,\s) \rangle_\s \propto
z^{-\Delta_{\sgbn} (T) }
\end{equation}
The solutions in \eqref{opf1} and \eqref{opf2} agree iff
\begin{equation}
\label{qstr}
C_+ (\T,\s)  {\cL}_{\sgb (\s)}^{0 aj ;0 b l} (\s)
\T_{0 aj } \T_{0 b l} = 0
\end{equation}
which also follows directly from Eq.~\eqref{glwiop}.
But in fact the quadratic structure in \eqref{qstr} is not in general zero
(see e.g. Subsec.~\ref{cyclfl})
\begin{equation}
{\cL}_{\sgb (\s)}^{0 aj ;0 b l} (\s)
\T_{0 aj } \T_{0 b l} \neq  0
\end{equation}
which implies that the twisted one-point correlators vanish
\begin{equation}
 C_+ (\T,\s) = 0 \quad \Rightarrow \quad\langle \hgp (\T,z,\s) \rangle_\s =0
 \sp \s = 0, \ldots ,N_c-1
\end{equation}
for all sectors of all WZW permutation orbifolds.

Another way to understand the vanishing of the one-point correlators is to
note that Eq.~\eqref{glwiop} has no non-trivial solution. To see this, one may
peel off unitary matrices from $\T$ in \eqref{repmatperm} to show that
\eqref{glwiop} is equivalent to the simpler problem for the untwisted
representation matrices $T$
\begin{equation}
c_+(T) ^\be (T_a)_\be{}^\a = 0 \sp  \, a = 1 \ldots {\rm dim}\,g
\end{equation}
which also has no non-trivial solution. This situation is familiar in the
case of the
{\it untwisted} one-point WZW correlators, where the global Ward identity
$c_+ T_a =0$, $\forall \,a$ has no non-trivial solution.

Similarly, one finds $\langle \hgm (\T,\bz,\s) \rangle_\s=0$ for the twisted
one-point right mover correlators.

\subsection{General Left Mover Two-Point Correlators \label{gtpc} }

In this subsection, we solve the twisted KZ equations \eqref{tlmkzeq}
for the left mover two-point correlators
\begin{equation}
\hat A_+ (1,2) \equiv \langle \hg (\T^{(1)},z_1,\s)  \hg (\T^{(2)},z_2,\s) \rangle_\s
\end{equation}
of the general WZW permutation orbifold. In this case the global
Ward identity reads
\begin{equation}
\hat A_+ (1,2) (\T_{0aj}^{(1)} + \T_{0aj}^{(2)} ) = 0 \sp
a =1 \ldots {\rm dim}\,\gbn \;,\forall \;j
\end{equation}
and this tells us that
\begin{equation}
\label{pergi}
{\cL}_{\sgb(\s)}^{0 aj; 0 bl} (\s) \T_{0 a j }^{(1)} \T_{0 b l }^{(1) } =
{\cL}_{\sgb(\s)}^{0 aj; 0 bl} (\s) \T_{0 a j }^{(2)} \T_{0 b l }^{(2) } \qquad
\mbox{when}\;\, \hat A_+ (1,2) \neq 0
\end{equation}
because of the symmetry \eqref{lgsym} of $\lr (\s)$.

For purposes of integration, the optimal form for the connections
of the two-point correlators is
\begin{subequations}
\label{tpcon}
\begin{eqnarray}
\hat W_1 \!\!\! \!\!\!\! & & \simeq  - \frac{2}{z_{12}}{\cL}_{\sgb(\s)}^{0 aj;0bl} (\s)
\T_{0 a j}^{(1)} \T_{0 b l }^{(1)} \nn \\
& & + {\cL}_{\sgb(\s)}^{n(r) aj; -n(r), bl} (\s)
\left[ \frac{2}{z_{12}} \left( \frac{z_2}{z_1}
\right)^{\srac{\bar n(r)}{\r(\s)} } \T_{n(r) a j}^{(2)}
\T_{-n(r), b l }^{(1)}
- \frac{1}{z_1} \T_{n(r) a j }^{(1)} \T_{-n(r), b l }^{(1)} \right]
(1 - \delta_{\bar n(r), 0} )  \hskip 1.3cm
\end{eqnarray}
\begin{eqnarray}
\hat W_2  \!\!\! \!\!\!\! & & \simeq - \frac{2}{z_{21}}{\cL}_{\sgb(\s)}^{0 aj;0bl} (\s)
\T_{0 a j}^{(1)} \T_{0 b l }^{(1)}  \nn \\
& & + {\cL}_{\sgb(\s)}^{n(r) aj; -n(r),bl} (\s)
\left[ \frac{2}{z_{21}} \left( \frac{z_1}{z_2}
\right)^{1-\srac{\bar n(r)}{\r(\s)} } \T_{n(r) a j}^{(2)}
\T_{-n(r), b l }^{(1)}
- \frac{1}{z_2} \T_{n(r) a j }^{(2)} \T_{-n(r), b l }^{(2)} \right]
(1 - \delta_{\bar n(r), 0} ) \hskip 1.3cm
\end{eqnarray}
\end{subequations}
where we have used \eqref{altkzcon}, \eqref{lgsym} and the global Ward identity
\eqref{gidt}, including the form
\eqref{pergi}. We have also used $n(r) \rightarrow -n(r)$ and \eqref{lgsym}
in the second term of $\hat W_2$. Using Eqs.~\eqref{permg}, \eqref{perml} and
\eqref{idjperm}, all the matrix
structures in \eqref{tpcon} are seen to commute.

This gives the solution for the general two-point correlators
\begin{subequations}
\label{tpf0}
\begin{eqnarray}
\hat A_+ (1,2) \!\!\!\!\!& =& \!\!\!\!\! C_+(\T,\s)
z_1^{- {\cL}_{\sgb(\s)}^{n(r) aj; -n(r),bl} (\s)
\T_{n(r) a j }^{(1)} \T_{-n(r), b l }^{(1)}(1 - \delta_{\bar n(r), 0}
)}z_2^{- {\cL}_{\sgb(\s)}^{n(r) aj; -n(r), bl} (\s)
\T_{n(r) a j }^{(2)} \T_{-n(r), b l }^{(2)}(1 - \delta_{\bar n(r), 0}
)} \nn \\
& & \times z_{12}^{ - 2 {\cL}_{\sgb(\s)}^{0 aj; 0 bl} (\s)
\T_{0 a j }^{(1)} \T_{0 b l }^{(1) }}
\exp \left( 2 {\cL}_{\sgb(\s)}^{n(r) aj; -n(r), bl} (\s)
I_{\srac{\bar n(r)}{\r(\s)}}  ( \srac{z_1}{z_2})
 \T_{n(r) a j }^{(2)} \T_{-n(r), b l }^{(1)}
(1 - \delta_{\bar n(r), 0} ) \right) \nn  \\ & &
\end{eqnarray}
\begin{eqnarray}
\label{tpf}
& = & C_+(\T,\s) z_1^{- \Delta_{\sgbn}( T^{(1)})}  z_2^{- \Delta_{\sgbn}( T^{(2)})}
\left( \frac{z_1 z_2}{z_{12}^2} \right)^{ {\cL}_{\sgb(\s)}^{0 aj; 0 bl} (\s)
\T_{0 a j }^{(1)} \T_{0 b l }^{(1) }}
\nn \\
& & \times \exp \left( 2 {\cL}_{\sgb(\s)}^{n(r) aj; -n(r), bl} (\s)
I_{\srac{\bar n(r)}{\r(\s)}}  ( \srac{z_1}{z_2})
 \T_{n(r) a j }^{(2)} \T_{-n(r), b l }^{(1)}
(1 - \delta_{\bar n(r), 0} ) \right)
\end{eqnarray}
\begin{equation}
C_+ (\T,\s)( \T_{0 a j }^{(1)} + \T_{0 a j }^{(2)} ) = 0
\sp \s = 0, \ldots ,N_c -1 \ .
\end{equation}
\end{subequations}
Here we have used \eqref{pergi} and \eqref{twcfm} to obtain the form in
\eqref{tpf}, and the integrals
\begin{equation}
I_{\srac{\bar n(r)}{\rho (\s)}}  (y) \equiv \int^y \frac{d x}{x-1}
x^{-\srac{\bar n(r)}{\rho (\s) }}
\end{equation}
are evaluated in App.~\ref{intapp}.

As a check, we note that the result \eqref{tpf0} satisfies the $L_\s (0)$ Ward
identity
\begin{equation}
\hat A_+ (1,2) \Big( \overleftarrow{\partial_{z_1}}z_1 +\overleftarrow{\partial_{z_2}} z_2
+ \Delta_{\sgbn} (T^{(1)}) +  \Delta_{\sgbn} (T^{(2)}) \Big)=0
\end{equation}
because $I_{\srac{\bar n(r)}{\r(\s)}}  ( \srac{z_1}{z_2})$ is scale invariant.

\subsection{More About the Cyclic Permutation Orbifolds \label{cyclfl} }

The twisted KZ systems of Subsecs.~\ref{twlmKZ} and \ref{twrmKZ} can be worked out more
explicitly by using the universal forms of $\T = T t$ and $\lr$ in
Eqs.~\eqref{Tcycl0} and \eqref{lgs00} respectively.
In this subsection we shall do so for the case of the general WZW cyclic
permutation orbifold $A_g(\z_\l)/\z_\l$, whose data is also given in
Subsec.~\ref{cyclorb}. We need in particular the twisted representation
matrices \eqref{st-in-AZl} in the factorized form
\begin{subequations}
\label{tcprop}
\begin{equation}
\T_{aj}^{(r)} (T,\s) \equiv \T_{aj}^{(r)} = T_a \otimes t_{rj} \equiv  T_a t_{r j}
\sp [T_a , T_b ] = i f_{ab}{}^c T_c
\end{equation}
\begin{equation}
(t_{rj})_{sl}{}^{tm } = \d_{jl}\d_l^m \d_{r+s-t, 0 \,\rmod \rho (\s)}
\end{equation}
\begin{equation}
t_{rj } t_{sl} = \delta_{jl} t_{r+s,j} \sp t_{r \pm \rho (\s),j} = t_{rj}
\end{equation}
\begin{equation}
 a=1 \ldots \text{dim }\gfrak,\quad \bar{r}=0,\ldots,\r(\s)-1,
 \quad j=0,\ldots,\srac{\l}{\r(\s)}-1,\quad \s=0,\ldots,\l-1
\end{equation}
\end{subequations}
which is a special case of Eq.~\eqref{Tcycl0}.

Using \eqref{Lcycl},   \eqref{id1} and \eqref{tcprop},
we may evaluate the quadratic forms
\begin{subequations}
\label{quafor}
\begin{equation}
\lr^{raj;-r,b l} (\s) \T_{aj}^{(r)} \T_{bl}^{(-r)} = \Delta_{\sgbn} (T) \one
\sp (\one)_{s \a l}{}^{t \beta m} \equiv \delta_{s-t,0 \rmod \r(\s)}
\delta_\a^\be \delta_l^m
\end{equation}
\begin{equation}
\label{qstr2}
\lr^{0 aj ;0 bl} (\s) \T_{aj}^{(0)} \T_{b l}^{(0)} =
\frac{1}{\rho (\s)} \Delta_\sgbn (T) \one \neq 0
\end{equation}
\begin{equation}
\label{confwt}
2 \lr^{raj;-r,b l} (\s) \frac{\bar r}{\rho (\s)} \T_{aj}^{(r)}
\T_{bl}^{(-r)}=
 \lr^{raj;-r,b l}  (\s) \T_{aj}^{(r)}
\T_{bl}^{(-r)} ( 1 - \de_{\bar n(r),0} ) =
\Delta_{\sgbn}(T ) \frac{\rho (\s) -1}{\rho (\s)} \one
\end{equation}
\begin{equation}
\Delta_{\sgbn} (T) \de_\a^\be =\frac{ (\eta^{ab}T_aT_b)_{\a}{}^{\be} }{2k+Q_{\sgbn} }
 \sp \a,\be = 1 \ldots {\rm dim}\,T
\end{equation}
\end{subequations}
where $\Delta_{\sgbn} (T)$ is the conformal weight of each irrep $T$ of each
simple $\gbn$. To evaluate \eqref{confwt}, we used the summation identities
\begin{equation}
2\sum_{r=0}^{\rho (\s)-1} \frac{\bar r}{\rho (\s)} =
\sum_{r=0}^{\rho (\s)-1} (1 - \delta_{r,0})  =   \rho (\s) -1 \ .
\end{equation}
Following the more general discussion of Subsec.~\ref{gopc}, we verify
that the one-point correlators of the WZW cyclic permutation orbifolds are zero
\begin{equation}
\langle \hg (\T,z,\s) \rangle_\s = 0 \sp \sp \s = 0, \ldots , \l -1
\end{equation}
because the quadratic structure \eqref{qstr2} is non-zero.

Using \eqref{tlmkzeq}, \eqref{tcprop} and \eqref{quafor}, a more explicit form of the twisted left mover KZ
equations for  $A_g(\z_\l)/\z_\l$ is obtained
\begin{subequations}
\begin{equation}
\partial_\m \hat A_+ (\s)= \hat A_+  (\s) \hat W_\m (\s) \sp
\hat A_+ (\s) \equiv \hat A_+(\T,z,\s) \sp \s = 0, \ldots, \l -1
\end{equation}
\begin{equation}
\label{azlprop}
\hat W_{\mu} (\s)  = \frac{2}{\rho (\s)} \frac{1}{2k + Q_{\sgbn}}
\sum_{\nu \neq \mu} \eta^{ab} \frac{T_a^{(\nu)} T_b^{(\mu)}}{z_{\mu
\nu}} \sum_{r=0}^{\rho (\s) -1} \sum_{j=0}^{\srac{\lambda}{\rho(\s)} -1}
\left( \frac{z_\nu}{z_\mu} \right)^{\srac{r}{\rho (\s )}}
t_{rj }^{(\nu)} t_{-rj}^{(\mu)} - \frac{\Delta_{\sgbn}
(T^{(\mu)})}{z_{\mu}} \frac{\rho (\s) -1}{\rho (\s)}
\end{equation}
\begin{equation}
\label{gwoc2}
\hat A_+ (\s) \hQ_{aj} (\s) = 0 \sp
 a = 1 \ldots {\rm dim}\,\gbn \sp j = 0, \ldots ,\srac{\lambda}{\rho (\s)}-1
\end{equation}
\begin{equation}
\hQ_{aj} (\s) = \sum_{\mu =1}^N T_a^{(\mu)} t_{0j}^{(\mu) }
\end{equation}
\begin{equation}
[ \hQ_{aj} (\s), \hQ_{bl} (\s) ] = i \de_{jl} f_{ab}{}^c \hQ_{cj} (\s)
\sp [ \hat W_\mu (\s), \hQ_{aj} (\s) ] = 0
\end{equation}
\end{subequations}
where we have included the global Ward identities of sector $\s$ in \eqref{gwoc2}.

In this case, we verify the abelian flatness \eqref{abfl2}
of $\hat W_\mu (\s)$ as follows. By direct computation using \eqref{azlprop} and
\begin{equation}
[T_a^{(\m)},T_b^{(\n)}] = [ t_{rj}^{(\m)},t_{sj}^{(\n)} ] = 0 \quad
{\rm when} \;\, \n \neq \m
\end{equation}
we find the proportionality
$$
[ \hat W_\m (\s) , \hat W_\n (\s) ]  \sim
\sum_{\epsilon \neq \m , \n} \sum_{r,s=0}^{\rho (\s) -1}
\sum_{j=0}^{\srac{\l}{\rho (\s)} -1}
\left\{ t_{-r,j}^{(\m)} t_{r-s,j}^{(\n)} t_{sj}^{(\epsilon)}
\left( \frac{z_\n}{z_\m} \right)^{\srac{r}{\rho (\s) }}
\left( \frac{z_\epsilon}{z_\n} \right)^{\srac{s}{\rho (\s) }}
\frac{1}{z_{\m \n} z_{\n \epsilon}} \hskip 1cm
\right.
$$
\begin{equation}
\label{wwcom}
 +t_{s-r,j}^{(\m)} t_{-s,j}^{(\n)}  t_{rj}^{(\epsilon)}
\left( \frac{z_\epsilon}{z_\m} \right)^{\srac{r}{\rho (\s) }}
\left( \frac{z_\m}{z_\n} \right)^{\srac{s}{\rho (\s) }}
\frac{1}{z_{\m \epsilon} z_{\n \m }}
\end{equation}
$$
\hskip 4cm \left.
- t_{-r,j}^{(\m)} t_{-s,j}^{(\n)}  t_{r+s,j}^{(\epsilon)}
\left( \frac{z_\epsilon}{z_\m} \right)^{\srac{r}{\rho (\s) }}
\left( \frac{z_\epsilon}{z_\n} \right)^{\srac{s}{\rho (\s) }}
\frac{1}{z_{\m \epsilon} z_{\n \epsilon }} \right\}
i f^{abc} T_a^{(\m)} T_b^{(\n)} T_c^{(\epsilon)} \ .
$$
To check that this commutator vanishes
\begin{equation}
[ \hat W_\m (\s) , \hat W_\n (\s) ] = 0
\end{equation}
one needs only the identities
\begin{equation}
\label{tperiod}
t_{r \pm \rho (\s),j }^{(\m)} = t^{(\m)}_{r j} \sp
\frac{1}{z_{\m \epsilon} z_{\n \epsilon}}  = \frac{1}{ z_{\m \n} z_{\n \epsilon}}
+\frac{1}{ z_{\m \epsilon} z_{\n \m}}
\end{equation}
and then one finds for each $j$ that each distinct configuration
$t^{(\m)}_{pj} t^{(\n)}_{qj} t^{(\epsilon)}_{rj}$ in \eqref{wwcom}
 is multiplied by a sum of three terms which cancel.
To see this analytically, the reader may find helpful the following intermediate
relations
\begin{subequations}
\begin{equation}
\label{suid1}
\sum_{s=0}^{\l -1} t_{r-s,j}^{(\n)} t_{sj}^{(\epsilon)} y^{s/\l}
=\sum_{s=0}^{\l -1} t_{-s,j}^{(\n)} t_{r+s,j}^{(\epsilon)} y^{(r+s)/\l}
+ \frac{1-y}{y} \sum_{s=\l -r}^{\l -1} t_{-s,j}^{(\n)} t_{r+s,j}^{(\epsilon)}
y^{(r+s)/\l} \sp \forall \, y
\end{equation}
\begin{equation}
\label{suid2}
\sum_{r=0}^{\l -1} t_{s-r,j}^{(\m)} t_{rj}^{(\epsilon)} x^{r/\l}
=\sum_{s=0}^{\l -1} t_{-r,j}^{(\m)} t_{r+s,j}^{(\epsilon)} x^{(r+s)/\l}
+ \frac{1-x}{x} \sum_{r=\l -s}^{\l -1} t_{-r,j}^{(\m)} t_{r+s,j}^{(\epsilon )}
x^{(r+s)/\l} \sp \forall \, x
\end{equation}
\begin{equation}
\sum_{s=0}^{\l-1} \sum_{r = \l -s} ^{\l-1} f_{rsj} =
\sum_{r=0}^{\l-1} \sum_{s = \l -r} ^{\l-1} f_{rsj} \sp \forall  \, f
\end{equation}
\end{subequations}
where \eqref{suid1} and \eqref{suid2} are consequences of the $t$ periodicity in
\eqref{tperiod}.

We continue with  $A_g(\z_\l)/\z_\l$ in the simple case $\l$=prime,
where we have for the twisted sectors $\s \neq 0$
\begin{subequations}
\begin{equation}
\rho (\s) = \lambda \sp \bar r, \bar s, \bar t = 0, \ldots ,\lambda -1 \sp
j,l , m= 0 \sp t_{rj} \rightarrow t_r \sp \s = 1, \ldots  ,\lambda -1
\end{equation}
\begin{equation}
\T_a^{(r)} = T_a t_r \sp t_r t_s  = t_{r+s} \sp t_0 = 1 \sp
(t_r)_s{}^t = \d_{r +s -t,\, \rmod \lambda} \ .
\end{equation}
\end{subequations}
In this case, we consider the twisted two-point
left mover correlators and their global Ward identities
\begin{subequations}
\begin{equation}
\hat A_+ (1,2)\equiv \langle \hgp (\T^{(1)} ,z_1,\s) \hgp (\T^{(2)},z_2,\s) \rangle_\s
\end{equation}
\begin{equation}
\hat A_+ (1,2)(T_a^{(1)} + T_a^{(2)} ) = 0 \sp a = 1 \ldots {\rm dim}\,\gbn
\end{equation}
\end{subequations}
which are a special case of the general discussion in Subsec.~\ref{gtpc}.
The global identities imply that $\hat A_+(1,2)=0$ unless
\begin{equation}
\Delta_{\sgbn} (T^{(1)} ) = \Delta_{\sgbn} (T^{(2)} )
\end{equation}
as in the untwisted case.

Then using the global Ward identities again, the twisted KZ connections
of the two-point correlators may be written
\begin{subequations}
\begin{equation}
\hat W_1 \simeq - \Delta_{\sgbn} (T^{(1)}) \left[
\frac{2}{\lambda z_{12}} \left( 1 + \sum_{r=1}^{\lambda-1}
\left( \frac{z_2}{z_1} \right)^{\srac{r}{\lambda}} t_r^{(2)}
 t_{-r}^{(1)}) \right) + \frac{\lambda -1}{\lambda z_1} \right]
 \end{equation}
\begin{equation}
\hat W_2 \simeq - \Delta_{\sgbn} (T^{(1)}) \left[
\frac{2}{\lambda z_{21}} \left( 1 + \sum_{r=1}^{\lambda-1}
\left( \frac{z_1}{z_2} \right)^{1-\srac{r}{\lambda}} t_r^{(2)}
 t_{-r}^{(1)}) \right) + \frac{\lambda -1}{\lambda z_2} \right]
 \end{equation}
\begin{equation}
[ t_r^{(2)}t_{-r}^{(1)}, t_s^{(2)}t_{-s}^{(1)}] = 0 \sp \forall \; r,s \ .
\end{equation}
 \end{subequations}
Integrating, we obtain the twisted two-point correlators%
\footnote{With \eqref{datacycl}, \eqref{qstr2} and the global Ward identity,
the result \eqref{fropc} can also be obtained as a special case of the more
 general two-point correlator in \eqref{tpf0}.}  of sector $\s$
\begin{subequations}
\label{fropc}
\begin{equation}
\label{twopp}
\hat A_+ (1,2) = C_+ (\T) \left( (z_{12})^{\srac{2}{\lambda}} (z_1
z_2)^{\srac{\lambda-1}{\lambda}} \prod_{r=1}^{\lambda -1}
e^{ \srac{2}{\lambda} I_{r/\lambda} (\srac{z_1}{z_2}) t_r^{(2)} t_{-r}^{(1)} }
\right)^{-\Delta_{\sgbn} (T^{(1)})}
\end{equation}
\begin{equation}
 \s = 1 ,  \ldots ,  \l -1 \sp \l = \mbox{prime}
 \end{equation}
\begin{equation}
I_{r/\l}(y)
 = \int^y \frac{d x}{x-1} x^{-r/\l} \sp
C_+ (\T) (T_a^{(1)} + T_a^{(2)} ) = 0
\end{equation}
\begin{equation}
\label{gwiip}
\hat A_+ (1,2)\Big( \overleftarrow{\partial_1} z_1 + \overleftarrow{\partial_2}
z_2 + 2 \Delta_{\sgbn} (T^{(1)}) \Big) = 0
\end{equation}
\end{subequations}
where the integrals $I_{r/\lambda}$ are evaluated in App.~\ref{intapp}.
Because $I_{r/\l}(z_1/z_2)$ is scale invariant, this result also
satisfies the $L_\s(0)$ Ward identity in \eqref{gwiip}.
Note that the result \eqref{fropc} is independent of the sector $\s$. This
can be traced back to the fact that, for prime $\l$, the twisted current
algebra \eqref{twisted{J,J}Z-lambda} of each twisted sector reduces to a left and right mover
copy of the same simple \cite{Borisov:1997nc} orbifold affine algebra at level
$\hat k = \l k$.

For $H= \z_2$, the result \eqref{fropc} reduces to
\begin{subequations}
\label{z2case}
\begin{equation}
\hat A_+(1,2) = C_+(\T) \left( \frac{1}{z_{12}} \frac{1}{\sqrt{z_1 z_2}}
\left( \frac{ \sqrt{\frac{z_1}{z_2}} + 1 }{\sqrt{\frac{z_1}{z_2}}
- 1} \right)^{t_1^{(2)} t_1^{(1)} } \right)^{\Delta_{\sgbn} (T^{(1)})}
\end{equation}
\begin{equation}
\label{pauli}
t_0^{(\mu)} = 1 \sp
t_1^{(\mu)} = \tau_1^{(\mu)} \sp \mu = 1,2 \sp
t_1^{(2)} t_1^{(1)} \in \{ 0 \; {\rm or} \; 1 \}
\end{equation}
\begin{equation}
C_+ (\T)  (T_a^{(1)} + T_a^{(2)} ) = 0
\end{equation}
\end{subequations}
for the single twisted sector $\sigma=1$. In this result, $\tau_1$ is the
first Pauli matrix.

For the case $\l$=prime and $\s \neq  0$, we have also worked out the right
 mover two-point correlators and the full nonchiral product
\begin{subequations}
\begin{equation}
\hat A(1,2) = \hat A_- (1,2) \hat A_+ (1,2)
= C_- C_+
\left( |z_{12}|^{\srac{4}{\lambda}} ( |z_1| |z_2|)^{\srac{2(\lambda-1)}{\lambda}}
\right)^{-\Delta_{\sgbn} (T^{(1)})}
e^{ -\srac{2}{\l} \Delta_{\sgbn} (T^{(1)}) F(1,2) }
\end{equation}
\begin{equation}
F(1,2) \equiv  \sum_{r=1}^{\l -1} [
 I_{\srac{r}{\lambda}} (\bar{y},y_0) t_r^{(1)} t_{-r}^{(2)}
 + I_{\srac{r}{\lambda}} (y,y_0) t_r^{(2)} t_{-r}^{(1)}] \sp y = \srac{z_1}{z_2}
\end{equation}
\begin{equation}
\label{Isym}
I_{\srac{\nb}{\rho (\s)}} (y,y_0) =
\int_{y_0}^{y} \frac{d x }{x-1} \ x^{-\srac{\nb  }{\rho (\s)}} =
 I_{\srac{\rho(\s) - \nb  }{\rho (\s)}}(y^{-1},y_0^{-1})
\end{equation}
\begin{equation}
C_+   (T_a^{(1)} + T_a^{(2)} ) =
(T_a^{(1)} + T_a^{(2)} ) C_-  = 0
\end{equation}
\end{subequations}
where $y_0$ is a fixed reference point for the integrals $I_{n/\rho}$
(see also App.~\ref{intapp}).
As expected, these correlators are $1 \leftrightarrow 2$ symmetric
$\hat A(2,1) = \hat A(1,2)$. To see this,
we note that $\Delta_{\sgbn} (T^{(1)})=\Delta_{\sgbn} (T^{(2)})$ and
\begin{subequations}
\begin{eqnarray}
F(2,1) & = & \sum_{r=1}^{\l -1} [
 I_{\srac{r}{\lambda}} (\bar{y}^{-1},y_0) t_r^{(2)} t_{-r}^{(1)}
 + I_{\srac{r}{\lambda}} (y^{-1},y_0) t_r^{(1)} t_{-r}^{(2)}]  \\
 & = &
\sum_{r=1}^{\l -1} [
 I_{\srac{\l-r}{\lambda}} (\bar{y},y_0^{-1}) t_{\l-r}^{(1)}t_{r-\l}^{(2)}
 + I_{\srac{\l-r}{\lambda}} (y,y_0^{-1}) t_{\l-r}^{(2)} t_{r-\l}^{(1)}]
 \label{lrcorb} \\
& = &
\sum_{r=1}^{\l -1} [
 I_{\srac{r}{\lambda}} (\bar{y},y_0^{-1}) t_{r}^{(1)}t_{-r}^{(2)}
 + I_{\srac{r}{\lambda}} (y,y_0^{-1}) t_{r}^{(2)} t_{-r}^{(1)}]
 \label{lrcorbc}  \\
& = &
F(1,2) + \Delta \\
\Delta & =  & \sum_{r=1}^{\l -1}
 I_{\srac{r}{\lambda}} (y_0,y_0^{-1})[ t_{r}^{(1)}t_{-r}^{(2)} +
 t_{r}^{(2)} t_{-r}^{(1)}] = 0 \ .
 \end{eqnarray}
\end{subequations}
Here we have used the symmetry relation \eqref{Isym} and $t$ periodicity to obtain
\eqref{lrcorb}, and a relabelling to obtain \eqref{lrcorbc}.
The same relations are used in the rearrangement
\begin{equation}
\sum_{r=1}^{\l -1}
 I_{\srac{r}{\lambda}} (y_0,y_0^{-1}) t_{r}^{(1)}t_{-r}^{(2)}
 =\sum_{r=1}^{\l -1}
 I_{\srac{r}{\lambda}} (y_0^{-1},y_0) t_{r}^{(2)}t_{-r}^{(1)}
= -\sum_{r=1}^{\l -1}
 I_{\srac{r}{\lambda}} (y_0,y_0^{-1}) t_{r}^{(2)}t_{-r}^{(1)}
\end{equation}
which shows that $\Delta =0$.

\section{The Inner-Automorphic WZW orbifolds \label{innerorb} }

As our final large example,
we work out the inner-automorphic WZW orbifolds \cite{Halpern:2000vj}
\begin{equation}
\frac{A_g(H(d))}{H(d)} \sp H(d) \subset {\rm Lie}\, G \subset {\rm Aut}(g)
\end{equation}
in further detail, where Lie $G$ is the Lie group whose Lie algebra is
compact simple Lie $g$. For these orbifolds, we concentrate
on the rectification problem and the twisted KZ systems which describe
the twisted left and right mover correlators.

\subsection{Basic Properties and Rectification \label{innerprop} }

Let $h_\s \in H(d) \subset {\rm Lie}\,G \subset {\rm Aut}(g)$ be an element of any group $H(d)$ of inner
automorphisms of simple $g$. Following Ref.~\cite{Halpern:2000vj},
we know the matrix action $w(h_\s)$
\begin{subequations}
\begin{equation}
\label{w(hs)of-inner-auto}
 w(h_\s)_\a{}^\be = \d_\a{}^\be e^{2\pi i\s \a\cdot d},
 \quad w(h_\s)_A{}^B =\d_A{}^B, \quad w(h_\s)_\a{}^A=w(h_\s)_A{}^\a =0
\end{equation}
\begin{equation}
 \a \in \Delta(g),\quad A=1...\text{rank }g, \quad \s = 0,\ldots,\r(1)-1,
 \quad N_c = \r(1)
\end{equation}
\end{subequations}
of $h_\s$ on the untwisted currents in the Cartan-Weyl basis. This is the action
of $h_\s$ in the adjoint $(T_a^{adj})_b{}^{c}= -if_{ab}{}^{c}$.
Refs.~\cite{Freericks:1988zg} and \cite{Halpern:2000vj} discuss the conditions on
the vector $d$ such that $H(d)$ is a group of finite order.

Then the matrix action $W(h_\s;T)$ of $h_\s$ in representation $T$ must solve
the linkage relation \eqref{eq:WTWwTa} in the form
\begin{equation}
 W\hc(h_\s;T) T_\a W(h_\s;T) = e^{2\pi i\s \a\cdot d}T_\a ,\quad W\hc(h_\s;T)
  T_A W(h_\s;T) = T_A \ . \label{WTW=eTa-of-inner-auto}
\end{equation}
To solve these equations, it is convenient to introduce the weight basis of
matrix representation $T$
\begin{subequations}
\label{algTcw}
\begin{equation}
 (T_A)_\l{}^{\l'} \equiv \;\langle T,\l | H_A(0) | T,\l' \rangle = \l_A\d_{\l,\l'} ,\quad
    (T_\a)_\l{}^{\l'} \equiv \;\langle T,\l | E_\a(0) | T,\l' \rangle
    \propto\; \d_{\l,\l'+\a} \label{rep-matrices-of-inner-auto}
\end{equation}
\begin{equation}
\label{algTcw1}
[ T_A, T_B] = 0 \sp [T_A,T_\alpha] = \alpha_A T_\alpha \sp
\end{equation}
\begin{equation}
\label{algTcw2}
[ T_\a , T_{\be} ] =
\left\{\begin{array}{lll}
N_\gamma(\alpha,\beta) T_\gamma  && \textrm{if}\  \alpha+\beta=\gamma\\
\alpha\cdot T  && \textrm{if}\ \alpha+\beta=0\\
0 && \textrm{otherwise}\\
\end{array} \right.
\end{equation}
\begin{equation}
\label{algTcw3}
N_\gamma (\a,\be) = - N_\gamma(\be,\a)
= N_{-\be }(-\gamma,\a) = - N_{-\gamma }(-\a,-\be)
\end{equation}
\end{subequations}
where $\{ \l \}$ are the weights of the representation. The solution of the
linkage relation
\begin{equation}
\label{linkrel}
 W(h_\s;T)_\l{}^{\l'} = \d_\l{}^{\l'} e^{-2\pi i\s \l\cdot d} ,
 \quad W\hc(h_\s;T)_\l{}^{\l'} = \d_\l{}^{\l'} e^{+2\pi i\s \l\cdot d}
\end{equation}
is obtained immediately in this basis. Note that the adjoint matrices of the
weight basis are similarity transformations of
$(T_a^{\rm adj})_b{}^{c}= -if_{ab}{}^{c}$, so that $W$ does not reduce to $w$
in this case.

We discuss the orders $\r(\s)$ and $R(T,\s)$ of $w(h_\s)$ and $W(h_\s;T)$,
which are the smallest integers satisfying
\begin{equation}
 e^{2\pi i \s \a \cdot d \r(\s)} =1, \quad e^{-2\pi i \s \lambda \cdot d R(T,\s)}
  =1 \ .
\end{equation}
These conditions imply that
\begin{equation}
  \left\{\begin{array}{lll}
        & R(T,\s)=\r(\s) &\text{when $T$ is in the conjugacy class of } \\
        &   & \text{ the root lattice} \\
        & e^{-2\pi i \s \l_{\rm min}\cdot d R(T,\s)} =1 &\text{when $T$ is in the
              conjugacy class of} \\
        &   &\text{  $\l_{\rm min}$ plus the root lattice} \\
\end{array} \right.
\end{equation}
where $\{\l_{\rm min}\}$ are the minimal weights of simple $g$, given in
Ref.~\cite{Humphreys:1972}. As the simplest example, we find
\vspace{-.03in}
\begin{equation}
\quad R(T,\s)  =\left\{\begin{array}{lll}
        & \r(\s) &\text{for integral spin }j \\
        & 2\r(\s)  &\text{for half-integral spin }j \\
\end{array} \right.
\end{equation}
for $g=SU(2)$. In what follows, we revert to our abbreviation
$R(\s) \equiv R(T,\s)$.

In \eqref{amno} and \eqref{linkrel}, the actions $w(h_\s)$ and $W(h_\s;T)$ are
diagonal, so the solutions of the $H$-eigenvalue problem and the extended
$H$-eigenvalue problem  can be taken as:
\begin{subequations}
\label{more-on-inner-auto}
\begin{equation}
 U\hc(h_\s) =\schisig=1, \quad U\hc(T,h_\s) = \one ,\quad \sG (\s)=G, \quad
 \scf (\s) = f, \quad \st (T,\s) = T
\end{equation}
\begin{equation}
\label{innerdata}
 n_A=0, \quad \srac{n_\a(r)}{\r(\s)} = -\s \a\cdot d  ,\quad \srac{N(r)}{R(\s)}
 = \s \l\cdot d \sp
 \sm(\st,\s) = M(k,T) = \frac{k}{y(T)} \one \ .
\end{equation}
\end{subequations}
This tells us that all the selection rules discussed above are satisfied
naturally by the untwisted objects in the Cartan-Weyl basis and weight basis.
 For example, the selection rule \eqref{sel-rule-for-st-1} for $\st$ takes the
 form
\begin{equation}
 e^{-2\pi i \s (\l-\l')\cdot d} (T_A)_{\l}{}^{\l'} = (T_A)_{\l}{}^{\l'} ,
 \quad  e^{-2\pi i \s (\l-\l')\cdot d} (T_\a)_{\l}{}^{\l'} =
 e^{-2\pi i \s \a\cdot d} (T_\a)_{\l}{}^{\l'} \label{eT=T}
\end{equation}
which is easily checked from \eqref{rep-matrices-of-inner-auto}.

The inner-automorphic data in \eqref{more-on-inner-auto} gives the
 monodromies and mode expansions of the twisted currents of $A_g(H(d))/H(d)$
\begin{subequations}
\begin{equation}
\hat{H}_A(z e^{2 \pi i}) = \hat{H}_A (z)\sp
\hat{E}_\a(z e^{2 \pi i})=e^{2 \pi i \sigma \a \cdot d} \hat{E}_\a(z)
\label{innermonodromies}
\end{equation}
\begin{equation}
\hat{\bar{H}}_A(\bz e^{-2 \pi i}) = \hat{\bar{H}}_A (\bz)\sp
\hat{\bar{E}}_\a(\bz e^{-2 \pi i})=e^{2 \pi i \sigma \a \cdot d}
\hat{\bar{E}}_\a(\bz)
\label{innermonodromiesr}
\end{equation}
\begin{equation}
\hat{H}_A(z)=\sum_{m \in \sz} \hat{H}_A(m) z^{-m-1} \sp
\hat{E}_\a (z)=\sum_{m \in \sz} \hat{E}_\a (m-\sigma \alpha \cdot d)
z^{-(m - \sigma \alpha \cdot d)-1} \label{innermodeexpansion}
\end{equation}
\begin{equation}
\hat{\bar{H}}_A(\bz)=\sum_{m \in \sz} \hat{\bar{H}}_A(m) \bz^{m-1} \sp
\hat{\bar{E}}_\a (\bz)=\sum_{m \in \sz} \hat{\bar{E}}_\a (m-\sigma \alpha \cdot d)
\bz^{(m - \sigma \alpha \cdot d)-1}  \label{innermodeexpansionr}
\end{equation}
\end{subequations}
as a special case of \eqref{jmod}. Here we have followed the notation of
Ref.~\cite{Halpern:2000vj} for the labelling of the twisted current modes.
Then one finds that  the twisted left mover current algebra is the standard
\cite{Kac:1984mq,Goddard:1986bp,Freericks:1988zg,Lerche:1989uy,Halpern:2000vj}
 inner-automorphically twisted affine Lie algebra
 \begin{subequations}
 \begin{equation}
 \gfrakh(h_\s) = \gfrakh( H(d) \subset {\rm Lie} \; G
 \subset {\rm Aut}(g); \s ), \quad \s=0,...,\r(1)-1
\end{equation}
\begin{equation}
 [\hat{H}_{A}(m), \hat{H}_{B}(n)] = km \delta_{AB} \delta_{m+n,0} 
\end{equation}
\begin{equation}
 [\hat{H}_{A}(m), \hat{E}_{\a }(n-\s \a \cdot d)]  =\a_A
 \hat{E}_{\a}(m+n-\s \a \cdot d) 
\end{equation}
\begin{equation}
 [\hat{E}_{\alpha}(m-\s \a \cdot d), \hat{E}_{\beta}(n-\s \beta \cdot d) ]
 =\left\{\begin{array}{lll}
N_\gamma(\alpha,\beta) \hat{E}_{\gamma}(m+n-\s \gamma \cdot d)  && \textrm{if}\  \alpha+\beta=\gamma\\
\alpha\cdot \hat{H} (m+n)+k(m-\s \a \cdot d)\delta_{m+n,0} && \textrm{if}\ \alpha+\beta=0\\
0 && \textrm{otherwise}\\
\end{array} \right.
\end{equation}
\begin{equation}
 A=1 \ldots {\textrm{rank}}\,g,
\quad \a \in\Delta(g),\quad \s=0,...,\r(1)-1
\end{equation}
\end{subequations}
 and the commuting twisted
right mover algebra $\gfrakh(h_\s^{-1})$ has the same
form except for the sign change of the central terms.

In this case also we have been able to solve the rectification problem for the
twisted right movers. Using \eqref{algTcw3}, one finds that the simplest
rectification%
\footnote{The rectification in \eqref{hh,he-rightmover} corresponds to the choice
$\theta (\s) =-1$ for all $\s$ in Eq.~\eqref{rmcot}, and $\theta (0) = -1$ differs by a
Chevalley involution from the trivial rectification $\theta (0) =1$ discussed
in Subsec.~\ref{quarg} for the untwisted sector. Unfortunately, $\theta (0)=1$ is not
a rectification beyond $\s =0$.}
 for all $\s$ is
\begin{subequations}
\label{hh,he-rightmover}
\begin{equation}
 [\hHbb_{A}(m), \hHbb_{B}(n)] = km \delta_{AB} \delta_{m+n,0}
\end{equation}
\begin{equation}
\label{hh,he-r2}
 [\hHbb_{A}(m), \hEbb_{\a }(n-\s \a \cdot d)]  =\a_A \hEbb_{\a}(m+n-\s \a \cdot d)
\end{equation}
\begin{equation}
 [\hEbb_{\alpha}(m-\s \a \cdot d), \hEbb_{\beta}(n-\s \beta \cdot d) ]
 =\left\{\begin{array}{lll}
N_\gamma(\alpha,\beta) \hEbb_{\gamma}(m+n-\s \gamma \cdot d)  && \textrm{if}\  \alpha+\beta=\gamma\\
\alpha\cdot \hHbb(m+n)+k(m-\s \a \cdot d)\delta_{m+n,0} && \textrm{if}\ \alpha+\beta=0\\
0 && \textrm{otherwise}\\
\end{array} \right.
\end{equation}
\begin{equation}
\label{rmcop}
 \hHbb_A(m) \equiv - \hhbtwo_A(-m), \quad \hEbb_\a(m-\s\a \cdot d) \equiv - \hEb_{-\a}(-m+\s\a \cdot d)
\end{equation}
\end{subequations}
so that $\gfrakh(h_\s^{-1}) \simeq \gfrakh(h_\s)$ for
all sectors $\s$ of the orbifolds $A_g(H(d))/H(d)$.

The stress tensors and twisted conformal weight matrices
of the inner-automorphic WZW orbifolds are also very simple
\begin{subequations}
\label{Linn}
\begin{equation}
\label{Linner}
\lr = L_g \sp \D_{\sgb (\s)}  (\T) = D_g (T) = \Delta_g (T)
\end{equation}
\begin{equation}
\hat T_\s (z) = \frac{1}{2k + Q_g} \left( \sum_A : \hat H_A(z) \hat H_A (z):
 + \sum_\a : \hat E_\a (z) \hat E_{-\a} (z) :  \right)
 \end{equation}
\begin{equation}
\hat{\bar{T}}_\s (\bz) = \frac{1}{2k + Q_g} \left( \sum_A : \hat{\bar{H}}_A(\bz)
\hat{\bar{H}}_A (\bz) :
 + \sum_\a : \hat{\bar{E}}_\a (\bz) \hat{\bar{E}}_{-\a} (\bz) :  \right)
 \end{equation}
 \end{subequations}
because the eigenvalue matrices $U$ and $U(T)$ are unity.
Here  $\Delta_g(T)$ is the conformal weight of (untwisted) irrep $T$ of
(untwisted) $g$.

Similarly, we obtain the commutators with the twisted affine primary fields
\begin{subequations}
\begin{equation}
\hg (T,\bz,z)_\l{}^{\l '} = ( \hgm (T,\bz) \hgp (T,z) )_\l{}^{\l '}
\end{equation}
\begin{equation}
\hgp (T,z) \equiv \hgp (T,z,\s) \sp \hgm (T,\bz) \equiv \hgm (T,\bz,\s)
\end{equation}
\begin{equation}
\label{comHEg}
[ \hat H_A (m), \hgp (T,z)] = \hgp (T,z) z^m T_A
\; , \;
[ \hat E_\a (m-\s \a \cdot d), \hgp (T,z)] =
 \hgp (T,z) z^{m-\s \a \cdot d}T_\a
\end{equation}
\begin{equation}
\label{comHEgr}
[ \hat{\bar{H}}_A (m), \hgm (T,\bz)] = -\bz^{-m} T_A \hgm (T,\bz)
\; , \;
[ \hat{\bar{E}}_\a (m-\s \a \cdot d), \hgm (T,\bz)] =
 - \bz^{-(m-\s \a \cdot d)}T_\a \hgm (T,\bz)
\end{equation}
\begin{equation}
\label{Lgui}
[L_\s (m), \hgp (T,z)] = \hgp(\T,z)  (\overleftarrow{\partial} z
+ (m+1)\Delta_{g} (T) )z^m
\end{equation}
\begin{equation}
\label{Lgbi}
[\bar L_\s (m), \hgm (T,\bz)] = \bz^m (\bz \bar{\partial}
+ (m+1)  \Delta_{g} (T) )\hgm (T,\bz)
\end{equation}
\end{subequations}
because $\T = T$ and $\D_{\sgb (\s)} (\T)= \Delta_{g} (T)$.

\subsection{The Untwisted Affine Vacuum and Another Mode Normal Ordering \label{amno}}

For inner-automorphic orbifolds the action of the general twisted currents
on the true ground state $ | 0\rangle_\s$ is not known
(see Refs.~\cite{Freericks:1988zg} and \cite{Halpern:2000vj}).
Here we provisionally study the action of the twisted
currents on the untwisted affine vacuum $|0 \rangle$
\begin{subequations}
\label{vacin}
\begin{equation}
\label{vacina}
0= ( \hat H_A (m)  - \d_{m,0}  \s k d_A ) |0 \rangle
= \hat E_\a (m-\s \a \cdot d)|0 \rangle \sp m \geq 0
\end{equation}
\begin{equation}
\label{vacinb}
0 = \langle 0 | (\hat H_A (m) - \d_{m,0}  \s k d_A )=
\langle 0 | \hat E_\a (m-\s \a \cdot d) \sp m \leq 0
\end{equation}
\begin{equation}
\label{vacinc}
0= ( \hat{\bar{H}}_A (m) + \d_{m,0}  \s k d_A ) |0 \rangle
= \hat{\bar{E}}_\a (m-\s \a \cdot d)|0 \rangle \sp m \leq 0
\end{equation}
\begin{equation}
\label{vacind}
0 = \langle 0 | (\hat{\bar{H}}_A (m) + \d_{m,0}  \s k d_A )=
\langle 0 | \hat{\bar{E}}_\a (m-\s \a \cdot d) \sp m \geq 0
\end{equation}
\begin{equation}
A = 1 \ldots {\rm rank}\,g \sp \forall \a \in \Delta (g) \sp
\s = 0 , \ldots , \rho (1) -1 \ .
\end{equation}
\end{subequations}
The left mover conditions \eqref{vacina}, \eqref{vacinb} were given in
Ref.~\cite{Halpern:2000vj} and the right mover conditions \eqref{vacinc},
\eqref{vacind} follow immediately from the right mover copies
\begin{subequations}
\begin{equation}
0 = ( \hHbb_A(m) - \delta_{m,0} \s k d_A) | 0 \rangle =
\hEbb_\a (m - \s \a \cdot d)| 0 \rangle \sp m \geq 0
\end{equation}
\begin{equation}
0 = \langle 0 | ( \hHbb_A(m) - \delta_{m,0} \s k d_A)  =
 \langle 0 |  \hEbb_\a (m - \s \a \cdot d) \sp m \leq 0
\end{equation}
\end{subequations}
of the left mover conditions, where $\hjbb$ are the rectified right mover
currents in \eqref{rmcop}.

As an application of the commutators \eqref{comHEg}, \eqref{comHEgr}
and the vacuum conditions \eqref{vacin}, we  give the global  Ward
identities for the residual symmetry associated to the Lie subalgebra \eqref{liesa}
of the twisted current algebra:
\begin{subequations}
\label{gidilr}
\begin{equation}
\hat A (\s) = \hat A_- (\s) \hat A_+  (\s) \sp
\hat A_{\pm} (\s) \equiv \langle 0|\hg_\pm (1,\s) \cdots \hg_\pm (N,\s)|0 \rangle
\equiv \langle \hg_\pm (1,\s) \cdots \hg_\pm (N,\s) \rangle
\end{equation}
\begin{equation}
\label{gidi}
\langle [\hat H_A(0), \hgp (1,\s) \cdots \hgp (N,\s) ] \rangle = 0 \quad
\Rightarrow \quad \hat A_+ (\s) Q_A = 0 \sp A = 1 \ldots
{\rm rank}\,g
\end{equation}
\begin{equation}
\label{gidir}
\langle [\hat{\bar{H}}_A(0), \hgm (1,\s) \cdots \hgm (N,\s) ] \rangle = 0 \quad
\Rightarrow \quad Q_A \hat A_- (\s)=  0 \sp A = 1 \ldots
{\rm rank}\,g
\end{equation}
\begin{equation}
Q_A \equiv \sum_{\mu =1}^N T_{A}^{(\mu)} \sp [Q_A, Q_B ] = 0 \ .
\end{equation}
\end{subequations}
In the permutation orbifolds (see Subsecs.~\ref{twlmKZ} and \ref{twrmKZ}),
the subalgebra \eqref{liesa} provided all the residual symmetries of the twisted sectors, but,
in this case, we also find further coordinate-dependent%
\footnote{Coordinate-dependent Ward identities have also been encountered
in affine-Virasoro theory \cite{Halpern:1995fy,Halpern:1996js}.} Ward identities
associated to the twisted root operators at $m=0$:
\begin{subequations}
\label{gidier}
\begin{equation}
\label{gidie}
\langle [\hE_\a (-\s \a \cdot d),\hgp (1,\s) \cdots \hgp (N,\s) ]\rangle
=0 \quad \Rightarrow \quad \hat A_+ (\s) Q_\a (z,\s)=0 \sp \forall \; \a \in \Delta (g)
\end{equation}
\begin{equation}
\langle [\hEb_\a (-\s \a \cdot d)\hgm (1,\s) \cdots \hgm (N,\s) ]\rangle
=0 \quad \Rightarrow \quad  \bar{Q}_\a (\bz,\s) \hat A_- (\s) =0 \sp
\forall \; \a \in \Delta (g) \ .
\end{equation}
\begin{equation}
 Q_\a (z,\s) \equiv \sum_{\m=1}^N z_\m^{-\s \a \cdot d} T_\a^{(\m)}\sp
\bar{Q}_\a (\bz,\s) \equiv \sum_{\m=1}^N \bz_\m^{\;\,\s \a \cdot d}T_\a^{(\m)}
\end{equation}
\begin{equation}
[Q_A, Q_\a (z,\s) ] = \a_A Q_\a (z,\s) \sp
[ Q_A, \bar{Q}_\a (\bz,\s) ] = \a_A \bar{Q}_\a (\bz,\s)
\end{equation}
\begin{equation}
[ Q_\a (z) , Q_{\be} (z) ] =
\left\{\begin{array}{lll}
N_\gamma(\alpha,\beta) Q_\gamma (z)  && \textrm{if}\  \alpha+\beta=\gamma\\
\alpha\cdot Q  && \textrm{if}\ \alpha+\beta=0\\
0 && \textrm{otherwise}\\
\end{array} \right.
\end{equation}
\begin{equation}
[ \bar{Q}_\a (\bz) , \bar{Q}_{\be} (\bz) ] =
\left\{\begin{array}{lll}
N_\gamma(\alpha,\beta) \bar{Q}_\gamma (\bz)  && \textrm{if}\
\alpha+\beta=\gamma\\
\alpha\cdot Q  && \textrm{if}\ \alpha+\beta=0\\
0 && \textrm{otherwise} \ . \\
\end{array} \right.
\end{equation}
\end{subequations}
Note that, taken together, the Ward identities \eqref{gidilr} and  \eqref{gidier}
 reduce to the usual untwisted global Ward identities
\begin{equation}
\hat A_+ (0)  \sum_{\mu =1}^N T_{a}^{(\mu)} =
\left(\sum_{\m=1}^N T_a^{(\m)} \right)  \hat A_- (0)
= 0 \sp a = 1 \ldots {\rm dim}\,g
\end{equation}
in the untwisted sector $\s =0$.

We also obtain the $L_\s (0)$ and $\bar L_\s (0)$ Ward identities
\begin{equation}
\label{L0wii}
 \hat A_+ (\s) \sum_{\m=1}^N \left(\overleftarrow{\partial_\m} z_\m
  + \Delta_{g} (T^{(\m)} ) \right) = 0 \sp
 \sum_{\m =1}^N \left( \bz_\m \bar \partial_\m + \Delta_{g} (T^{(\m)} )
 \right) \hat A_- (\s) =0
\end{equation}
from \eqref{L0wi} and \eqref{Linner} or \eqref{Lgui} and \eqref{Lgbi}.

We continue with the $M'$ and $\bM '$ normal orderings of the twisted
currents
\begin{subequations}
\label{pmmbno}
\begin{eqnarray}
:\hj_{n(r) \mu}(m+\srac{n(r)}{\r(\s)}) \hj_{n(s) \nu}
(n+\srac{n(s)}{\r(\s)}):_{M'} \
 & \equiv &
 \theta(m \geq 0 ) \ \hj_{n(s)
\nu}(n+\srac{n(s)}{\r(\s)}) \hjb_{n(r) \mu}(m+\srac{n(r)}{\r(\s)})
\nn \\
 & & + \theta(m < 0) \
\hj_{n(r) \mu}(m+\srac{n(r)}{ \r(\s)})
\hj_{n(s) \nu}(n+\srac{n(s)}{ \r(\s)}) \nn \\\label{pmmbnol}
\end{eqnarray}
\begin{eqnarray}
:\hjb_{n(r) \mu}(m+\srac{n(r)}{\r(\s)}) \hjb_{n(s) \nu}
(n+\srac{n(s)}{\r(\s)}):_{\bM '} \
& \equiv &
 \theta(m\leq 0 ) \ \hjb_{n(s)
\nu}(n+\srac{n(s)}{\r(\s)}) \hjb_{n(r) \mu}(m+\srac{n(r)}{\r(\s)}) \nn \\
  & & + \theta(m > 0) \
\hjb_{n(r) \mu}(m+\srac{n(r)}{ \r(\s)})
\hjb_{n(s) \nu}(n+\srac{n(s)}{ \r(\s)}) \nn \\
\end{eqnarray}
\begin{equation}
\hj = \{ \hat{H}_A, \hat{E}_\a \} \sp
\hjb =\{ \hat{\bar{H}}_A, \hat{\bar{E}}_\a \} \sp n_A = 0 \sp
\srac{n_\a (r) }{\rho (\s) } = - \s \a \cdot d
\end{equation}
\end{subequations}
which are tailored for the untwisted affine vacuum $| 0 \rangle$. The
$M$ ordering in \eqref{pmmbnol} was studied in Ref.~\cite{Halpern:2000vj}.

This gives the relation between the normal ordered products
\begin{subequations}
\begin{eqnarray}
: \hh_A(z) \hh_B (z) : & = &  : \hh_A(z) \hh_B (z) :_{M'} \\
: \hE_\a (z) \hE_{-\a} (z) : & = & : \hE_\a (z) \hE_{-\a} (z) :_{M'}
+ \frac{\s \a \cdot d}{z} \a \cdot \hh (z) - \frac{k \s \a \cdot d}{2 z^2}
(1 + \s \a \cdot d) \quad  \;\;\; \;\; \;\;\;\;\\
 : \hhb_A(\bz) \hhb_B (\bz) : & = &  : \hhb_A(\bz) \hhb_B (\bz) :_{\bM '} \\
 : \hEb_\a (\bz) \hEb_{-\a} (\bz) : & = & : \hEb_\a (\bz) \hEb_{-\a} (\bz) :_{\bM '}
- \frac{\s \a \cdot d}{\bz} \a \cdot \hhb (\bz) + \frac{k \s \a \cdot d}{2 \bz^2}
(1 - \s \a \cdot d) \;\;\; \;\;\;\;\;\;
\end{eqnarray}
\end{subequations}
and hence the mode ordered form of the left and right mover Virasoro generators
\begin{subequations}
\label{Viropinlr}
\begin{eqnarray}
(2k+Q_g)L_\s (m) \!\!\!&= &\sum_{p\in \sz}
\{ :\sum_{\a}\hat{E}_\a(p-\s \a \cdot d)
\hat{E}_{-\a}(m-p+\s \a \cdot d):_{M \p} \nn \\
 & + &\!\!\!\!  \sum_{A} :\hat{H}_A(p) \hat{H}_A (m-p):_{M \p} \} +
 Q_g \{\s  d \cdot \hat{H}(m)-
 \delta_{m,0} \frac{k}{2} \s^2 d^2 \}  \hskip 1.5cm
\label{Viropin}\end{eqnarray}
\begin{eqnarray}
(2k+Q_g)\bar L_\s (m) \!\!\!&= &\sum_{p\in \sz}
\{ :\sum_{\a}\hEb_\a(p-\s \a \cdot d)
\hEb_{-\a}(-m-p+\s \a \cdot d):_{\bM \p} \nn \\
 & + & \!\!\!\! \sum_{A} :\hhb_A(p) \hhb_A (-m-p):_{\bM \p} \} -
 Q_g \{\s  d \cdot \hhb (-m) +
 \delta_{m,0} \frac{k}{2} \s^2 d^2 \}  \hskip 1.5cm
\label{Viropinr}\end{eqnarray}
\begin{equation}
\label{said}
\sum_\a \a_A \a_B = Q_g \delta_{AB} \ .
\end{equation}
\end{subequations}
The left mover generators \eqref{Viropin} were given in Ref.~\cite{Halpern:2000vj}.

Then using the vacuum conditions \eqref{vacin} we verify that the untwisted
affine vacuum is indeed primary under Vir$\oplus$Vir
\begin{subequations}
\label{lmcwin0}
\begin{equation}
L_\s (m \geq 0) | 0 \rangle = \de_{m,0} \hat\Delta_0 (\s) | 0 \rangle \sp
\bar L_\s (m \geq 0) | 0 \rangle = \de_{m,0} \hat{\bar{\Delta}}_0 (\s) | 0 \rangle
\end{equation}
\begin{equation}
\label{lmcwin}
 \hat \Delta_0 (\s) = \hat{\bar{\Delta}}_0 (\s) =\frac{k}{2} \s^2 d^2
\end{equation}
\end{subequations}
 with the same left and right mover conformal weights. The left mover
 result in \eqref{lmcwin} is of course very old \cite{Freericks:1988zg}.

Consistent with \eqref{lmcwin0}, it is not difficult to verify that the set of right mover Virasoro
generators in \eqref{Viropinr} can be expressed as a copy
\begin{eqnarray}
(2k+Q_g)\bar L_\s (m) \!\!\!&= &\sum_{p\in \sz}
\{ :\sum_{\a}\hEbb_\a(p-\s \a \cdot d)
\hEbb_{-\a}(m-p+\s \a \cdot d):_{M \p} \nn \\
 & + & \!\!\!\! \sum_{A} :\hHbb_A(p) \hHbb_A (m-p):_{M \p} \} +
 Q_g \{\s  d \cdot \hHbb (m) -
 \delta_{m,0} \frac{k}{2} \s^2 d^2 \}  \hskip 1.2cm
\label{Viropinr2}\end{eqnarray}
of the left mover Virasoro generators in \eqref{Viropin}. Here, $\{\hjbb \}$
are the rectified right mover current modes in \eqref{hh,he-rightmover}.
As noted in Subsec.~\ref{unrec}, this  situation is expected when the
twisted right mover currents are rectifiable.

We will also need the corresponding $M'$ and $\bM'$ normal orderings of the
twisted currents with the twisted affine primary fields
\begin{subequations}
\label{Mpord}
\begin{equation}
: \hat H_A (m) \hgp (T,z) :_{M'} \equiv
\theta (m \geq 0) \hgp (T,z) \hat H_A (m) + \theta (m<0) \hat H_A
(m) \hgp (T,z)
\end{equation}
\begin{equation}
: \hat E_\a (m - \s \a \cdot d) \hgp (T,z) :_{M'} \hskip 9cm
\end{equation}
$$
\equiv
\theta (m \geq 0) \hgp (T,z) \hat E_\a (m- \s \a \cdot d)
+ \theta (m<0) \hat E_\a(m -\s \a \cdot d) \hgp (T,z)
$$
\begin{equation}
: \hat{\bar{H}}_A (m) \hgm (T,\bz) :_{\bM^\prime} \equiv
\theta (m \leq 0) \hgm (T,\bz) \hat{\bar{H}}_A (m) + \theta (m>0) \hat{\bar{H}}_A
(m) \hgm (T,\bz)
\end{equation}
\begin{equation}
: \hat{\bar{E}}_\a (m - \s \a \cdot d) \hgm (T,\bz) :_{\bM^\prime} \hskip 9cm
\end{equation}
$$
\equiv
\theta (m \leq 0) \hgm (T,\bz) \hat{\bar{E}}_\a (m- \s \a \cdot d)
+ \theta (m>0) \hat{\bar{E}}_\a(m -\s \a \cdot d) \hgm (T,\bz) \ .
$$
\end{subequations}
>From these definitions and the commutators in \eqref{comHEg}, \eqref{comHEgr}
one computes the exact operator products
\begin{subequations}
\begin{equation}
\hat H_A (z) \hgp (T,w) =  \frac{1}{z-w} \hgp (T,w) T_A
+ : \hat H_A (z) \hgp (T,w) :_{M'}
\end{equation}
\begin{equation}
\hat E_\a (z) \hgp (T,w) = \left( \frac{w}{z}
\right)^{-\s \a \cdot d} \frac{1}{z-w} \hgp (T,w) T_\a
+ : \hat E_\a (z) \hgp (T,w) :_{M'}
\end{equation}
\begin{equation}
\hat{\bar{H}}_A (\bz) \hgm (T,\bw) =  - \frac{1}{\bz-\bw} T_A \hgm (T,\bw)
+ : \hat{\bar{H}}_A (\bz) \hgm (T,\bw) :_{\bM'}
\end{equation}
\begin{equation}
\hat{\bar{E}}_\a (\bz) \hgm (T,\bw) = - \left( \frac{\bw}{\bz}
\right)^{\s \a \cdot d} \frac{1}{\bz-\bw}  T_\a \hgm (T,\bw)
+ : \hat{\bar{E}}_\a (\bz) \hgm (T,\bw) :_{\bM'} \ .
\end{equation}
\end{subequations}
Then the relations
\begin{subequations}
\label{opemprel}
\begin{equation}
:\hat H_A (z) \hgp (T,z) :=  : \hat H_A (z) \hgp (T,z) :_{M'}
\end{equation}
\begin{equation}
: \hat E_\a (z) \hgp (T,z): =  : \hat E_\a (z) \hgp (T,z) :_{M'}
+ \frac{\s \a \cdot d}{z}  \hgp (T,z) T_\a
\end{equation}
\begin{equation}
:\hat{\bar{H}}_A (\bz) \hgm (T,\bz) :=: \hat{\bar{H}}_A (\bz) \hgm (T,\bz) :_{\bM'}
\end{equation}
\begin{equation}
: \hat{\bar{E}}_\a (\bz) \hgm (T,\bz): =  : \hat{\bar{E}}_\a (\bz) \hgm (T,\bz) :_{\bM'}
+ \frac{\s \a \cdot d}{\bz}   T_\a \hgm (T,\bz)
\end{equation}
\end{subequations}
are obtained among the normal ordered products.

\subsection{Twisted Vertex Operator Equations in Mode-Ordered Form}

For inner-automorphic WZW orbifolds, the twisted
vertex operator equations \eqref{tvoe} reduce to
\begin{subequations}
\begin{eqnarray}
\partial \hgp (T,z) \!\! & = & 2 L_g^{ab}
 : \hat{J}_{a}(z) \hgp (T,z) : T_{b} \nn \\
& = &
\frac{2}{2k + Q_g} \left( \sum_A : \hat{H}_{A}(z) \hgp (T,z) : T_{A}
+ \sum_\a : \hat{E}_{\a}(z) \hgp (T,z) : T_{-\a} \right) \hskip 1cm
\label{difrelti}
\end{eqnarray}
\begin{eqnarray}
\bar \partial \hgm (T,\bz) \! \! & = &  -2 L_g^{ab}
  : \hat{\bJ}_{a}(\bz) T_{b} \hgm (T,\bz) : \nn \\
& = &
-\frac{2}{2k + Q_g} \left( \sum_A : \hat{\bar{H}}_{A}(\bz) T_{A} \hgm (T,\bz) :
+  \sum_\a: \hat{\bar{E}}_{\a}(\bz)  T_{-\a} \hgm (T,\bz) : \right)\! .  \hskip 1cm
\label{difreltir}
\end{eqnarray}
\end{subequations}
Using \eqref{opemprel}, the corresponding forms in terms of
$M{}^\prime$ and $\bM'$ ordered products are
\begin{subequations}
\begin{eqnarray}
\partial \hgp (T,z)
& = & \frac{2}{2k + Q_g} \Big(\sum_A : \hat{H}_{A}(z) \hgp (T,z) :_{M'} T_{A}
+ \sum_\a: \hat{E}_{\a}(z) \hgp (T,z) :_{M'} T_{-\a} \nn \\
  & & \hskip 1.5cm +\sum_\a \frac{\s \a \cdot d}{z}  \hgp (T,z) T_\a T_{-\a} \Big)
\label{difreltiM}
\end{eqnarray}
\begin{eqnarray}
\bar \partial \hgm (T,\bz)
& =&  -\frac{2}{2k + Q_g} \Big( \sum_A: \hat{\bar{H}}_{A}(\bz) T_{A} \hgm (T,\bz) :_{\bM'}
+ \sum_\a: \hat{\bar{E}}_{\a}(z) T_{-\a} \hgm (T,\bz) :_{\bM'} \nn \\
& &  \hskip 1.7cm +\sum_\a\frac{\s \a \cdot d}{\bz}  T_{-\a}T_\a \hgm (T,\bz) \Big) \ .
\label{difreltiMr}
\end{eqnarray}
\end{subequations}
The explicit forms of these mode ordered products
\begin{subequations}
\label{gJinner}
\begin{equation}
: \hj_a (z) \hgp (T,z)  :_{M \p} \ =
\hj_a^- (z) \hgp (T,z) + \hgp (T,z)  \hj_a^+ (z) \sp a = 1\ldots {\rm dim}\,g
\end{equation}
\begin{equation}
\hat H_A^- (z) = \sum_{m \leq -1} \hat H_A (m) z^{-m-1} \sp
\hat H_A^+ (z) = \sum_{m \geq 0} \hat H_A (m) z^{-m-1}
\end{equation}
\begin{equation}
\hat E_\a^- (z) = \sum_{m \leq -1} \hat E_\a (m-\s \a \cdot d)
z^{-(m-\s \a \cdot d)-1} \sp
\hat E_\a^+ (z) = \sum_{m \geq 0} \hat E_\a (m-\s \a \cdot d)
z^{-(m-\s \a \cdot d)-1}
\end{equation}
\begin{equation}
 \hat H_A^- (z) +\hat H_A^+ (z) = \hat H_A (z) \sp
\hat E_\a^- (z) + \hat E_\a^+ (z) = \hat E_\a (z)
\end{equation}
\end{subequations}
\begin{subequations}
\label{gJbinner}
\begin{equation}
: \hjb_a (\bz) \hgm (T,z)  :_{\bM \p} \ =
\hjb_a^+ (\bz) \hgm (T,\bz) + \hgm (T,\bz)  \hjb_a^- (\bz)
\sp a = 1 \ldots {\rm dim}\,g
\end{equation}
\begin{equation}
\hat{\bar{H}}_A^+ (\bz) = \sum_{m  \geq 1} \hat{\bar{H}}_A (m) \bz^{m-1} \sp
\hat{\bar{H}}_A^- (\bz) = \sum_{m \leq 0} \hat{\bar{H}}_A (m) \bz^{m-1} \sp
\end{equation}
\begin{equation}
\hat{\bar{E}}_\a^+ (\bz) = \sum_{m \geq 1} \hat{\bar{E}}_\a (m-\s \a \cdot d)
\bz^{(m-\s \a \cdot d)-1} \sp
\hat{\bar{E}}_\a^- (\bz) = \sum_{m \leq 0} \hat{\bar{E}}_\a (m-\s \a \cdot d)
\bz^{(m-\s \a \cdot d)-1} \sp
\end{equation}
\begin{equation}
\hat{\bar{H}}_A^- (\bz) + \hat{\bar{H}}_A^+ (\bz)  = \hat{\bar{H}}_A (\bz) \sp
\hat{\bar{E}}_\a^- (\bz) + \hat{\bar{E}}_\a^+ (\bz) = \hat{\bar{E}}_\a (\bz)
\end{equation}
\end{subequations}
follow from Eq.~\eqref{Mpord}.

The twisted partial currents in \eqref{gJinner}, \eqref{gJbinner}
satisfy the commutation relations
\begin{subequations}
\label{comi}
\begin{equation}
[ \hat H_A^+ (z), \hgp (T,w) ] = \frac{1}{z-w} \hgp (T,w) T_A \sp
|z| > |w|
\end{equation}
\begin{equation}
[ \hat H_A^- (z), \hgp (T,w) ] = -\frac{1}{z-w} \hgp (T,w) T_A \sp
|w| > |z|
\end{equation}
\begin{equation}
[ \hat E_\a^+ (z), \hgp (T,w) ] = \left( \frac{w}{z}
\right)^{-\s \a \cdot d}
\frac{1}{z-w} \hgp (T,w) T_\a \sp
|z| > |w|
\end{equation}
\begin{equation}
[ \hat E_\a^- (z), \hgp (T,w) ] = -\left( \frac{w}{z}
\right)^{-\s \a \cdot d}
\frac{1}{z-w} \hgp (T,w) T_\a \sp
|w| > |z|
\end{equation}
\end{subequations}
\begin{subequations}
\label{lcomi}
\begin{equation}
[ \hat{\bar{H}}_A^- (\bz), \hgm (T,\bw) ] = - \frac{1}{\bz-\bw} T_A \hgm (T,\bw)
\sp |\bz| > |\bw|
\end{equation}
\begin{equation}
[ \hat{\bar{H}}_A^+ (\bz), \hgm (T,\bw) ] = \frac{1}{\bz-\bw} T_A \hgm (T,\bw)
 \sp |\bw| > |\bz|
\end{equation}
\begin{equation}
[ \hat{\bar{E}}_\a^- (\bz), \hgm (T,\bw) ] = -
 \left( \frac{\bw}{\bz} \right)^{\s \a \cdot d}\frac{1}{\bz-\bw}
 T_\a \hgm (T,\bw)  \sp | \bz| > |\bw|
\end{equation}
\begin{equation}
[ \hat{\bar{E}}_\a^+ (\bz), \hgm (T,\bw) ] =
\left( \frac{\bw}{\bz} \right)^{\s \a \cdot d}\frac{1}{\bz-\bw}
 T_\a  \hgm (T,\bw) \sp |\bw| > |\bz|
\end{equation}
\end{subequations}
and the vacuum conditions
\begin{subequations}
\label{vaci}
\begin{equation}
  \langle 0 | \hat H_A^- (z)  = 0 \sp \hat H_A^+ (z)|0 \rangle =
\frac{1}{z} \s k d_A  |0 \rangle \sp
\hat E_\a^+ (z)|0 \rangle  = \langle 0 | \hat E_\a^- (z) = 0
\end{equation}
\begin{equation}
\label{vacib}
\langle : \hat H_A (z)  \hgp (T,z) :_{M \p}\rangle =\frac{1}{z} \s k d_A
\langle \hgp (T,z) \rangle \sp
\langle : \hat E_\a (z) \hgp (T,z)  :_{M \p} \rangle = 0
\end{equation}
\end{subequations}
\begin{subequations}
\label{vacir}
\begin{equation}
\langle 0 | \hat{\bar{H}}_A^+ (\bz) = 0 \sp
\hat{\bar{H}}_A^- (\bz)|0 \rangle =
- \frac{1}{\bz} \s k d_A  |0 \rangle \sp
\hat{\bar{E}}_\a^- (\bz)|0 \rangle  = \langle 0 | \hat{\bar{E}}_\a^+ (\bz) = 0
\end{equation}
\begin{equation}
\langle : \hat{\bar{H}}_A (\bz) \hgm (T,\bz)  :_{\bM \p}\rangle
=-\frac{1}{\bz} \s k d_A
\langle \hgm (T,\bz) \rangle \sp
\langle : \hat{\bar{E}}_\a (\bz) \hgm (T,\bz) :_{\bM \p} \rangle = 0
\end{equation}
\end{subequations}
which follow from Eqs.~\eqref{gJinner}, \eqref{gJbinner} and
\eqref{vacin}.

\subsection{Consistency Check  \label{ddin} }

We  consider the consistency check of Subsec.~\ref{consvoe}, this time for
the twisted vertex operator equations of the inner-automorphic orbifolds.
In the case of the twisted left mover equation,
we start from the differential equation
\begin{equation}
\part \hgp (T,z) =[ L_\s (-1), \hgp (T,z) ]
\end{equation}
and the $M'$ normal ordered form \eqref{Viropin} of the left mover Virasoro
generators. Then using these and the commutators \eqref{comHEg} of the twisted
currents with the twisted primary fields, we find a somewhat different form of
the twisted left mover vertex operator equation
\begin{eqnarray}
\partial \hgp (T,z)
&=& \frac{1}{2k + Q_g} \Big( 2 \sum_A : \hat{H}_{A}(z) \hgp (T,z) :_{M'} T_{A}
+ 2 \sum_\a : \hat{E}_{\a}(z) \hgp (T,z) :_{M'} T_{-\a} \nn \\
 &  & +\frac{\hgp (T,z)}{z}  Q_g \s d\cdot T \Big) \ .
\label{difreltiMd}\end{eqnarray}
This form of the equation agrees with the form in \eqref{difreltiM} iff
\begin{equation}
\label{taid}
\sum_\a \a \cdot d T_\a T_{-\a} = \frac{1}{2} Q_g d \cdot T
\end{equation}
and this relation is indeed an identity according to \eqref{said}
and the algebra of $T$ in   \eqref{algTcw2}.

As a simple application of the twisted vertex operator equation \eqref{difreltiMd}
and the vacuum conditions \eqref{vacib}, we obtain the equations for the
twisted left mover one-point correlators
\begin{subequations}
\begin{equation}
\part \langle \hgp (T,z,\s) \rangle = \langle \hgp (T,z,\s) \rangle
\frac{1}{z} \s d \cdot T
\end{equation}
\begin{equation}
\langle \hgp (T,z,\s) \rangle T_A =0 \sp A = 1 \ldots {\rm rank}\,g
\sp \langle \hgp (T,z,\s) \rangle T_\a =0 \sp \forall \,\a\in \Delta (g)
\end{equation}
\end{subequations}
in sector $\s$ of each
inner-automorphic WZW orbifold. The second relation includes the global
Ward identities in this case. The solution of this system is
\begin{subequations}
\label{solopc}
\begin{equation}
\label{solopca}
\langle \hgp (T,z,\s) \rangle = C_+ (T,\s) z^{\s d \cdot T}
\end{equation}
\begin{equation}
\label{solopcb}
 C_+  (T,\s) T_a =0\sp a = 1 \ldots {\rm dim}\,g
\end{equation}
\end{subequations}
where we have combined both parts of the global Ward identities in
\eqref{solopcb}. But this equation has no non-trivial solution, so we find
that
\begin{equation}
\langle \hgp (T,z,\s) \rangle = 0 \sp \s = 0 ,\ldots , \rho (1) -1
\end{equation}
in each  sector of all inner-automorphic
WZW orbifolds. The same conclusion can be obtained by comparing the solution
\eqref{solopca} with the $L_\s (0)$ Ward identities in \eqref{L0wii}.

\subsection{The Twisted KZ Equations of $A_g (H(d))/H(d)$}

Using the vacuum conditions \eqref{vaci}, the twisted vertex operator equation
 \eqref{difreltiMd} and commutation relations \eqref{comi}, one  finds after
 some algebra the  {\it twisted   KZ equations}
\begin{subequations}
\label{twkzi}
\begin{equation}
\hat A_+ (\s) \equiv \hat A_+ (T,z,\s) =
 \langle  \hgp (T^{(1)},z_1,\s) \cdots \hgp (T^{(N)},z_N,\s) \rangle
\end{equation}
\begin{equation}
\partial_\mu  \hat A_+ (\s)= \hat A_+ (\s)  \hat W_\mu (\s)  \sp
\hat W_\mu (\s) \equiv \hat W_\mu (T,z,\s)\sp \s = 0 ,\ldots ,\rho (1) -1
\end{equation}
\begin{equation}
\hat W_\mu (\s) = \frac{2}{2k+Q_g} \sum_{\nu \neq \mu} \frac{1}{z_{\mu \nu}}
\left[ \sum _A T_A^{(\nu)} T_A^{(\mu)} + \sum_\a \left( \frac{z_\nu}{z_\mu} \right)^{-\s
\a \cdot d}
T_\a^{(\nu)} T_{-\a}^{(\mu)} \right] + \frac{\s d \cdot T^{(\mu)}}{z_\mu}
\end{equation}
\end{subequations}
for each twisted left mover sector $\s$ of all inner-automorphic WZW orbifolds.
This twisted  KZ connection is flat
\begin{equation}
\label{flatin}
\part_\mu \hat W_\nu (\s) - \part_\nu \hat W_\mu (\s) +
[\hat W_\mu (\s),\hat W_\nu (\s) ] = 0
\end{equation}
and we have checked explicitly that this connection is {\it non-abelian} flat
\begin{equation}
\label{fld}
\part_\mu \hat W_\nu (\s) - \part_\nu \hat W_\mu (\s) = -
[\hat W_\mu (\s)  , \hat W_\nu (\s) ] =
- \frac{1}{z_{\mu} z_{\nu}}
\frac{2}{2k+Q_g}
\sum_\a  \left( \frac{z_\mu}{z_\nu} \right)^{-\s \a \cdot d}
\s \a \cdot d
T_\a^{(\mu)} T_{-\a}^{(\nu)} \ .
\end{equation}
To our knowledge, all previous KZ-like connections in conformal field theory
have been abelian flat. This remark includes the
KZ-like connections recently obtained for open WZW strings \cite{Giusto:2001sn}.

We have also checked explicitly that
\begin{subequations}
\begin{equation}
[Q_A,\hat W_\mu (\s) ] = 0 \sp A = 1 \ldots {\rm rank}\,g
\end{equation}
\begin{equation}
\label{flconi}
\partial_\m Q_\a (z,\s) - [ Q_\a (z,\s),\hat W_\m (\s) ] = 0
\sp \forall , \a \in \Delta (g)
\end{equation}
\end{subequations}
which guarantees the consistency of the Ward identities \eqref{gidi} and
\eqref{gidie} with the twisted KZ equations.
Finally, we have also checked that the $L_\s (0)$ Ward identity in \eqref{L0wii} is
satisfied when the twisted KZ equations \eqref{twkzi} and the global Ward identities
\eqref{gidi}, \eqref{gidie} are satisfied.

To obtain the corresponding results for the right mover sector of the
inner-automorphic orbifolds, we begin with the alternate form of the twisted
right mover vertex operator equation
\begin{eqnarray}
\bar \partial \hgm (T,\bz)
& = & -\frac{1}{2k + Q_g} \Big( 2 \sum_A : \hat{\bar{H}}_{A}(\bz)  T_{A} \hgm (T,\bz) :_{\bM'}
+ 2\sum_\a : \hat{\bar{E}}_{\a}(\bz) T_{-\a} \hgm (T,\bz) :_{\bM'} \nn \\
& &  \hskip 2cm -Q_g \s d\cdot T \frac{\hgm (T,\bz)}{\bz}
\Big) \label{difreltiMdr}
\end{eqnarray}
which is related to \eqref{difreltiMr} by the identity \eqref{taid}.

Then, using \eqref{difreltiMdr}, the vacuum conditions \eqref{vacir} and
the commutation relations \eqref{comHEgr}, one finds the
twisted right mover KZ equations
\begin{subequations}
\label{twrmkzin}
\begin{equation}
\hat A_- (\s) \equiv  \langle\hgm (T^{(1)},\bz_1,\s) \cdots \hgm (T^{(N)},
\bz_N,\s) \rangle \sp \s = 0 , \ldots ,\rho (1) -1
\end{equation}
\begin{equation}
\bar \partial_\mu \hat A_- (\s) = \hat{\bar{W}}_\mu (\s) \hat A_- (\s)
\end{equation}
\begin{equation}
\hat{\bar{W}}_\mu (\s) = \frac{2}{2k+Q_g} \sum_{\nu \neq \mu} \frac{1}{\bz_{\mu \nu}}
\left[ \sum_A T_A^{(\mu)} T_A^{(\nu)} + \sum_\a \left( \frac{\bz_\nu}{\bz_\mu} \right)^{-\s
\a \cdot d}
T_\a^{(\mu)} T_{-\a}^{(\nu)} \right] + \frac{\s d \cdot T^{(\mu)}}{\bz_\mu}
\end{equation}
\begin{equation}
\bar \partial_\mu {\hat{\bar W}}_{\nu}(\s)-
\bar \partial_\nu {\hat{\bar W}}_{\mu}(\s)
- [  {\hat{\bar W}}_{\mu}(\s),{\hat{\bar W}}_{\nu}(\s) ] = 0
\end{equation}
\end{subequations}
for each twisted right mover sector $\s$ of the inner-automorphic WZW orbifolds.
As in the case of the left movers, the twisted right mover connections
$\hat{\bar{W}}_\m$ are non-abelian flat and the relations
\begin{subequations}
\begin{equation}
[Q_A,\hat{\bar{W}}_\mu (\s) ] = 0 \sp A = 1 \ldots {\rm rank}\,g
\end{equation}
\begin{equation}
\label{flconir}
\bar \partial_\m \bar Q_\a (\bz,\s) + [ \bar Q_\a (\bz,\s),\hat{\bar{W}}_\m (\s) ]
 = 0 \sp \forall \, \a \in \Delta (g)
\end{equation}
\end{subequations}
guarantee the consistency of the Ward identities \eqref{gidir} and
\eqref{gidier} with the twisted right mover KZ equations.
Moreover, the $\bar L_\s (0)$ Ward identity
in \eqref{L0wii} is satisfied when the KZ equations \eqref{twrmkzin}
and the Ward identities \eqref{gidir}, \eqref{gidier} are satisfied.
Finally, one also finds that
 $\langle \hgm (T,\bz,\s) \rangle = 0$ for the twisted right mover one-point
 correlators.

\subsection{Relation to the Untwisted Sector \label{RUS} }

Our results for
the inner-automorphic WZW orbifolds are simpler than they appear. In particular,
we are able to relate the twisted KZ systems \eqref{gidilr}, \eqref{gidier},
\eqref{twkzi} and \eqref{twrmkzin} in twisted sector $\s \neq 0$ to the
corresponding ordinary KZ systems \cite{Knizhnik:1984nr}
of the untwisted sector $\s =0$. The relations are
\begin{subequations}
\label{uteq}
\begin{equation}
\hat A_+ (T,z,\s) = \hat A_+ (T,z,0) U(T,z,\s)
\end{equation}
\begin{equation}
\hat A_- (T,\bz,\s) = \bar U(T,\bz,\s)\hat A_- (T,\bz,0)
\end{equation}
\begin{equation}
U(T,z,\s) \equiv \prod_{\m =1}^N z_\m{}^{\s d \cdot T^{(\m)}} \sp
\bar U(T,\bz,\s) \equiv \prod_{\m =1}^N \bz_\m{}^{\s d \cdot T^{(\m)}}
\end{equation}
\begin{equation}
\partial_\m \hat A_+ (T,z,0) = \hat A_+ (T,z,0) \hat W_\m (T,z,0) \sp
\bar \partial_\m \hat A_- (T,\bz,0) = \hat{\bar{W}}_\m (T,\bz,0) \hat A_- (T,\bz,0)
\end{equation}
\begin{equation}
\hat W_\m (T,z,0) = \frac{2 \eta^{ab}}{2k + Q_g} \sum_{\n \neq \m}
\frac{T_a^{(\n)} T_b^{(\m)}}{z_{\m\n}} \sp
\hat{\bar{W}}_\m (T,\bz,0) = \frac{2 \eta^{ab}}{2k + Q_g} \sum_{\n \neq \m}
\frac{T_a^{(\m)} T_b^{(\n)}}{\bz_{\m\n}}
\end{equation}
\begin{equation}
\hat A_+ (T,z,0) Q_a (T,0) = Q_a (T,0) \hat A_- (T,\bz,0) = 0 \sp
Q_a (T,0) = \sum_{\m =1}^N T_a^{(\m)} \sp a = 1 \ldots {\rm dim}\,g \ .
\end{equation}
\end{subequations}
In checking this equivalence one uses the identity
\begin{equation}
x^{\, \s d \cdot T} T_\a x^{ - \s d \cdot T} = x^{\, \s \a \cdot d} T_\a \sp
\forall \, x , \a \in \Delta (g)
\end{equation}
 and one also finds the relations
\begin{subequations}
\begin{equation}
\hat W_\m (T,z,\s) = U^{-1} (T,z,\s) ( \hat W_\mu (T,z,0) + \partial_\m) U(T,z,\s)
\end{equation}
\begin{equation}
\hat{\bar{W}}_\m (T,\bz,\s) = \bar U (T,\bz,\s) ( \hat{\bar{W}}_\mu (T,\bz,0) -
 \bar \partial_\m) \bar U^{-1}(T,\bz,\s)
\end{equation}
\begin{equation}
U(T,z,\s) Q_\a (z,\s) U^{-1} (T,z,\s) = \bar U^{-1} (T,\bz,\s) \bar Q_\a (\bz,\s)
\bar U(T,\bz,\s) = Q_\a (T,0)
\end{equation}
\end{subequations}
among the connections and generators of the Ward identities.

The equivalence relations \eqref{uteq} tell us that the twisted affine primary
fields have the simple form
\begin{equation}
\label{twap}
\hgp (T,z,\s) = \hgp (T,z,0) z^{\, \s d \cdot T} \sp
\hgm (T,\bz,\s) = \bz^{\, \s d \cdot T} \hgm (T,\bz,0)
\end{equation}
in terms of the untwisted affine primary fields at $\s =0$. As a consequence
of \eqref{uteq} and \eqref{twap}, we also obtain
the relations
\begin{subequations}
\begin{equation}
\hat A (T,\bz,z,\s) = \bar U (T,\bz,\s) \hat A(T,\bz, z , 0) U (T,z,\s)
\end{equation}
\begin{equation}
\label{ncin}
\hg (T,\bz,z,\s) = \bz^{\, \s d \cdot T} \hg (T,\bz,z,0) z^{\, \s d \cdot T}
\end{equation}
\end{subequations}
for the full orbifold correlators $\hat A = \hat A_- \hat A_+$ and primary
fields $\hg = \hgm \hgp$.

As a check on these results, we may use the classical limit of \eqref{ncin}
 to compute the classical monodromies  (see Subsec.~\ref{clarg}) of the
 group orbifold elements in this case
 \begin{subequations}
 \label{inmon}
\begin{equation}
\hg (T,\bz e^{- 2 \pi i}, z e ^{2 \pi i},\s )
= e^{-2 \pi i \s d \cdot T } \hg (T,\bz,z,\s) e^{2 \pi i \s d \cdot T }
\end{equation}
\begin{equation}
\hg (T,\bz e^{- 2 \pi i}, z e ^{2 \pi i}, \s )_\l{}^{\l '}
= e^{- 2 \pi i \s d \cdot (\l - \l')} \hg (T,\bz,z,\s)_\l{}^{\l '}
\end{equation}
\end{subequations}
where we have used \eqref{rep-matrices-of-inner-auto} and the fact that the
classical group elements $\hg (T,\bz,z,0)$
 have trivial monodromy. Using the inner-automorphic
data \eqref{innerdata}, we see that the monodromies \eqref{inmon}
 are in agreement with the general formula \eqref{clmon} for the monodromies of
 the group orbifold elements.

We finally give the description of the left and right mover
inner-automorphically twisted current algebra in terms of spectral flow.
In our formulation the correct flows are
\begin{subequations}
\label{specflow}
\begin{equation}
\label{specflowa}
\hat H_A(m) = H_A (m) + \delta_{m,0} k \s d_A
\sp
\hat E_\a (m - \s \a \cdot d) = E_\a (m)
\end{equation}
\begin{equation}
\label{specflowb}
\hat{\bar{H}}_A(m) = \bar{H}_A (-m) - \delta_{m,0} k \s d_A
\sp
\hat{\bar{E}}_\a (m - \s \a \cdot d) = \bar{E}_\a  (-m)
\end{equation}
\end{subequations}
where $\{J,\bar{J} \}$ are the ordinary untwisted modes of affine $(g \oplus g)$
(which satisfy \eqref{jjalg}, \eqref{jbjbalg}, \eqref{jjbalg} and \eqref{Jgut}).
The left mover flow in \eqref{specflowa} is well known
\cite{Kac:1984mq,Goddard:1986bp,Freericks:1988zg,Lerche:1989uy,Halpern:2000vj},
while the right mover flow in \eqref{specflowb} is obtained from
\eqref{vacinc} and comparison of \eqref{Jgut} with \eqref{comHEgr} at $\s = 0$.
The right mover flow \eqref{specflowb} also gives the spectral flow
\begin{subequations}
\label{chev}
\begin{equation}
\hHbb_A(m) = \bar{H}_A (m)' + \delta_{m,0} k \s d_A
\sp
\hEbb_\a (m - \s \a \cdot d) = \bar{E}_\a  (m)'
\end{equation}
\begin{equation}
\label{chevb}
 \bar{H}_A (m)'  \equiv - \bar H_A (m) \sp
 \bar{E}_\a  (m)' \equiv  - \bar{E}_{-\a}  (m)
\end{equation}
\end{subequations}
for the rectified right mover modes in \eqref{hh,he-rightmover}.
Relative to the left mover flow \eqref{specflowa}, the extra
Chevalley involution in \eqref{chevb} is the same as that
discussed near \eqref{hh,he-rightmover}.

The spectral flows \eqref{specflow} and \eqref{chev} are
algebraic results which are independent of our provisional choice
of the untwisted affine vacuum $| 0 \rangle$ for the twisted
correlators. Moreover, given the flows \eqref{specflow} and
\eqref{chev}, we find that the simple operator results
\eqref{twap} can be obtained directly by comparison of
\eqref{Jgut} with \eqref{comHEg}, \eqref{comHEgr}. It follows
that the simple operator results \eqref{twap} and \eqref{ncin}
are also independent of our provisional choice of the untwisted
affine vacuum.

\vskip 1cm
\noindent {\bf Acknowledgements}

For helpful discussions, we thank R. Dijkgraaf, J. Evslin, R. Littlejohn,
C. Park, M. Porrati,
C. Schweigert, H. Verlinde, J. Wang and N. Warner. For hospitality, MBH and NO thank the
Niels Bohr Institute, and JdB and NO
thank the University of California, Berkeley and Lawrence Berkeley National
Laboratory.
The work of JdB  was supported in part by the
National Science Foundation under
Grant No PHY99-07949.
The work of MBH was supported in part by the Director, Office of Energy Research,
Office of High Energy and Nuclear Physics, Division of High Energy Physics of
the U.S. Department of Energy under Contract DE-AC03-76SF00098 and in part by
the National Science Foundation under grant PHY95-14797.
The work of NO is supported in part by the stichting FOM and
the European Commission RTN programme
HPRN-CT-2000-00131.

\appendix

\section{New Braid Relations \label{braid} }

The braid relations
\begin{equation}
 W\hc(h_\s;T)T_\nrm(\s) = e^{-2\pi i\frac{n(r)}{\r(\s)}} T_\nrm(\s)
 W\hc(h_\s;T) ,\; T_\nrm(\s) \!\equiv\! \schisig_\nrm U(\s)_\nrm{}^a T_a
\end{equation}
follow easily (start with \mbox{$W\hc T_\nrm W$}) from the linkage relation
\eqref{WTWwTa} and the $H$-eigenvalue problem \eqref{HEigen}. It is expected that
the linkage relation leads to other connections of this type between the
$H$-eigenvalue problem and its extended form.

\section{A Criterion for Rectification \label{recpro}}

In Subsecs.~\ref{cyclorb}, \ref{snorb}, \ref{unrec} and  \ref{innerprop}, we have argued
that twisted right mover current algebras can be rectified for each sector of
all permutation orbifolds and all inner-automorphic orbifolds. In this appendix,  we find
a group-theoretic sufficient condition for right mover rectification in any
sector $\s$ of any current-algebraic  orbifold.

We will present our argument in terms of the currents $\sjh$, $\sjbh$
with twisted boundary conditions \cite{deBoer:1999na}
\begin{equation}
\label{twbcmon}
 \sjh_a(z e^{2\pi i}) = w(h_\s)_a{}^{b}\sjh_b(z), \quad \sjbh_a(\bz e^{-2\pi i})
  = w(h_\s)_a{}^{b}\sjbh_b(\bz) \ .
\end{equation}
These are mixed monodromy objects which are related to the untwisted currents
by local isomorphisms,
\begin{subequations}
\label{twbcope}
\begin{equation}
 \sjh_a(z)\sjh_b(w) = \frac{G_{ab}}{(z-w)^2} \+
 \frac{if_{ab}{}^{c}}{z-w}\sjh_c(w) \+
 \Ord(z-w)^0
 \end{equation}
 \begin{equation}
 \sjbh\!_a(\bz) \sjbh\!_b(\bw) = \frac{G_{ab}}{(\bz-\bw)^2} \+
     \frac{if_{ab}{}^{c}}{\bz-\bw}\sjbh\!_c(\bw) \+
\Ord(\bz-\bw)^0
\end{equation}
\end{subequations}
as shown in the commuting diagram \cite{deBoer:1999na} of Fig.~1.

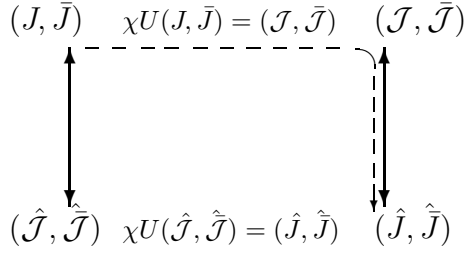
\begin{figure}[h]
\begin{picture}(248,188)(0,0)
\put(93,165){$(J,\bJ)$}
\put(122,158){\line(1,0){5}}
\put(143,158){\line(1,0){5}}
\put(163,158){\line(1,0){5}}
\put(183,158){\line(1,0){5}}
\put(203,158){\line(1,0){5}}
\put(132,158){\line(1,0){5}}
\put(152,158){\line(1,0){5}}
\put(173,158){\line(1,0){5}}
\put(193,158){\line(1,0){5}}
\put(213,158){\line(1,0){5}}
\put(224,150){\oval(15,15)[tr]}
\put(231,147){\line(0,-1){5}}
\put(231,139){\line(0,-1){5}}
\put(231,131){\line(0,-1){5}}
\put(231,123){\line(0,-1){5}}
\put(231,114){\line(0,-1){5}}
\put(231,106){\vector(0,-1){10}}
\put(136,165){$\foot{\schi U (J,\bJ) = (\sj,\sjb)}$}
\thicklines
\put(116,158){\vector(0,-1){60}}
\put(116,98){\vector(0,1){60}}
\put(93,86){$(\sjh,\sjbh)$}
\put(231,165){$(\sj,\sjb)$}
\put(136,86){$\foot{\schi U (\sjh,\sjbh) =(\hj,\hjb)} $}
\put(235,158){\vector(0,-1){60}}
\put(235,98){\vector(0,1){60}}
\put(231,86){$(\hj,\hjb)$}
\put(70,70) {{\footnotesize Each vertical double arrow is a local isomorphism }}
\put(45,59) {$\foot{J,\bJ}$}
\put(70,59) {{\footnotesize = currents: trivial monodromy, mixed under automorphisms}}
\put(45,47) {$\foot{\sj,\sjb}$}
\put(70,47) {{\footnotesize = eigencurrents: trivial monodromy, diagonal under automorphisms}}
\put(45,33) {$\foot{\hj,\hjb}$}
\put(70,33) {{\footnotesize = twisted currents with definite monodromy}}
\put(45,21) {$\foot{\sjh,\sjbh}$}
\put(70,21) {{\footnotesize = currents with twisted boundary conditions}}
\put(98,4) {Fig.1\,: Currents and orbifold currents}
\end{picture}
\vspace{.2in}
\end{figure}
\noindent The currents with twisted boundary conditions are related to
the twisted currents as follows
\begin{subequations}
\begin{equation}
 \sjh_a(z,\s) = U\hc(\s)_a{}^{\nrm}\schisig^{-1}_\nrm \hj_\nrm(z,\s),
 \; \sjbh_a(\bz,\s) = U\hc(\s)_a{}^{\nrm}\schisig^{-1}_\nrm \hjb_\nrm(\bz,\s)
 \end{equation}
 \begin{equation}
 \hj_\nrm(z,\s) = \schisig_\nrm U(\s)_\nrm{}^{a} \sjh_a(z,\s) ,\;
  \hjb_\nrm(\bz,\s) = \schisig_\nrm U(\s)_\nrm{}^{a} \sjbh_a(\bz,\s)
  \label{jhat=schiUsjhat}
\end{equation}
\end{subequations}
so that, according to \eqref{jhat=schiUsjhat}, the monodromy decompositions
of $\sjh$, $\sjbh$ give the twisted currents of the text.

In twisted sector $\s$,
the sufficient condition for rectification of the twisted right mover current
algebra is that we can find a  $g$ automorphism $A(\s)$
which relates the action
$w(h_\s^{-1})=w\hc(h_\s)$ to the action $w(h_\s)$:
\begin{equation}
 A\hc(\s)w\hc(h_\s)A(\s) = w(h_\s), \quad A\hc(\s)A(\s)=1, \quad A(\s) \in
 {\rm Aut}(g)
  \ . \label{Awhcw=w}
\end{equation}
Then, using the right mover forms of \eqref{twbcmon}, \eqref{twbcope}
and \eqref{Awhcw=w}, we find that the linear combinations $\sjbbh$ of
the right movers
\begin{subequations}
\begin{equation}
 \sjbbh_a(\bz) \equiv A(\s)_a{}^{b} \sjbh(\bz)_b
\end{equation}
\begin{equation}
 \sjbbh_a(\bz) \sjbbh\!_b(\bw) = \frac{G_{ab}}{(\bz-\bw)^2} +
 \frac{if_{ab}{}^{c}}{\bz-\bw}\sjbbh_c(\bw) + \Ord(\bz-\bw)^0  ,
 \quad \sjbbh_a(\bz e^{2\pi i}) = w(h_\s)_a{}^{b} \sjbbh_b(\bz)
\end{equation}
\end{subequations}
satisfy the same OPEs and twisted boundary conditions as the left movers
$\sjh_a(z)$. It follows that the monodromy decompositions and mode algebras of
both are identical.

\section{Summation Identities \label{sumid} }

The following summation identities are useful in evaluating exact OPEs
\begin{subequations}
\begin{equation}
\frac{1}{z} \left( \frac{w}{z} \right)^{\srac{n(r)}{\rho (\s)}}
\sum_m  \left( \frac{w}{z} \right)^{m}
\theta ( m + \srac{n(r)}{\rho (\s)} \geq 0)
=  \left( \frac{w}{z} \right)^{\srac{\bar n(r)}{\rho (\s)}}
\frac{1}{z-w}
\end{equation}
\begin{equation}
\frac{1}{z w } \left( \frac{w}{z} \right)^{\srac{n(r)}{\rho (\s)}}
\sum_m  (m + \srac{n(r)}{\rho (\s)}) \left( \frac{w}{z} \right)^{m}
\theta ( m + \srac{n(r)}{\rho (\s)} \geq 0)
=  \left( \frac{w}{z} \right)^{\srac{\bar n(r)}{\rho (\s)}}
\left\{ \frac{1}{(z-w)^2} + \frac{ \bar n(r)/\rho (\s)}{ w(z-w)}
\right\}\end{equation}
\begin{equation}
\frac{1}{\bz} \left( \frac{\bz}{\bw} \right)^{\srac{n(r)}{\rho (\s)}}
\sum_m  \left( \frac{\bz}{\bw} \right)^{m}
\theta ( m + \srac{n(r)}{\rho (\s)} \leq 0)
=  \left( \frac{\bw}{\bz} \right)^{\srac{\mnb}{\rho (\s)}}
\frac{1}{\bz-\bw}
\end{equation}
\begin{equation}
\frac{1}{\bz \bw } \left( \frac{\bz}{\bw} \right)^{\srac{n(r)}{\rho (\s)}}
\sum_m  (m + \srac{n(r)}{\rho (\s)}) \left( \frac{\bz}{\bw} \right)^{m}
\theta ( m + \srac{n(r)}{\rho (\s)} \leq 0)
=  -\left( \frac{\bw}{\bz} \right)^{\srac{\mnb}{\rho (\s)}} \!
\left\{ \frac{1}{(\bz-\bw)^2} + \frac{ \mnb/\rho (\s) }{ \bw(\bz-\bw)}\right\}
\end{equation}
\begin{equation}
\sum_r f( - n(r)) = \sum_r f ( n(r)) \sp \forall\, f\ .
\end{equation}
\end{subequations}
The first two relations hold for $|z| > |w |$ and the second two hold
for $ |\bz| > |\bw |$. For the first four relations,
we have used the change of variable
$m ' = m + \lfloor n(r)/\rho (\s) \rfloor $, where $\lfloor x \rfloor$ is the floor of $x$ (see
Ref.~\cite{Halpern:2000vj}).

\section{The General Current-Algebraic Orbifold ${\boldmath A(H)/H}$ \label{gcaob} }

The twisted left mover sectors of the general current-algebraic orbifold
\begin{equation}
\frac{A(H)}{H} \supset \frac{A_g(H)}{H} \sp H \subset {\rm Aut}(g)
\end{equation}
were studied in Refs.~\cite{deBoer:1999na,Halpern:2000vj}. Here, we supplement
that discussion by including the corresponding results for the twisted right
mover sectors.

The stress tensors of the general current-algebraic orbifold are:
\begin{subequations}
\label{gcao}
\begin{equation}
\hat T_\s (z) = {\cL}^{n(r) \mu ; - n(r),\nu} (\s) :
\hat J_{ n(r) \mu} (z) \hat J_{ -n(r), \nu} (z) : \hskip 7cm
\end{equation}
\begin{equation}
=\sum_{r,\mu,\nu} {\cL}^{n(r) \mu;-n(r), \nu}(\s)
\{ :\hat{J}_{n(r) \mu}(z) \hat{J}_{-n(r), \nu}(z):_M -
 \frac{i \bar{n}(r)}{\rho(\s)} \F_{n(r) \mu;-n(r), \nu}{}^{ 0 \delta}(\s)
 \frac{\hat{J}_{0 \delta}(z)}{z}
 \end{equation}
 $$
\hsp{1.3} + \frac{1}{z^2}   \frac{\bar{n}(r)}{2 \r(\s)}
(1-\frac{\bar{n}(r)}{\r(\s)}) \G_{n(r) \mu;-n(r), \nu}(\s) \}
$$
\begin{equation}
{\hat{\bar T}}_\s (\bz) = {\cL}^{n(r) \mu ; - n(r),\nu} (\s) :
{\hat{\bJ}}_{ n(r) \mu} (\bz) {\hat{\bJ}}_{ -n(r), \nu} (\bz) : \hskip 7cm
\end{equation}
\begin{equation}
\label{rtMord2}
=\sum_{r,\mu,\nu} {\cL}^{n(r) \mu;-n(r), \nu}(\s)
\{ :\hjb_{n(r) \mu}(\bz) \hjb_{-n(r), \nu}(\bz):_{\bM}
- \frac{i}{\bz} \frac{ \mnb }{\r(\s)}
\F_{n(r) \mu;-n(r), \nu}{}^{0 \delta}(\s) \hjb_{0 \delta}(\bz)
\end{equation}
$$
+ \frac{1}{\bz^2}
 \frac{\mnb }{2 \r(\s)}(1-\frac{\mnb }{\r(\s)})
 \G_{n(r) \mu;-n(r), \nu}(\s) \}
 $$
\begin{equation}
{\cal L}^{\nrm;\nsn}(\s) = \schisig^{-1}_\nrm \schisig^{-1}_\nsn L_H^{ab}
U\hc(\s)_a{}^\nrm U\hc(\s)_b{}^\nsn, \quad \srange \label{inv-inertia-tensor}
\end{equation}
\begin{equation}
= \de_{n(r) + n(s), 0 \rmod \,\rho (\s)} {\cal L}^{\nrm; \mnrn}(\s) \ .
\end{equation}
\end{subequations}
Here $M$ and $\bM$ normal ordering is defined in the text,
$L_H^{ab}$ is the $H$-invariant
 solution \cite{Halpern:1992gb,Halpern:1996js} of the Virasoro master
equation \cite{Halpern:1989ss,Morozov:1990uu,Halpern:1996js} corresponding to the
$H$-invariant CFT $A(H)$,
and $\{ {\cL} (\s) \} $ is the set of twisted inverse inertia tensors
\cite{deBoer:1999na,Halpern:2000vj} of the orbifold $A(H)/H$.

We also give the OPEs of the general stress tensors with the twisted currents
\begin{subequations}
\label{genTJ}
\begin{equation}
\label{genTJl}
  \hat T_\s(z)\hj_\nrm(w) = \sm_\nrm{}^\nrm(\s)[\frac{1}{(z-w)^2}+
  \frac{\pl_w}{(z-w)}]\hj_\nrm(w)\hskip 2cm
  \end{equation}
  $$
\hskip 2cm
+\frac{{\cal N}_\nrm{}^{\nsn;n(r)-n(s),\d} (\s) :\hj_\nsn(w)\hj_{n(r)-n(s),\d}(w):}{z-w}
               +\Ord(z-w)^0
$$
\begin{equation}
  \hat{\bar{T}}_\s(\bz)\hjb_\nrm(\bw) = \sm_\nrm{}^\nrm(\s)[\frac{1}{(\bz-\bw)^2}+
  \frac{\pl_{\bw}}{(\bz-\bw)}]\hjb_\nrm(\bw) \hskip 2cm
 \end{equation}
 $$
\hskip 2cm
+\frac{{\cal N}_\nrm{}^{\nsn;n(r)-n(s),\d} (\s) :\hjb_\nsn(\bw)\hjb_{n(r)-n(s),\d}(\bw):}{\bz-\bw}
               +\Ord (\bz-\bw)^0
$$
\end{subequations}
where \eqref{genTJl} and
the twisted tensors ${\cal{M}} (\s) $ and ${\cal{N}} (\s)$ are given
as duality transformations in \cite{Halpern:2000vj}. For the special case of
WZW orbifolds $A_g(H)/H$, one finds \cite{Halpern:2000vj}
\begin{equation}
\sm_\nrm{}^\nsn(\s)=\d_\nrm{}^\nsn, \quad {\cal N}_\nrm{}^{\nsn;\ntd}(\s) =0
\sp \s = 0 , \ldots , N_c -1
\end{equation}
so Eq.~\eqref{genTJ} reduces to the results given in \eqref{orbTJ}.

We finally comment on the general permutation orbifold $A(H)/H$,
$H ({\rm permutation}) \subset {\rm Aut}(g)$. In this case, we may
use \eqref{gcao} and the ground state conditions
\eqref{vacc}  to compute  the ground state conformal weights
(conformal weights of the twist fields)
\begin{subequations}
\begin{equation}
L_\s (m \geq 0) | 0 \rangle_\s = \delta_{m,0} \hat{\Delta}_0 (\s)
| 0 \rangle_\s
\end{equation}
\begin{equation}
\bar{L}_\s (m \geq 0) | 0 \rangle_\s = \delta_{m,0} \hat{\bar{\Delta}}_0
(\s) | 0 \rangle_\s
\end{equation}
\begin{eqnarray}
\hat{\bar{\Delta}}_0 (\s) & = &  {\cL}^{n(r)aj; -n(r),bl} (\s) \
\frac{\mnb}{2 \rho (\s)} \left( 1 - \frac{\mnb}{\rho (\s)} \right)
\G_{n(r)a j , -n(r), bl} (\s) \nn \\
&= & {\cL}^{n(r)aj; -n(r), bl} (\s) \
\frac{\bar n(r)}{2 \rho (\s)} \left( 1 - \frac{\bar n(r)}{\rho (\s)} \right)
\G_{n(r)a j; -n(r),b l} (\s)
=\hat \Delta_0 (\s) \label{gpocw}
\end{eqnarray}
\end{subequations}
where we have also used Eq.~\eqref{viriden} to complete \eqref{gpocw}. A more
explicit expression for $\hat \Delta_0 (\s)$ is given in
Ref.~\cite{Halpern:2000vj}. The equality \eqref{gpocw}
of the right and left mover ground
state conformal weights is a consistency check on the form
\eqref{mode-current-algebra} of the general twisted current algebra.
We further expect (but have not yet tried to prove) that the Virasoro
generators $L_\s (m)$, $\bar L_\s (m)$ of the general permutation orbifolds are
copies when $\bar L_\s (m)$ is expressed in terms of the rectified right
mover current modes in Eq.~\eqref{recjperm0}.

\section{The Factorized Form of $\T$(permutation) \label{Jperm}}

In this appendix, we work out further properties of the twisted representation
matrices $\T$ of the general permutation orbifold. In this development,
we suppress the sector label $\s$ and we
follow the convention \cite{deBoer:1999na,Evslin:1999ve,Halpern:2000vj}
that the normalization constants $\chi_{n(r) aj} = \chi_{n(r) j}$ are chosen
independent of $a$.

Then we find that the twisted representation matrices of all permutation orbifolds
can be written in the factorized form
\begin{subequations}
\begin{equation}
\T_{n(r) aj } = T_{a} \otimes t_{n(r) j} \equiv T_{a} t_{n(r) j}
\sp [T_a , T_b ] = i f_{ab}{}^c
T_c\sp t_{n(r) \pm \rho (\s),j} = t_{n(r)j}
\end{equation}
\begin{equation}
\label{tperme}
(t_{n(r) j})_{n(s)l}{}^{n(t) m} \equiv \chi_{n(r) j} \sum_I
U_{n(r)j}{}^I U_{n(s)l}{}^I (U^\dagger)_I{}^{n(t) m}
\end{equation}
\begin{equation}
\label{tpermc}
(t_{n(r) j} t_{n(s) l}) _{n(t)p}{}^{n(u) q}
=\chi_{n(r) j} \chi_{n(s) l} \sum_I
U_{n(r)j}{}^I U_{n(s)l}{}^I   U_{n(t)p}{}^I  (U^\dagger)_I{}^{n(u) q}
\end{equation}
\begin{equation}
\label{tcom}
[t_{n(r) j}, t_{n(s) l} ] = 0
\end{equation}
\end{subequations}
where $T$ is a matrix irrep of $\gbn$, \eqref{tperme} is equivalent to
\eqref{repmatperm}, \eqref{tpermc} follows from \eqref{tperme}, and
\eqref{tcom} follows from \eqref{tpermc}.
Examples of the matrices $t$ can be read from \
Eqs.~\eqref{st-in-AZl} and \eqref{st-in-SN}, and the factorized form is used
explicitly in Subsec.~\ref{cyclfl}.

To learn more about the matrices $t$, we recall the algebra of $\T$
\begin{subequations}
\label{E2}
\begin{equation}
\label{fper0}
[ \T_{n(r)a j} ,  \T_{n(s)b l }] = i \scf_{n (r) aj ;n(s) b l}
 {}^{n(r)+n(s),cm}  \T_{n(r)+n(s),cm} ,
\end{equation}
\begin{equation}
\label{fper1}
\F_{n (r) aj ;n(s) b l } {}^{n(r)+n(s),cm} = f_{ab}{}^c
\theta_{n(r)j;n(s)l}{}^{n(r)+n(s),m}
\end{equation}
\begin{equation}
\label{fper2}
\theta_{n(r)j;n(s)l}{}^{n(r)+n(s),m} =
\theta_{n(s)l;n(r)j}{}^{n(r)+n(s),m} \hskip 4cm
\end{equation}
\begin{equation}
\label{fper3}
\hskip 4cm \equiv
\chi_{n(r)j} \chi_{n(s)l} \chi^{-1}_{n(r)+n(s),m} \sum_I
U_{n(r)j}{}^I U_{n(s)l}{}^I (U^\dagger)_I{}^{n(r)+n(s),m}
\end{equation}
\end{subequations}
where \eqref{fper1}, \eqref{fper2} and \eqref{fper3} follow from \eqref{fper0}
and \eqref{scperm}.
Comparing \eqref{E2} with the factorized form $T t$ of $\T$, we find the
multiplication law for $t$:
\begin{equation}
\label{multlaw}
t_{n(r) j} t_{n(s) l} = \theta_{n(r)j;n(s)l}{}^{n(r)+n(s),m}
t_{n(r)+n(s),m } \ .
\end{equation}
Then we find for example that
\begin{subequations}
\label{idjperm}
\begin{equation}
\label{idjperm1}
[\T_{n(r)a j} ,\eta^{bc}  \T_{n(s)b l }   \T_{-n(s),cm } ] = 0 \sp \forall \; n(s)
\end{equation}
\begin{equation}
[\eta^{ab}  \T^{(\nu)}_{n(r)a j }   \T^{(\mu)}_{-n(r),b l },
\eta^{cd}  \T^{(\nu)}_{n(s)c m }   \T^{(\mu)}_{-n(s), d n } ] = 0
\sp \nu \neq \mu \sp \forall \; n(r), n(s) \ .
\end{equation}
\end{subequations}
are implied by \eqref{multlaw} and the algebra of $T$. These results tell us
that all the matrix structures in \eqref{tpcon} commute.


For permutation orbifolds (see Subsecs.~\ref{cyclorb}, \ref{snorb} and
\ref{unrec})) we can choose the normalization constants $\chi_{n(r) j}$ such that
\begin{equation}
\F_{n(r)a j;n(s)b l }{}^{n(t)cm} (\s)  = f_{ab}{}^c \delta_{jl}\delta_{l}^m
\delta_{n(r) + n(s) - n(t), 0 \rmod \rho(\s) }
\end{equation}
where  $j,l$ and $m$ are the semisimplicity indices of the general
orbifold affine algebra \eqref{coalg}. In this basis, one finds from \eqref{fper1} that
\begin{subequations}
\label{theta}
\begin{equation}
\label{trep0}
\theta_{n(r) j; n(s)l}{}^{n(r) + n(s),m} = \delta_{jl}
\delta_{l}^m \quad \rightarrow \quad
t_{n(r) j} t_{n (s) l } = \de_{jl} t_{n(r) +n (s),j} \ .
\end{equation}
\begin{equation}
\label{trep}
(t_{n(r)j})_{n(s)l}{}^{n(t)m} = \de_{jl} \de_l^m
\de_{n(r)+n(s)-n(t) ,0 \,\rmod \rho(\s)} \ .
\end{equation}
\end{subequations}
The explicit representation of $t$ in \eqref{trep} follows from \eqref{trep0},
and the result \eqref{theta} is stated in Eq.~\eqref{Tcycl0}.
As an application of \eqref{theta} we compute for all $n(r)$
\begin{equation}
\label{idlast}
[\eta^{ab}\sum_j  \T^{(2)}_{n(r)a j }   \T^{(1)}_{-n(r),b j },
\T^{(1)}_{0cl}  + \T^{(2)}_{0cl} ]
\end{equation}
$$
= i \eta^{ab} T_a^{(2)} f_{bc}{}^d T_d^{(1)}
\sum_j ( t_{n(r)j}^{(2)}  t_{-n(r),j}^{(1)}t_{0l}^{(1)}-
 t_{n(r)j}^{(2)}  t_{-n(r),j}^{(1)}t_{0l}^{(2)} )
$$
$$
= i T_a^{(2)} f^{a} {}_c{}^d T_d^{(1)} t_{n(r)l}^{(2)}  t_{-n(r),l}^{(1)}
(\theta_{-n(r), l; 0l}{}^{-n(r),l }-\theta_{n(r) l; 0l}{}^{n(r)l }
) = 0 \ .
$$
Generalization of this identity leads to the result \eqref{wTcom} of the text.

\section{The Integrals $I_{\bar n/\rho}$ \label{intapp}}

In this appendix, we evaluate the integrals $I_{\srac{\bar n}{\rho}}$
which are encountered for the permutation orbifolds in Sec.~\ref{correl}.
The integrals are defined up to additive constants as
\begin{subequations}
\begin{equation}
y \equiv \frac{z_1}{z_2} \sp \bar n \in \{ 0 , \ldots , \rho -1 \}
\end{equation}
\begin{equation}
\label{idef}
I_{\srac{\bar n}{\rho}}(y) \simeq \int^y \frac{d x}{x-1} x^{-
\srac{\bar n}{\rho}} \simeq \rho \int^{y^{1/\rho}} \frac{dz}{z^\rho
-1} z^{\rho - 1 - \bar n} \simeq \rho
\int^{y^{-1/\rho}} \frac{dz}{z^\rho
-1} z^{\bar n- 1}
\end{equation}
\begin{equation}
\partial_1 I_{\srac{\bar n}{\rho}}(\srac{z_1}{z_2}) = \frac{1}{z_{12}} \left(
\frac{z_2}{z_1} \right)^{\srac{\bar n}{\rho}} \sp
\partial_2 I_{\srac{\bar n}{\rho}} (
\srac{z_1}{z_2} )= \frac{1}{z_{21}} \left(
\frac{z_1}{z_2} \right)^{1-\srac{\bar n}{\rho}}
\end{equation}
\begin{equation}
\label{isym}
I_{\srac{\bar n}{\rho}} (y)\simeq I_{\srac{\rho -\bar n}{\rho}} (y^{-1})
\end{equation}
\end{subequations}
where the symmetry relation \eqref{isym} follows by comparing the last two
integral representations in \eqref{idef}.

The $z$ integrals can be simplified by the identities
\begin{subequations}
\begin{equation}
z^\rho -1 = \prod_{r=0}^{\rho-1} (z - z_{r/\rho})
\sp
z_{r/\rho} \equiv e^{2 \pi i \srac{r}{\rho}} \ .
\end{equation}
\begin{equation}
\frac{1}{z^\rho -1 } = \sum_{r=0}^{\rho-1} \frac{a_r(\rho)}{z - z_{r/\rho} }
\sp a_r (\rho) = \left( \prod_{s \neq r} (z_{r/\rho} - z_{s/\rho} ) \right)^{-1}
\end{equation}
\end{subequations}
so that we obtain for the first $z$ integral
\begin{equation}
I_{\srac{\bar n}{\rho}} (y) \simeq \rho
\sum_{r=0}^{\rho -1} a_r (\rho) \int^{y^{1/\rho}-z_{r/\rho}}
\frac{du}{u} (u + z_{r/\rho} )^{\rho - 1 -\bar n}
\end{equation}
after a change of variable to $u$. This can be evaluated via
the binomial theorem, with the result
\begin{equation}
I_{\srac{\bar n}{\rho}} (y) \simeq \rho \sum_{r=0}^{\rho -1} a_r(\rho)
(z_{r/\rho})^{\rho -1 - \bar n}
\left[
\ln ( y^{\srac{1}{\rho}} -z_{r/\rho} )
+ \sum_{l=1}^{\rho - 1 - \bar n  } \frac{1}{l}
\Big( \smal{ \begin{array}{c} \rho -1 - \bar n  \\ l \end{array} } \Big)
(  y^{\frac{1}{\rho}}  z_{r/\rho}^{-1} - 1 )^l \right] \ .
\end{equation}
This form is simplest for $\bar n$ near $\rho-1$, and \eqref{isym} gives the
corresponding results for low $\bar n$. As examples we list the results
\begin{subequations}
\begin{eqnarray}
& e^{I_0 (y)} &  \propto \;  y-1 \\
&  e^{I_{\srac{\rho -1}{\rho}} (y) }  & \propto \;
e^{I_{\srac{1}{\rho}}(y^{-1})} \propto \;
\prod_{r=0}^{\rho -1} ( y^{\srac{1}{\rho}} -
z_{r/\rho} )^{\rho a_r (\rho)} \\
& e^{I_{\srac{1}{2}} (y) } & \propto \;  e^{I_{\srac{1}{2}} (y^{-1}) }
\propto \; \frac{y^{\frac{1}{2}} -1}{y^{\frac{1}{2}} + 1} \\
& e^{I_{\srac{\rho -2 }{\rho}} (y)} & \propto \;
e^{I_{\srac{2 }{\rho}} (y^{-1})} \propto \;
  \prod_{r=0}^{\rho -1} ( y^{\srac{1}{\rho}} -z_{r/\rho}
)^{\rho a_r (\rho) z_{r/\rho}} \ e^{\rho a_r (\rho)
(y^{1/\rho} - z_{r/\rho} ) }
\end{eqnarray}
\end{subequations}
which are given up to multiplicative constants.

\vskip .5cm
\addcontentsline{toc}{section}{References}

\renewcommand{\baselinestretch}{.4}\rm
{\footnotesize


\begin{thebibliography}{10}

\bibitem{Kac:1967}
V.~Kac, ``Simple graded {Lie} algebras of finite growth,'' {\em Funct. Anal.
  Appl.} {\bf 1} (1967) 328.

\bibitem{Moody:1967gf}
R.~V. Moody, ``Lie algebras associated with generalized {Cartan} matrices,''
  {\em Bull. Am. Math. Soc.} {\bf 73} (1967) 217--221.

\bibitem{Bardakci:1971nb}
K.~Bardakci and M.~B. Halpern, ``New dual quark models,'' {\em Phys. Rev.} {\bf
  D3} (1971) 2493.

\bibitem{Halpern:1996js}
M.~B. Halpern, E.~Kiritsis, N.~A. Obers, and K.~Clubok, ``Irrational conformal
  field theory,'' {\em Phys. Rept.} {\bf 265} (1996) 1--138,
  \href{http://xxx.lanl.gov/abs/hep-th/9501144}{{\tt hep-th/9501144}}.

\bibitem{Halpern:1989ss}
M.~B. Halpern and E.~Kiritsis, ``General {V}irasoro construction on affine
  {$g$},'' {\em Mod. Phys. Lett.} {\bf A4} (1989) 1373.

\bibitem{Morozov:1990uu}
A.~Y. Morozov, A.~M. Perelomov, A.~A. Roslyi, M.~A. Shifman, and A.~V.
  Turbiner, ``Quasiexactly solvable quantal problems: One-dimensional analog of
  rational conformal field theories,'' {\em Int. J. Mod. Phys.} {\bf A5} (1990)
  803.

\bibitem{Halpern:1990zy}
M.~B. Halpern, E.~Kiritsis, N.~A. Obers, M.~Porrati, and J.~P. Yamron, ``New
  unitary affine -{V}irasoro constructions,'' {\em Int. J. Mod. Phys.} {\bf A5}
  (1990) 2275.

\bibitem{Halpern:1971ay}
M.~B. Halpern, ``The two faces of a dual pion-quark model,'' {\em Phys. Rev.}
  {\bf D4} (1971) 2398.

\bibitem{Dashen:1975hp}
R.~Dashen and Y.~Frishman, ``Four fermion interactions and scale invariance,''
  {\em Phys. Rev.} {\bf D11} (1975) 2781.

\bibitem{Knizhnik:1984nr}
V.~G. Knizhnik and A.~B. Zamolodchikov, ``Current algebra and {Wess--Zumino}
  model in two dimensions,'' {\em Nucl. Phys.} {\bf B247} (1984) 83--103.

\bibitem{Segal}
G.~Segal {\em unpublished}.

\bibitem{Goddard:1985vk}
P.~Goddard, A.~Kent, and D.~Olive, ``Virasoro algebras and coset space
  models,'' {\em Phys. Lett.} {\bf B152} (1985) 88.

\bibitem{Halpern:1992gb}
M.~B. Halpern, E.~B. Kiritsis, and N.~A. Obers, ``The {Lie} {$h$} invariant
  conformal field theories and the {Lie} {$h$} invariant graphs,'' {\em Int. J.
  Mod. Phys.} {\bf A7} (1992) s339,
  \href{http://xxx.lanl.gov/abs/hep-th/9110001}{{\tt hep-th/9110001}}.

\bibitem{Halpern:1971qj}
M.~B. Halpern and C.~B. Thorn, ``Two faces of a dual pion-quark model. 2.
  {Fermions} and other things,'' {\em Phys. Rev.} {\bf D4} (1971) 3084--3088.

\bibitem{Corrigan:1975sn}
E.~Corrigan and D.~B. Fairlie, ``Off-shell states in dual resonance theory,''
  {\em Nucl. Phys.} {\bf B91} (1975) 527.

\bibitem{Lepowsky:1978jk}
J.~Lepowsky and R.~L. Wilson, ``Construction of the affine {L}ie algebra
  {A1(1)},'' {\em Commun. Math. Phys.} {\bf 62} (1978) 43--53.

\bibitem{Frenkel:1984}
I.~B. Frenkel, J.~Lepowsky, and A.~Meurman, {\em Proc. Nat. Acad. Sci.} {\bf 81}
  (1984) 3256.

\bibitem{Kac:1984mq}
V.~G. Kac and D.~H. Peterson, ``Infinite dimensional {Lie} algebras, theta
  functions and modular forms,'' {\em Adv. Math.} {\bf 53} (1984) 125--264.

\bibitem{Dixon:1985jw}
L.~Dixon, J.~A. Harvey, C.~Vafa, and E.~Witten, ``Strings on orbifolds,'' {\em
  Nucl. Phys.} {\bf B261} (1985) 678--686.

\bibitem{Dixon:1986jc}
L.~Dixon, J.~A. Harvey, C.~Vafa, and E.~Witten, ``Strings on orbifolds. 2,''
  {\em Nucl. Phys.} {\bf B274} (1986) 285--314.

\bibitem{Dixon:1987qv}
L.~Dixon, D.~Friedan, E.~Martinec, and S.~Shenker, ``The conformal field theory
  of orbifolds,'' {\em Nucl. Phys.} {\bf B282} (1987) 13--73.

\bibitem{Hamidi:1987vh}
S.~Hamidi and C.~Vafa, ``Interactions on orbifolds,'' {\em Nucl. Phys.} {\bf
  B279} (1987) 465.

\bibitem{Dixon:1988qd}
L.~Dixon, P.~Ginsparg, and J.~A. Harvey, ``Beauty and the beast: Superconformal
  symmetry in a monster module,'' {\em Commun. Math. Phys.} {\bf 119} (1988)
  221--241.

\bibitem{Freericks:1988zg}
J.~K. Freericks and M.~B. Halpern, ``Conformal deformation by the currents of
  affine {$g$},'' {\em Ann. Phys.} {\bf 188} (1988) 258.

\bibitem{Dijkgraaf:1989hb}
R.~Dijkgraaf, C.~Vafa, E.~Verlinde, and H.~Verlinde, ``The operator algebra of
  orbifold models,'' {\em Commun. Math. Phys.} {\bf 123} (1989) 485.

\bibitem{Klemm:1990df}
A.~Klemm and M.~G. Schmidt, ``Orbifolds by cyclic permutations of tensor
  product conformal field theories,'' {\em Phys. Lett.} {\bf B245} (1990)
  53--58.

\bibitem{Fuchs:1992vu}
J.~Fuchs, A.~Klemm, and M.~G. Schmidt, ``Orbifolds by cyclic permutations in
  {Gepner} type superstrings and in the corresponding {Calabi--Yau}
  manifolds,'' {\em Ann. Phys.} {\bf 214} (1992) 221--257.

\bibitem{Dijkgraaf:1997xw}
R.~Dijkgraaf, G.~Moore, E.~Verlinde, and H.~Verlinde, ``Elliptic genera of
  symmetric products and second quantized strings,'' {\em Commun. Math. Phys.}
  {\bf 185} (1997) 197--209, \href{http://xxx.lanl.gov/abs/hep-th/9608096}{{\tt
  hep-th/9608096}}.

\bibitem{Kac:1997nq}
V.~G. Kac and I.~T. Todorov, ``Affine orbifolds and rational conformal field
  theory extensions of w(1+infinity),'' {\em Commun. Math. Phys.} {\bf 190}
  (1997) 57--111, \href{http://xxx.lanl.gov/abs/hep-th/9612078}{{\tt
  hep-th/9612078}}.

\bibitem{Birke:1999ik}
L.~Birke, J.~Fuchs, and C.~Schweigert, ``Symmetry breaking boundary conditions
  and {WZW} orbifolds,'' {\em Adv. Theor. Math. Phys.} {\bf 3} (1999) 671--726,
  \href{http://xxx.lanl.gov/abs/hep-th/9905038}{{\tt hep-th/9905038}}.

\bibitem{deBoer:1999na}
J.~de~Boer, J.~Evslin, M.~B. Halpern, and J.~E. Wang, ``New duality
  transformations in orbifold theory,'' {\em Int. J. Mod. Phys.} {\bf A15}
  (2000) 1297, \href{http://xxx.lanl.gov/abs/hep-th/9908187}{{\tt
  hep-th/9908187}}.

\bibitem{Halpern:2000vj}
M.~B. Halpern and J.~E. Wang, ``More about all current-algebraic orbifolds,''
  {\em Int. J. Mod. Phys.} {\bf A16} (2001) 97,
  \href{http://xxx.lanl.gov/abs/hep-th/0005187}{{\tt hep-th/0005187}}.

\bibitem{Borisov:1997nc}
L.~Borisov, M.~B. Halpern, and C.~Schweigert, ``Systematic approach to cyclic
  orbifolds,'' {\em Int. J. Mod. Phys.} {\bf A13} (1998) 125,
  \href{http://xxx.lanl.gov/abs/hep-th/9701061}{{\tt hep-th/9701061}}.

\bibitem{Evslin:1999qb}
J.~Evslin, M.~B. Halpern, and J.~E. Wang, ``General {Virasoro} construction on
  orbifold affine algebra,'' {\em Int. J. Mod. Phys.} {\bf A14} (1999) 4985,
  \href{http://xxx.lanl.gov/abs/hep-th/9904105}{{\tt hep-th/9904105}}.

\bibitem{Evslin:1999ve}
J.~Evslin, M.~B. Halpern, and J.~E. Wang, ``Cyclic coset orbifolds,'' {\em Int.
  J. Mod. Phys.} {\bf A15} (2000) 3829--3860,
  \href{http://xxx.lanl.gov/abs/hep-th/9912084}{{\tt hep-th/9912084}}.

\bibitem{Halpern:1996et}
M.~B. Halpern and N.~A. Obers, ``New semiclassical nonabelian vertex operators
  for chiral and nonchiral {WZW} theory,'' {\em Int. J. Mod. Phys.} {\bf A12}
  (1997) 4317, \href{http://xxx.lanl.gov/abs/hep-th/9610081}{{\tt
  hep-th/9610081}}.

\bibitem{Novikov:1982ei}
S.~P. Novikov, ``The {Hamiltonian} formalism and a many valued analog of
  {Morse} theory,'' {\em Usp. Mat. Nauk} {\bf 37} (1982) 3--49.

\bibitem{Witten:1984ar}
E.~Witten, ``Nonabelian bosonization in two dimensions,'' {\em Commun. Math.
  Phys.} {\bf 92} (1984) 455--472.

\bibitem{Bantay:1998ek}
P.~Bantay, ``Characters and modular properties of permutation orbifolds,'' {\em
  Phys. Lett.} {\bf B419} (1998) 175--178,
  \href{http://xxx.lanl.gov/abs/hep-th/9708120}{{\tt hep-th/9708120}}.

\bibitem{Humphreys:1972}
J.~E. Humphreys, {\em Introduction to {Lie} algebras and representation
  theory}.
\newline \newblock Springer-Verlag, 1972.

\bibitem{Goddard:1986bp}
P.~Goddard and D.~Olive, ``Kac--{M}oody and {V}irasoro algebras in relation to
  quantum physics,'' {\em Int. J. Mod. Phys.} {\bf A1} (1986) 303.

\bibitem{Lerche:1989uy}
W.~Lerche, C.~Vafa, and N.~P. Warner, ``Chiral rings in {N=2} superconformal
  theories,'' {\em Nucl. Phys.} {\bf B324} (1989) 427.

\bibitem{Halpern:1995fy}
M.~B. Halpern and N.~A. Obers, ``Flat connections and nonlocal conserved
  quantities in irrational conformal field theory,'' {\em J. Math. Phys.} {\bf
  36} (1995) 1080--1110, \href{http://xxx.lanl.gov/abs/hep-th/9312050}{{\tt
  hep-th/9312050}}.

\bibitem{Giusto:2001sn}
S.~Giusto and M.~B. Halpern, ``Hamiltonian formulation of open {WZW} strings,''
  \href{http://xxx.lanl.gov/abs/hep-th/0101220}{{\tt hep-th/0101220}}.

\end{thebibliography}

\providecommand{\href}[2]{#2}\begingroup\raggedright\endgroup

}

\end{document}